\documentclass{aa}

\usepackage[T1]{fontenc}
\usepackage{ae,aecompl}
\usepackage{natbib,twoopt}
\usepackage{booktabs}
\usepackage{ulem}

\usepackage{graphicx}   
\usepackage{amsmath}    
\usepackage{amssymb}    
\usepackage{gensymb}    
\usepackage{multirow}
\usepackage{mathtools}
\usepackage{IEEEtrantools}
\usepackage{float}
\usepackage{rotating}
\usepackage{bm}
\usepackage{adjustbox}             
\usepackage{subcaption}            
\usepackage{threeparttable}
\usepackage{txfonts} 
\usepackage{footmisc}
\usepackage{hyperref}
\hypersetup{
    colorlinks=true,
    linkcolor=blue,
    filecolor=blue,
    urlcolor=blue,
    citecolor=blue,
}

\usepackage{pdflscape}
\usepackage{placeins}
\usepackage{scalefnt}
\usepackage{xcolor}                 

\usepackage{relsize}

\newcommand\tstrut{\rule{0pt}{2.4ex}}
\newcommand\bstrut{\rule[-1.0ex]{0pt}{0pt}}

\newcommand{\electrons}{\hbox{$\mathrm{e}^{-}$}}
\newcommand{\ergs}{\hbox{$\mathrm{erg\,s}^{-1}$}}
\newcommand{\Zsun}{\hbox{${Z}_{\odot}$}}
\newcommand{\Msun}{\hbox{$\mathrm{M}_{\rm{\odot}}$}}
\newcommand{\angstrom}{\text{\normalfont\AA}}

\newcommand{\sfr}{\hbox{$\mathrm{SFR}$}}
\newcommand{\sfruv}{\hbox{$\mathrm{SFR}_{\mathrm{UV}}$}}
\newcommand{\sfruvcorr}{$\mathrm{SFR}_{\mathrm{UV, corr}}$}
\newcommand{\sfrtot}{$\mathrm{SFR}_{\mathrm{UV}+\mathrm{IR}}$}
\newcommand{\mass}{\hbox{$\mathrm{M}_{\ast}$}}
\newcommand{\lmass}{\hbox{$\log(\mathrm{M}_{\ast} / \mathrm{M}_{\odot})$}}
\newcommand{\re}{\hbox{$R_{\rm{e}}$}}
\newcommand{\reMass}{\hbox{$R_{\rm{e}}$-$M_{\rm{*}}$}}
\newcommand{\deltare}{\hbox{$\Delta (R_{\rm{e}}$-$M_{\rm{*}})$}}
\newcommand{\sigmass}{\hbox{$\Sigma_{\mathrm{M}}$}}
\newcommand{\deltasigmass}{\hbox{$\Delta (\Sigma_{\rm{M}}$-$M_{\rm{*}}$})}
\newcommand{\sigmassMass}{\hbox{$\Sigma_{\rm{M}}$-$M_{\rm{*}}$}}
\newcommand{\sigsfr}{\hbox{$\Sigma_{\mathrm{SFR}}$}}
\newcommand{\deltasigsfr}{\hbox{$\Delta (\Sigma_{\rm{SFR}}$-$M_{\rm{*}}$})}
\newcommand{\sigsfrMass}{\hbox{$\Sigma_{\rm{SFR}}$-$M_{\rm{*}}$}}

\newcommand{\pks}{\hbox{$P_{\mathrm{KS}}$}}
\newcommand{\pvalue}{\hbox{$p-$value}}
\newcommand{\halpha}{\hbox{$H_{\alpha}$}}
\newcommand{\zlow}{\hbox{$0.3 < z < 1.1$}}
\newcommand{\zmed}{\hbox{$1 < z < 2$}}
\newcommand{\zhigh}{\hbox{$2 < z < 3.1$}}
\newcommand{\Zemiss}{\hbox{$Z_{\mathrm{emiss}}$}}
\newcommand{\Zabs}{\hbox{$Z_{\mathrm{abs}}$}}

\newcommand{\python}{\hbox{\textsc{Python}}}
\newcommand{\galfit}{\hbox{\texttt{GALFIT}}}

\newcommand{\sx}{\hbox{\texttt{SExtractor}}}
\newcommand{\galapagos}{\hbox{\texttt{GALAPAGOS}}}
\newcommand{\ks}{\hbox{K-S}}

\newcommand{\Spitzer}{\hbox{\textit{Spitzer}/MIPS}}
\newcommand{\Herschel}{\hbox{\textit{Herschel}/PACS}}

\newcommand{\BATsix}{\hbox{BAT6}}
\newcommand{\SHOALS}{\hbox{SHOALS}}

\newcommand{\SDSS}{\hbox{\textit{SDSS}}}

\newcommand{\orcid}[1]{\href{https://orcid.org/#1}{\includegraphics[width=8pt]{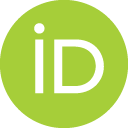}}}

\graphicspath{{Figures/}} 

\begin{document}
    \title{Are the host galaxies of Long Gamma-Ray Bursts more compact than star-forming galaxies of the field?}

    \author{B.~Schneider
    \inst{1}\orcid{0000-0003-4876-7756}
    \and E.~Le Floc'h \inst{1}
    \and M.~Arabsalmani \inst{2,3}\orcid{0000-0001-7680-509X}
    \and S.~D.~Vergani \inst{4}
    \and J.~T.~Palmerio \inst{4}\orcid{0000-0002-9408-1563}
    }

    \institute{CEA, IRFU, DAp, AIM, Universit\'e Paris-Saclay, Universit\'e Paris Cit\'e, Sorbonne Paris Cit\'e, CNRS, 91191 Gif-sur-Yvette, France \\
    \email{benjamin.schneider@cea.fr}
    \and
    Excellence Cluster ORIGINS, Boltzmannstra{\ss}e 2, 85748 Garching, Germany
    \and
    Ludwig-Maximilians-Universit\"at, Schellingstra{\ss}e 4, 80799 M\"unchen, Germany
    \and
    GEPI, Observatoire de Paris, PSL University, CNRS, 5 Place Jules Janssen, F-92190 Meudon, France
    }
    \date{Accepted XXX. Received YYY; in original form ZZZ}

  \abstract
   {Long Gamma-Ray Bursts (GRBs) offer a promising tool to trace the cosmic history of star formation, especially at high redshift where conventional methods are known to suffer from intrinsic biases. Previous studies of GRB host galaxies at low redshift showed that high surface densities of stellar mass and star formation rate (SFR) can potentially enhance the GRB production. Evaluating the effect of such stellar densities at high redshift is therefore crucial to fully control the ability of long GRBs for probing the activity of star formation in the distant Universe.}
   {We assess how the size, the stellar mass and star formation rate surface densities of distant galaxies affect their probability to host a long GRB, using a sample of GRB hosts at $z > 1$ and a control sample of star-forming sources from the field.}
   {We gather a sample of 45 GRB host galaxies at $1 < z < 3.1$ observed with the Hubble Space Telescope WFC3 camera in the near-infrared. Our subsample at $1 < z < 2$ has cumulative distributions of redshift and stellar mass consistent with the host galaxies of already-known unbiased GRB samples, while our GRB host selection at $2 < z < 3.1$ has lower statistics and is probably biased to the high end of the stellar mass function. Using the GALFIT parametric approach, we model the GRB host light profile with a Sérsic component and derive the half-light radius for 35 GRB hosts, which we use to estimate the star formation rate and stellar mass surface densities of each object. We compare the distribution of these physical quantities to the SFR-weighted properties of a complete sample of star-forming galaxies from the 3D-HST deep survey at comparable redshift and stellar mass.}
   {We show that, similarly to $z < 1$, GRB hosts are smaller in size and they have higher stellar mass and star formation rate surface densities than field galaxies at $1 < z < 2$. Interestingly, this result is robust even when considering separately the hosts of GRBs with optically-bright afterglows and the hosts of dark GRBs, as the two subsamples share similar size distributions.
   At $z > 2$ though, GRB hosts appear to have sizes and stellar mass surface densities more consistent with those characterizing the field galaxies. This may reveal an evolution with redshift of the bias between GRB hosts and the overall population of star-forming sources, although we cannot exclude that our result at $z > 2$ is also affected by the prevalence of dark GRBs in our selection.}
   {In addition to a possible trend toward low metallicity environment, other environmental properties such as stellar density appears to play a role in the formation of long GRBs, at least up to $z \sim 2$. This might suggest that GRBs require special environments to be produced.}

   \titlerunning{Are the host galaxies of LGRB more compact than star-forming galaxies of the field?}
   \authorrunning{B. Schneider et al.}
   \maketitle
%

\section{Introduction}
\label{sec:Introduction}
    Long-duration Gamma-Ray Bursts (GRBs) are extremely luminous ($\sim 10^{53}\ \ergs$) and powerful explosions with a typical prompt emission duration longer than 2 seconds. Two types of progenitors have been proposed to explain these extreme phenomena: a single massive star \citep{woosley1993a, woosley2006a, yoon2006a} well-known as the collapsar model or a binary system of massive stars \citep{fryer2005a, cantiello2007a, chrimes2020a}. In both cases, they connect long GRBs to the death of massive ($> 40\ \Msun$) and fast rotating stars. The strongest support for this association lies in multiple observations of the spatial and temporal coincidence between a type Ic-BL supernova (SN) and a GRB \citep{hjorth2003a, stanek2003a, xu2013a}. Observations of their host galaxies also support this connection by identifying that actively star-forming galaxies favor GRBs production \citep{sokolov2001a, bloom2002a, lefloch2003a, perley2013b, hunt2014a, greiner2015a, palmerio2019a} and that GRBs mostly occur in the UV-bright regions of their hosts \citep{fruchter2006a, blanchard2016a, lyman2017a}. Due to the short lifetime of massive stars ($<50$ Myr), long GRBs are linked to recent star formation activity in their host environment. The rate of GRBs could thus offer a unique opportunity to constrain the cosmic star formation rate history (CSFRH), especially at high redshifts ($z>5$) where GRBs are still detectable \citep{salvaterra2009a, tanvir2009a} and where the uncertainties affecting the estimates from UV-selected galaxies become predominant. The comparison of the two approaches reveals that at high redshifts the GRB rate predicts a substantially higher star formation rate (SFR) density than the one inferred from UV-selected galaxies \citep{kistler2008a, robertson2011a}. Recent discovery of massive dusty star-forming galaxies at $z>3$ \citep{wang2019a} points out that UV-selected galaxy samples miss these galaxies and may indeed underestimate the CSFRH at high-$z$. On the other hand, long GRBs might require specific conditions to form, depending on e.g., metallicity or local density, which could also introduce biases in the CSFRH determination.

    At the end-life of the progenitor, a high angular momentum is needed to launch the GRB jet. In order to have this critical requirement, the collapsar model requires a low metallicity ($Z < 0.3\ \Zsun$, \citealt{yoon2006a}). Indeed, stars with higher metallicity produce stronger stellar winds that remove angular momentum and inhibit GRB production. For binary system models, tidal interaction and mass transfer in binaries can spin up the system \citep{petrovic2005a} to produce the relativistic jet and thus require a lower constraint on the metallicity \citep{chrimes2020a}.
    Because GRB progenitors are not directly observable, the characterization of GRB host (GRBH) galaxies offer an indirect but precious tool to explore the GRB environment and further constrain the physical conditions which favor their formation. 
    Studies based on GRB host galaxies showed that GRBs tend to avoid high metallicity galaxies \citep{vergani2015a, perley2016b, palmerio2019a} and support the hypothesis of a bias toward low metallicity environment. For instance, \cite{palmerio2019a} found that GRB production is significantly reduced for galaxies with $Z \gtrsim 0.7\ \Zsun$.
    Spatially resolved spectroscopic studies of nearby GRB host galaxies \citep{levesque2011a, kruhler2017a} show that the integrated host metallicity may differ from the GRB site metallicity by about $0.1-0.3$~dex, which could reconcile the apparent discrepancies between the theoretical predictions of the collapsar model and the current observational constraints. On the other hand, several studies reported GRB host galaxies with super-solar metallicity \citep{levesque2010c, savaglio2012a, heintz2018a}, which questions the existence of a hard metallicity cap. Although there is a consensus that metallicity plays a role, the precise way it affects the GRB occurrence rate thus remains unclear.

    In a cosmological context, such metallicity condition would not affect the relation between GRB and cosmic star formation rate at $z\gtrsim 3$, because sub-solar metallicities are typical of galaxies in the early universe. Hence, long GRBs may trace the star formation rate in an unbiased way assuming that no other biases are involved. However, the discrepancies on the metallicity constraints reported above may also suggest other possible influences, as discussed in several studies of GRB hosts. For instance, \cite{perley2015a} suggested that GRB explosions are enhanced in intense starburst galaxies \citep[see also][]{arabsalmani2020a} in addition to a trend toward low metallicity environment.
    \cite{michalowski2016a} focused on GRB~980425 and observed clues of a possible recent atomic gas inflow toward its host that may have triggered the formation of massive stars able to produce a GRB.
    \cite{arabsalmani2015b, arabsalmani2019a} also reported evidences of a companion dwarf galaxy interacting with the host of GRB~980425. They suggested that the interaction of galaxies can favor GRB formation.
    Moreover, GRB hosts show a higher specific star formation rate (star formation rate per unit mass) compared to field galaxies \citep{salvaterra2009a, schulze2018a}. Finally, GRB hosts are found to be more compact and smaller than field galaxies \citep{conselice2005a, fruchter2006a, wainwright2007a}. In particular, \cite{kelly2014a} showed that at $z < 1$ GRBs tend to occur in compact and dense environments, that is, in galaxies with star formation and stellar mass surface densities higher than observed in field galaxies at comparable stellar mass and redshift.
    However, the existence of this trend toward more compact environments, and its link to metallicity, if any, has not been explored at $z > 1$. This remains a crucial aspect to establish the link between the long GRB rate and the SFR in the distant Universe. Furthermore, determining the influence of stellar density on the GRB occurrence rate could also shed indirect lights into our understanding of the main drivers and/or the relative importance of progenitor models in the formation of long GRBs.

    In this work, we quantify the stellar mass surface density (\sigmass) and the star formation surface density (\sigsfr) in GRB host galaxies up to $z \sim 3$ and assess how these physical properties compare with those observed in field galaxies at similar redshift. We present results based on a sample of long GRB host galaxies observed with the Hubble Space Telescope (HST) and more particularly with the Wide Field Camera 3 (WFC3) instrument in the infrared (IR) band. The high resolution images of the HST provide the possibility to measure precisely the galaxies size, avoiding contamination by nearby galaxies. Our analysis is mostly based on images obtained with the $F160W$ filter ($\lambda_{mean} \sim 1.54\, \mu$m) where the observed emission is more sensitive to the bulk of the galaxy stellar mass compared to data at shorter wavelengths, and which also minimizes the effect from dust obscuration.
    This paper is organized as follows. In Sect.~\ref{sec:Data} we introduce the GRB host galaxy sample, the control star-forming galaxy sample and the limit of completeness of both samples. In Sect.~\ref{sec:Methods} we describe the methods to derive structural and physical parameters for the GRB hosts galaxies. Section~\ref{sec:Results} presents our results and their comparison with the field galaxies. Section~\ref{sec:Discussion} discusses our results more broadly and their implications. Finally the conclusions are presented in Sect.~\ref{sec:Conclusions}. Throughout the paper, we use the $\Lambda$CDM cosmology from \cite{planckcollaboration2020a} with $\Omega_{\rm m} = 0.315$, $\Omega_{\Lambda} = 0.685$, $H_{0} = 67.4$~km~s$^{-1}$~Mpc$^{-1}$. Stellar masses (\mass) and SFRs are reported assuming a \cite{chabrier2003a} initial mass function (IMF).

\section{Data}
\label{sec:Data}
    \subsection{Sample selection}
    \label{subsec:Sample Selection}
        We consider all long GRBs with a redshift measurement (spectroscopic or photometric) in $1 < z < 4$ from J. Greiner’s database\footnote{\url{https://www.mpe.mpg.de/~jcg/grbgen.html}}. This page gathers all GRBs detected and localized since 1996 by high-energy space observatories such as HETE, INTEGRAL, Fermi and \textit{Swift}. For each GRB, the page provides a collection of information (localization, error box, $T_{90}$ and redshift if available) collected from GCN (Gamma-ray Coordinates Network) messages and refereed publications.
        We find a total of 317 GRBs in the range of redshift considered. We query these objects in the Mikulski Archive for Space Telescopes (MAST) database and select the ones performed with the WFC3/IR instrument of the HST. We extract the enhanced data products available in the Hubble Legacy Archive (HLA) database\footnote{\url{http://hla.stsci.edu/}}. These products are generated with the standard HST pipeline (\texttt{AstroDrizzle} software) which corrects geometric distortion, removes cosmic rays and combines multiple exposures. The images are north up aligned and have a final pixel scale of 0.09 arcsec. Most of the sample is located at $z < 3.1$, with one single source lying at $z = 3.5$. For this reason, we restrain our study at $1 < z < 3.1$. \\
        To verify that all HST observations have been included in the HLA database, we cross-check standard products available in the HST archive with the enhanced HLA products. Two additional observations have been found in the HST archive (GRB~060512 and GRB~100414A). However, HST observations of GRB~060512 have poor quality with visible star trails, and data of GRB~100414A correspond to another object (NGC~4698) because no observations were performed for this GRB field. We exclude these two objects in our analysis. \\ 
        Our sample is composed of 42 long GRB host galaxies observed in the $F160W$ filter.
        At $1 < z < 3.1$, we additionally find in the HLA database a total of two GRB hosts solely observed in the $F110W$ filter (GRB~070125 and GRB~080207). We include them in the final sample because the wavelength probed by this filter ($\lambda_{\rm{mean}} \sim 1.18\, \mu$m) is close to the one of $F160W$ filter ($\lambda_{\rm{mean}} \sim 1.54\, \mu$m). Therefore, we do not expect significantly different size measurements between these filters.
        We also include the peculiar GRB~090426, classified as a short GRB based on its $T_{90} < 2s$ \citep{levesque2010a} but as a long one regarding the properties of the host galaxy \citep{thone2011a}.
        Finally, we exclude the unsecured case of GRB~140331A due to multiple candidate hosts and a photometric redshift close to one. \\
        All HST observations (except for GRB~160509A) were taken at a late time after the GRB detection, when the afterglow has faded significantly. For GRB~160509A, two HST observations were performed after 35.3 and 422.1 days in the $F160W$. In the HLA database, only products for the observations at 35.3 days are available. Because of the short delay between the detection and the observations, a possible contamination of the afterglow cannot be excluded. \cite{kangas2020a} showed that the remaining afterglow at 35.3 days is very weak ($H_{F160W} = 26.07$ mag) compared to the host galaxy. We conclude for this object that the HST observations considered are not strongly affected by the GRB afterglow and that the host galaxy is assumed dominant. \\
        The final GRB host sample is shown in Table~\ref{tab:grb_sample}. It is composed of 44 bursts mainly ($\sim 90 \%$) detected by \textit{Swift}. Among them, only two (GRB~090404 and GRB~111215A) have a photometric redshift estimated from the spectral energy distribution (SED) of the host galaxy. These redshifts are less reliable than spectroscopic determinations, but since they represent only a small fraction ($< 5\%$) of the full sample, we do not expect a significant impact on our results.
        In our analysis, we divide the sample into two bins of redshift, \zmed\ and \zhigh\ to enclose the cosmic noon at $z \sim 2$ \citep{madau2014a, forsterschreiber2020a} where the cosmic star formation rate volume density has reached its maximum.

    \subsection{Host assignment}
    \label{subsec:Host_assignment}
        Since the launch of \textit{Swift} in 2004, GRB positions are often determined with an accuracy of $\sim 1$". Because long GRBs are associated with the death of massive stars \citep{hjorth2003a}, the GRB site is expected to be close to the center or the brightest region of its host galaxy \citep{fruchter2006a,blanchard2016a,lyman2017a}.
        An unambiguous way to assign a host galaxy to a GRB is to match the redshift measured from the fine-structure lines of the GRB afterglow with the redshift obtained from the emission lines of the host candidate. Unfortunately, this is not always possible, especially for dark GRBs where faint or no optical counterpart is detected. In this case, to assign the burst to its host galaxy, a standard approach is to use the probability of chance coincidence ($P_{\mathrm{cc}}$). The $P_{\mathrm{cc}}$ can be estimated from the Poisson probability of finding a galaxy in a given radius around the transient event localization \citep[see][]{bloom2002a}. Another possible approach relies on a Bayesian inference framework \citep{aggarwal2021a}. The majority of our GRBs have already been well studied in the literature. \cite{blanchard2016a} and \cite{lyman2017a} assigned host galaxies using $P_{\mathrm{cc}}$ on HST images but the coordinates of the identified hosts are not reported. As a starting point, we extract the best GRB coordinates available in the literature \citep[e.g.,][]{perley2016b}. We then let \sx\ \citep{bertin1996a} find the closest object to the best GRB position. We cross-check the objects found by \sx\ with images provided in \cite{blanchard2016a} and \cite{lyman2017a}. We successfully identify the host galaxies for the majority of our sample. Only the hosts of GRB~150314A and GRB~160509A have not yet been reported in the literature. For these two cases, we consider the Bayesian formalism of \cite{aggarwal2021a} and use the provided \python\ package \textit{astropath}. For each GRB, we first extract the best afterglow position and errors from the literature. Then, all objects within or crossing the error circle of the afterglow position are considered as possible host galaxies. Following the recommendations of \cite{aggarwal2021a}, we estimate the galaxy centroids, magnitudes and angular sizes of objects with a nonparametric approach (i.e., \sx).
        It is common to consider that the host is undetected when HST observations reveal either a blank region with no obvious source or if the detected object has a larger projected offset than typically observed for previous GRBs \citep[$>10$ kpc,][]{bloom2002a,blanchard2016a,lyman2017a}.
        For GRB~150314A and GRB~160509A, one or more extended objects with $H_{F160W} \lesssim 24$~mag can be seen within the 1.5" \textit{Swift} error box region. Based on Hubble's UDF (Ultra Deep Field), the number of sources with a limiting H-band magnitude of 24 within 1.5" is estimated to be about 0.075 \citep{rafelski2015a}.
        We therefore assume a probability of zero ($P(U) = 0$) that the host galaxy is not detected. It means that the GRB host is necessarily one of the objects detected by the HST near the afterglow position. This hypothesis is then supported by the $F160W$ magnitudes (Table~\ref{tab:grb_sample}) determined by \galfit\ \citep{peng2002a, peng2010a}, which are consistent with the other magnitudes of GRB hosts at similar redshift and stellar mass.
        For the prior probability that the object $i$ is the host galaxy, $P(O_i)$, we consider the “inverse” prior. The formalism is inspired from the $P_{\mathrm{cc}}$ calculation and gives higher prior probability to brighter candidates. In addition, the angular distance of the object from the GRB position is taken into account by the $p(\omega | O_i)$ prior and set to the “exponential” model. We assign as the host galaxy the object with the highest posterior probability. We find for GRB~150314A and GRB~160509A a probability of 0.92 and 0.54, respectively.
        Finally, a total of six GRBs are rejected because no host galaxies are detected on the HST observations. The $3 \sigma$ $F160W$ magnitude limits found by \cite{blanchard2016a, lyman2017a} reveal extremely faint hosts ($\ge 26.7$~mag). These galaxies would lie below the stellar mass completeness limit of the 3D-HST survey that we further use for our control sample. \\
        We finally note that \cite{kruhler2015a} quantified the possible number of misidentifications in their sample of 96 targets. They found a probability of $\sim 30\%$ for having 2 over 96 sources with a wrong association. Our sample is similarly composed of well-localized GRBs from \textit{Swift}, we therefore do not expect a larger number of misidentified objects nor a large impact on our results.

    \subsection{Control sample}
    \label{subsec:control_sample}
        \begin{figure}
            \centering
            \includegraphics[width=\hsize]{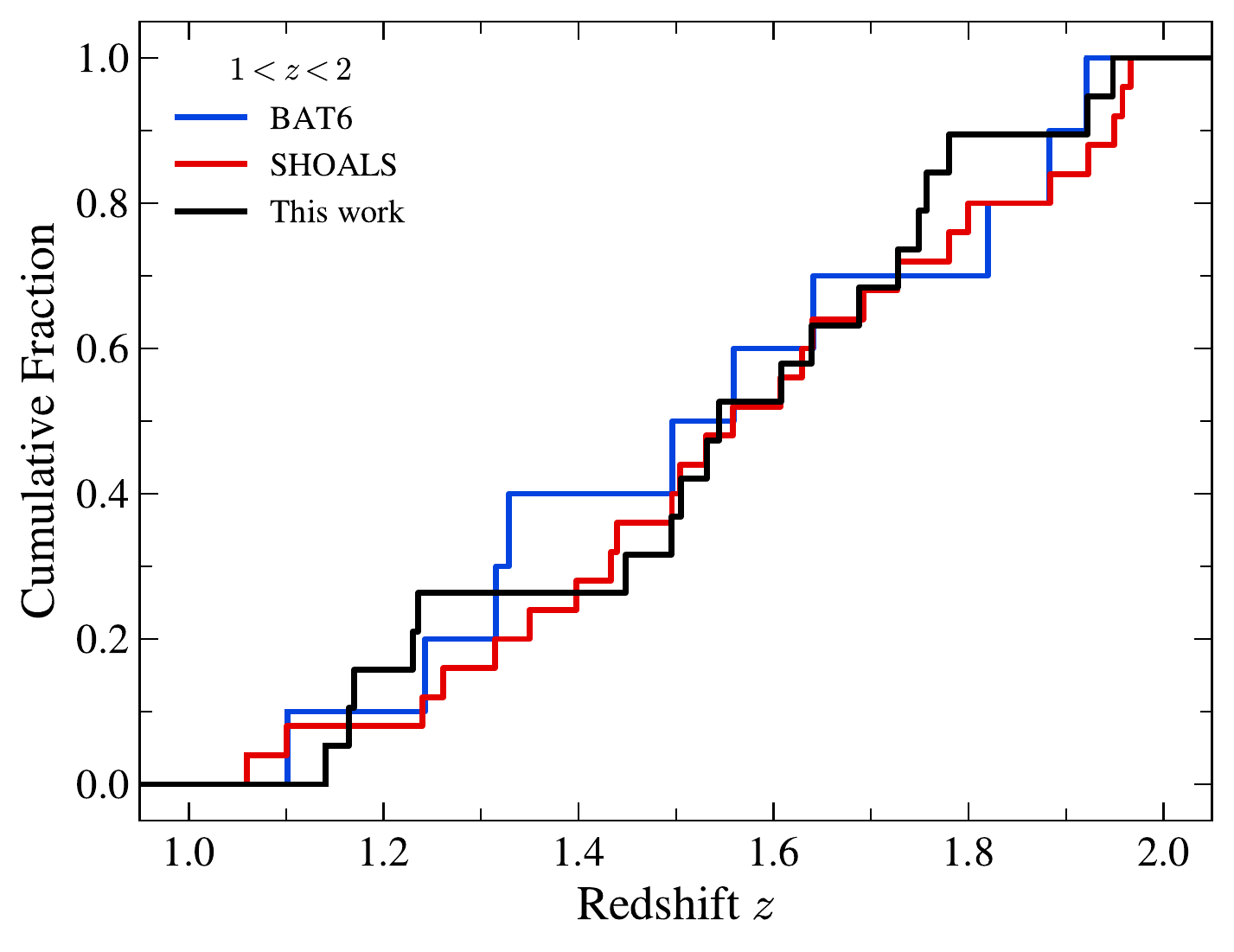} \\
            \includegraphics[width=\hsize]{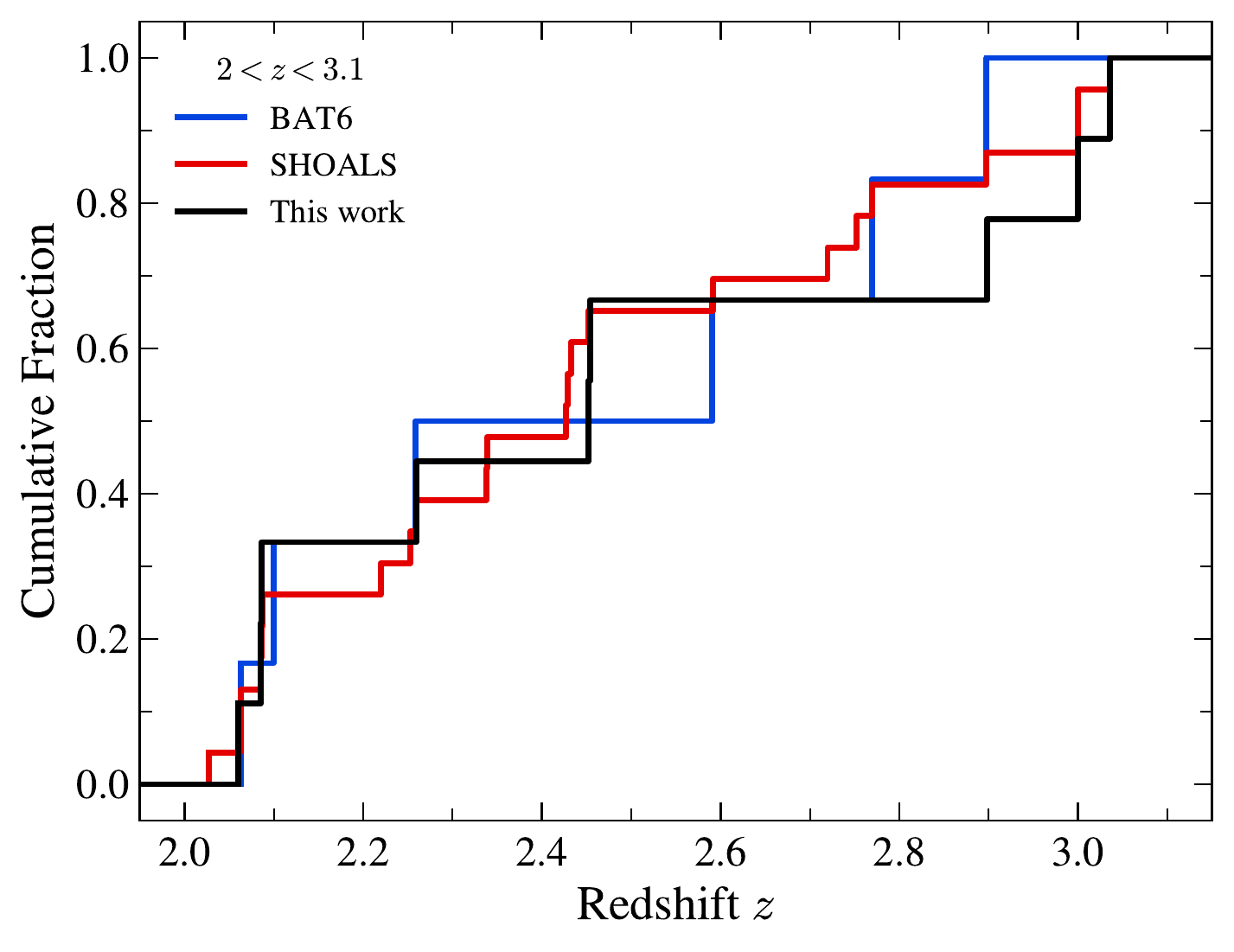}
            \caption{Redshift cumulative distributions of our GRB host sample, compared to the host galaxies of unbiased \SHOALS\ and \BATsix\ samples. Objects are divided into two bins of redshifts, \textit{top panel} with GRB hosts at \zmed\ and \textit{bottom panel} with GRB hosts at \zhigh.}
            \label{fig:cumul_unbiased_sample_z}
        \end{figure}

       To compare the properties of GRB host galaxies with those of field galaxies, we use a population of star-forming galaxies from the Cosmic Assembly Near-infrared Deep Extragalactic Legacy Survey (CANDELS) and 3D-HST surveys.
       CANDELS\footnote{\url{http://arcoiris.ucolick.org/candels/index.html}} \citep{grogin2011a, koekemoer2011a} is a deep near-infrared imaging survey carried out with the near-infrared WFC3 and optical Advanced Camera for Surveys (ACS) instruments on board the HST. The survey targets five well-known extragalactic fields (AEGIS, COSMOS, GOODS-N, GOODS-S, and UDS) and represents a total area of $\sim0.25$ degree$^2$ with more than 250,000 galaxies.
       The 3D-HST\footnote{\url{https://archive.stsci.edu/prepds/3d-hst/}} survey is a near-infrared spectroscopic survey with the WFC3 and ACS grisms on board the HST \citep{brammer2012a, momcheva2016a}. The survey provides a third dimension (i.e., redshift) for approximately 70\% of the CANDELS survey. The photometric analysis of the resulting CANDELS~+~3D-HST mosaic plus other wavelengths from ground- and space-based observatories is presented in \cite{skelton2014a}. \\
       Within the 3D-HST catalog, we select the star-forming galaxies with the rest-frame $U-V$ and $V-J$ colors method \citep{wuyts2007a, williams2009a}. For the resulting objects, the stellar masses and star formation rates considered are described in Appendix~\ref{app:main_sequence_of_galaxies}. We find consistent values with the well-established main sequence of star-forming galaxies \citep[e.g.,][]{whitaker2014a}. The structural parameters used are extracted from \cite{vanderwel2014a}. The light profile modeling is based on a single Sérsic model \citep{sersic1963a, sersic1968a} fit by \galfit\ \citep{peng2002a, peng2010a} and \galapagos\ \citep{barden2012a}. Details of the methodology are presented in \cite{vanderwel2012a}. This paper uses the data products released in the version 4.1.5, available through the 3D-HST website and described in \cite{momcheva2016a}.

    \subsection{Completeness of the samples}
    \label{subsec:Completeness of the samples}
        In deep imaging surveys, the number of sources detected is limited by the depth of images and instrument performances. At a given redshift, the resulting sample is only a subsample of all existing galaxies at that age of the Universe.
        The stellar mass completeness of the 3D-HST/CANDELS survey is discussed in \cite{tal2014a}. In our analysis, we combine for the star-forming galaxies several physical quantities such as stellar mass, SFR and half-light radius extracted from various studies \citep{momcheva2016a, whitaker2014a, vanderwel2014a}. Consequently, each object does not necessarily have an estimate for all its properties (e.g., the size if \galfit\ has not successfully converged) and it would not be fair to consider the same mass-completeness limits as determined by \cite{tal2014a}. We combine the SFR estimates obtained by adding the UV and IR light (\sfrtot) with the UV-SFR corrected from dust extinction (\sfruvcorr) to have at least one SFR value for objects having a stellar mass (see Appendix~\ref{app:main_sequence_of_galaxies} for more details) and thus preserve the mass-completeness limits determined by \cite{tal2014a}.
        Hence, the most limiting factor lies in the galaxy size measurements. \cite{vanderwel2012a} showed that accurate and precise measurements of galaxy sizes can be obtained down to a magnitude of $H_{F160W} = 24.5$~mag, corresponding to a 95\% magnitude completeness \citep{skelton2014a}. Based on that, \cite{vanderwel2014a} provided the equivalent stellar mass completeness limits per bin of 0.5 redshift. We consider the mean value of their mass-completeness limits included in our redshift bins. For \zmed\ and \zhigh, we obtain a completeness limit of $ 10^{9}$ \Msun\ and $ 10^{9.5}$ \Msun, respectively.

\begin{figure}
        \centering
            \includegraphics[width=\hsize]{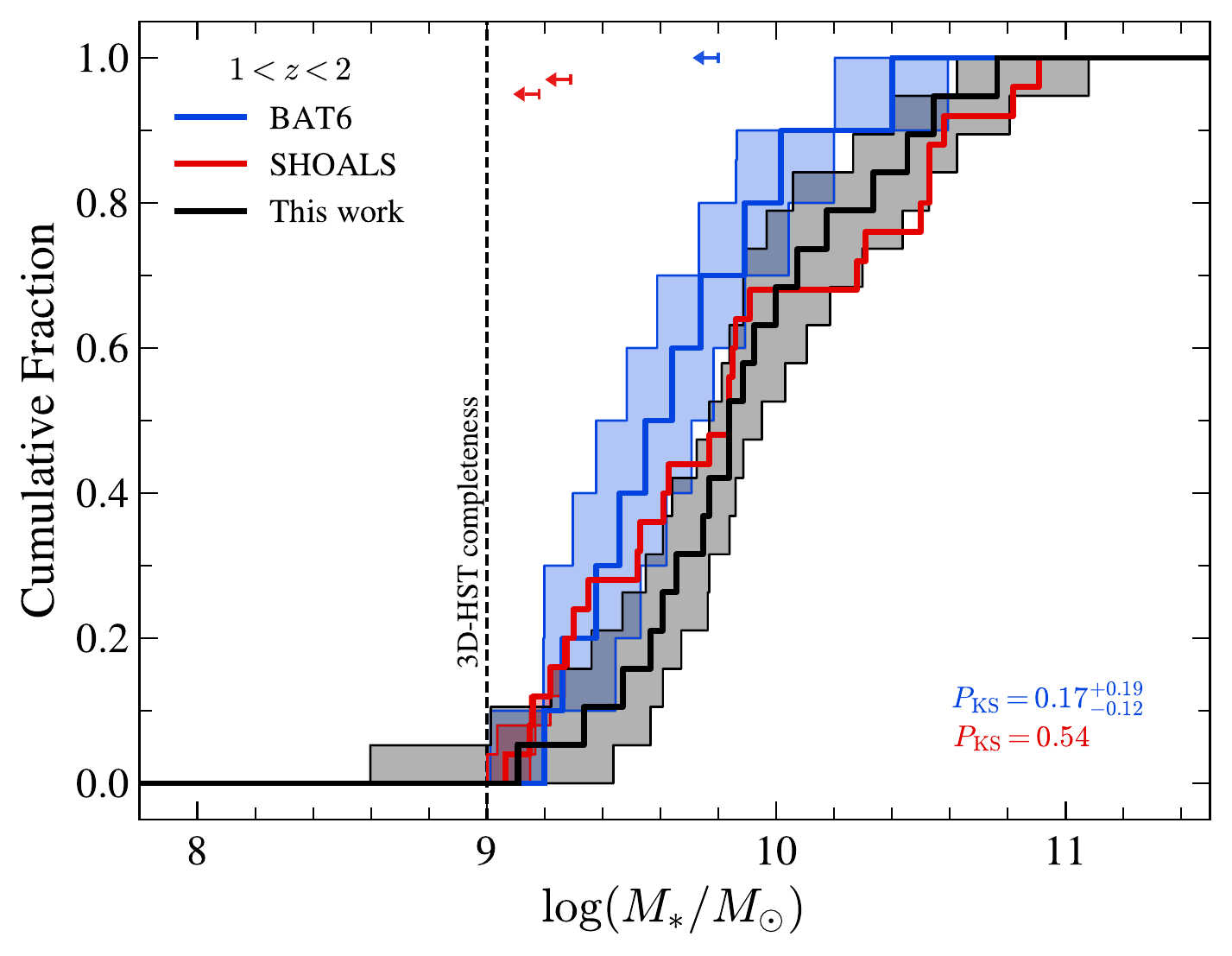} \\
            \includegraphics[width=\hsize]{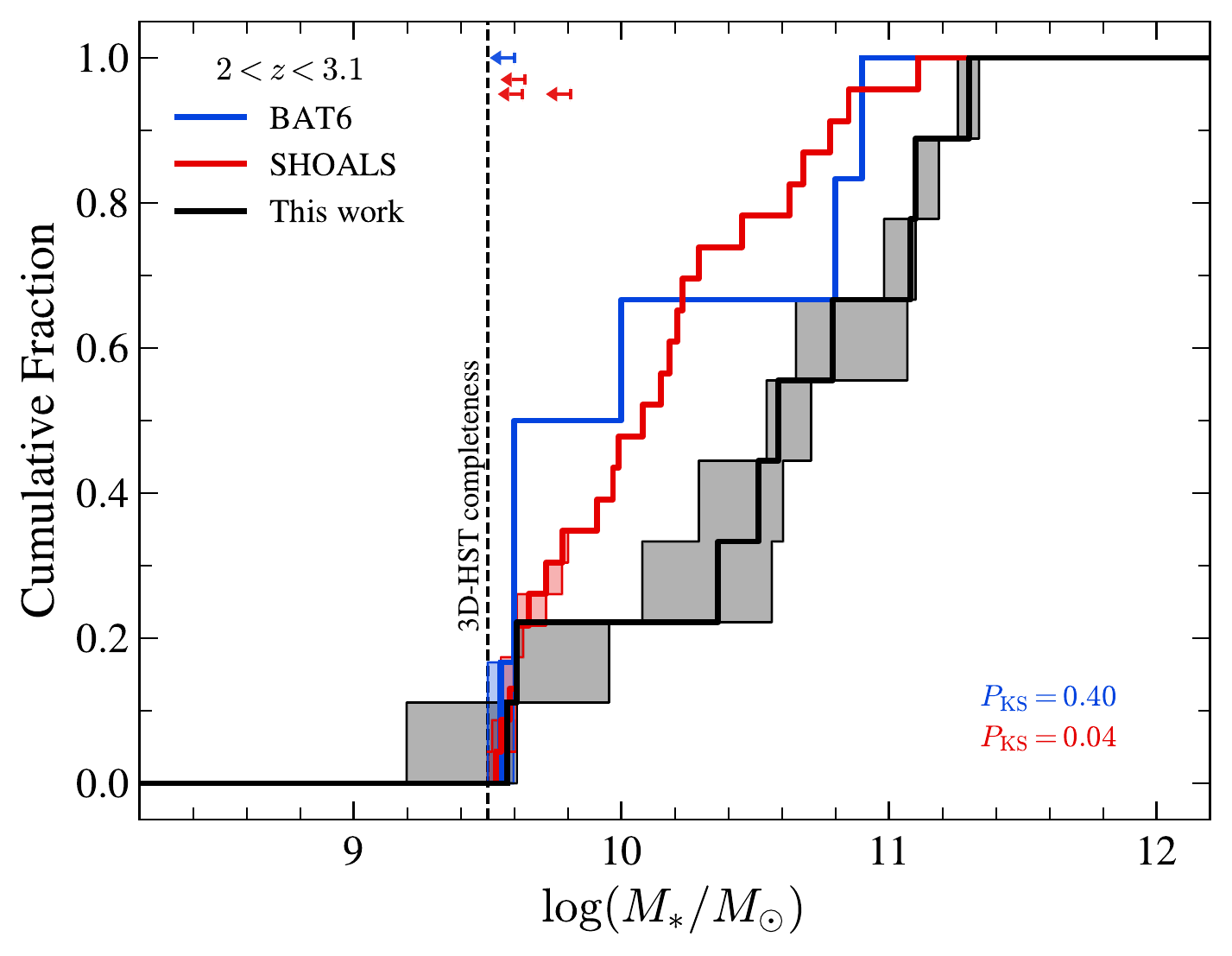}
            \caption{Stellar mass cumulative distributions of our GRB host sample, compared to the host galaxies of unbiased \SHOALS\ and \BATsix\ samples. Two bins of redshift are distinguished, \textit{top panel} with GRB hosts at \zmed\ and \textit{bottom panel} with objects at \zhigh. Upper limits are represented as arrows at the top of the plots. The $1\sigma$ uncertainty on the cumulative distribution is given by the shaded region around the curve. The \pvalue\ returned by the two-sided \ks\ test is provided in the right bottom part of both panels and color-coded according to the unbiased sample selected to compute the test. The vertical dashed line symbolizes the stellar mass completeness limit of the 3D-HST survey.}
            \label{fig:cumul_unbiased_sample_mass}
        \end{figure}

        Regarding the GRB samples, their host galaxies so far observed with HST/WFC3 at $z > 1$ represent only a small fraction of all GRBs currently identified at these redshifts. In addition, these observations result from different HST programs with distinct objectives, with a clear trend toward dark GRB host galaxies.
        For this reason, the selection function is not simple to model. To get an insight into the effect that our selection method introduces, we compare the redshift and stellar mass cumulative distribution functions (CDF) of our GRB host sample to the host galaxies of complete unbiased GRB samples of \BATsix\ \citep{salvaterra2012a} and \SHOALS\ \citep{perley2016a}.
        The CDFs are computed at \zmed\ and \zhigh\ with a method similar to the one developed by \cite{palmerio2019a} and described further in Sect.~\ref{subsec:statistical_tests}. We extract stellar masses from \cite{perley2016b} for the \SHOALS\ sample and from \cite{palmerio2019a} and \cite{perley2016b} for the \BATsix\ sample. We note for the \BATsix\ sample that all objects at \zhigh\ are included in the \SHOALS\ sample. The choice of stellar masses used for our GRB host sample is discussed later in Sect.~\ref{subsec:grb_hosts_properties}. In the further analysis, we only consider GRB hosts with a stellar mass above the mass-completeness limit of the 3D-HST sample described earlier. We therefore perform a comparison between the different GRB host samples using a similar constraint. \\
        The distributions of redshifts and stellar masses at \zmed\ and \zhigh\ are shown in Figs.~\ref{fig:cumul_unbiased_sample_z} and \ref{fig:cumul_unbiased_sample_mass}. In the literature, the stellar masses of the \SHOALS\ sample are provided without uncertainties. The uncertainties associated with the CDF, red shaded area in Fig.~\ref{fig:cumul_unbiased_sample_mass}, are only produced by the upper mass limit present in the sample.
        At \zmed, the results show good agreement between the hosts associated with the two unbiased GRB samples and ours. We note however a small offset between the \BATsix\ sample and the two other samples. At this redshift bin, the majority of GRB hosts (8/10) in the \BATsix\ sample are also included in \SHOALS. For these sources, the stellar masses reported by \citet{perley2016b} are on average larger than the ones derived by \citet{palmerio2019a} (see also Fig.~\ref{fig:SED_dispersion}). However, only 3 (over 22) stellar masses from \citet{perley2016b} are used in our own sample. This might suggest that the small offset observed at \zmed\ between our sample and \BATsix\ has a different origin than the one observed between \SHOALS\ and \BATsix.
        At \zhigh, our sample appears to be biased toward more massive GRB host galaxies.
        We note that our samples are composed of $\sim 60\%$ and $100\%$ of dark GRBs ($\beta_{\rm{ox}} < 0.5$, \citealp{jakobsson2004a}) at \zmed\ and \zhigh, respectively. The estimated fraction of dark GRBs in the overall GRB population is not well constrained but it seems that approximately $25-40\%$ of \textit{Swift} GRBs are dark \citep{fynbo2009a, greiner2011a}, and that the fraction likely increases with the host stellar mass \citep{perley2016a}.
        In our sample, the large number of dark bursts is probably due to an important part of HST programs (proposals ID: 11840, 12949, 13949) dedicated to dark GRB host galaxies.
        This population appears to be more massive, more luminous, redder and dustier than the hosts of optically-bright GRBs \citep[e.g.,][]{kruhler2011b, svensson2012a, perley2013a, perley2016b, chrimes2019a}. This likely explains why GRB hosts at the high-mass end are over-represented in our sample at \zhigh, compared to the mass distribution of host galaxies drawn from unbiased GRB samples. \\
        To summarize, we conclude that our two subsamples are globally consistent with unbiased populations of GRBs previously studied, although we note a trend for GRB hosts with larger stellar masses in our highest redshift bin.

\section{Methods and measurements}
\label{sec:Methods}

  \subsection{\galfit\ modeling of GRB hosts}
  \label{subsec:Galaxy_structural_parameters}

    \subsubsection{Profile fitting}
    \label{subsub:profile_fitting}

        We determine the structural parameters of the GRB host galaxies using a parametric approach based on \galfit\ \citep{peng2002a, peng2010a}. \galfit\ is a software using a 2D fitting algorithm to model the surface brightness of galaxies. It allows the fitting of commonly used astronomical brightness profiles including exponential, Sérsic, Nuker, Gaussian, King, Moffat, and PSF. We fit the GRB host galaxies using a unique single Sésic profile to have a similar approach to the 3D-HST sample \citep{vanderwel2012a, vanderwel2014a}. Because we have more than a few galaxies to analyze, we automate the following process in \python. First, we create a cutout of $200 \times 200$ pixels around the host galaxy position. We then run \sx\ \citep{bertin1996a} on the resulting cutout to detect all objects present in the image. We use the segmentation map returned by \sx\ to mask all unnecessary sources. Close, large and/or bright sources to the target object are fit simultaneously to reduce their possible contamination. However, fitting many objects increases the number of free parameters and can make \galfit\ converge to a local minimum. To choose the neighboring objects to be included in the analysis, we use similar conditions described in \cite{vikram2010a}. We observe that their conditions based on the isophotal surface and semi-major axis of objects give good results.
        For the remaining unmasked objects, we model them by using single Sérsic profile. We initialize the parameters of each component to the values returned by \sx\ through MAG\_AUTO, FLUX\_RADIUS, ELONGATION, and THETA\_IMAGE. We note that the empirical formula, $\re = 0.162\times\mathrm{FLUX\_RADIUS}^{1.87}$, determined by \cite{haussler2007a} can help to converge in some cases. Finally, we start the Sérsic index at an exponential profile ($n = 1$). If \galfit\ does not converge, we progressively increase the value.

        ~\sx\ tends to overestimate the sky level \citep{haussler2007a}. For this reason, we set the input sky level at the \sx\ value and let \galfit\ optimize simultaneously the sky value and the other components.
        We also let \galfit\ determine internally its own sigma image (noise map). The calculation is described in the \galfit\ user's manual (Eq. 33). It takes into account the Poisson source noise in addition to the uncertainty on the sky estimation.
        HLA products are given in electrons/s and have to be converted in electrons before computing the noise map (\galfit\ requirement). We multiply the image by the EXPTIME keyword from the fits header to go back in \electrons\ unit.
        ~\galfit\ considers two additional keywords from the fits header: GAIN (detector gain) and NCOMBINE (number of combined images). As we are already in electrons unit (not in counts), we set the GAIN keyword to 1. For HLA products, the EXPTIME value already includes the total exposure time from each individual frame. We thus set the NCOMBINE keyword to 1.

        ~\galfit\ needs the instrumental response, also known as point spread function (PSF), to convolve its models and improve the fitting process. To create a PSF model for the HST, three methods are possible: use an empirical model by stacking isolated and bright point-like sources from observations, use a synthetic model from \texttt{TinyTim} modeling software or use a combination of the two.
        Models created by \texttt{TinyTim} \citep{krist2011a} are often not adapted for data analysis due to instrumental effects such as spacecraft jitter or instrument breathing. They need to be corrected for a better matching.
        Moreover, it is not feasible to derive an empirical PSF model for each GRB host image. Some GRB host fields are very poor in stars (e.g., GRB~060719). The resulting PSF models would have a low signal to noise ratio (S/N) and may introduce artifact in the \galfit\ model. To obtain a PSF model with a high S/N, we extract and combine the stars from all GRB host fields. We isolate a total of 35 stars that we provide to \texttt{PSFEx} \citep{bertin2011a} to generate a PSF.
        We investigate the possible effects of the PSF modeling. To do that, we apply our wrapper with different PSF models on all GRB hosts at \zmed. We use two PSFs derived in a rich-stars and poor-stars fields in addition to the one combining stars from all fields. The three PSFs have a similar radius profile but the S/N is progressively degraded as the number of stars used to generate the PSF decreases.
        We find a good agreement for all parameters, only the Sérsic index varies with the PSF used. It tends to increase as the S/N of the PSF decreases. As our study is mainly focused on the half-light radius of GRB host galaxies. We conclude that using the PSF combining stars from multiple fields would not significantly affect our results.

        We investigate if our values determined by \galfit\ are consistent with those inferred by \cite{vanderwel2014a}. Our measurements on the randomly selected objects from the \hbox{3D-HST} catalog show good agreement with their estimates (see Appendix~\ref{app:3D-HST_comparison} for details). The half-light radii are recovered within 10\% at a magnitude of 21.5 (\textit{bottom panel} of Fig.~\ref{fig:comparison_3D-HST}). We then tend to progressively overestimate the \re\ as the magnitude increases until reaching 50\% at a $F160W$ magnitude of 26. Given that our GRB hosts above the 3D-HST mass-completeness limit have magnitudes below 25, we conclude that our fitting procedure is consistent with the one of \cite{vanderwel2014a} and that the comparison between GRB hosts and 3D-HST objects does not suffer from a strong systematic bias.

    \subsubsection{Uncertainties}
    \label{subsub:uncertainties}
        It is well known that \galfit\ tends to underestimate the uncertainties associated with the model parameters \citep{haussler2007a}. To improve the uncertainty estimates for GRB host models, we use a Monte Carlo (MC) approach. First, we consider the best-fitting models returned by \galfit\ to create artificial sources. We then inject these sources into randomly selected \hbox{50-pixels}\footnote{We cut all models when the flux goes below $0.5 \%$ of the maximum and note that $\sim 80 \%$ objects have a final size lower than $50 \times 50$ pixels.} empty regions of the science image. For all objects, the box size is maintained constant to probe environments with similar neighboring objects. We perform one hundred realizations for each object. Finally, the uncertainties are given by the standard deviation between the realizations and the best model.
        This method mainly captures the uncertainty from the sky estimation. We find for most of the objects a higher uncertainties than those of \galfit, especially for the magnitudes and half-light radii. In some cases when the S/N becomes small or the neighbor contamination is dominant, our MC method determines an error lower than the one derived by \galfit. For this reason, we consider in our analysis the largest uncertainty returned by \galfit\ or the MC approach.

    \subsubsection{Alternative approach}
    \label{subsub:alternative_approach}
        If \galfit\ does not converge, we use an alternative procedure to obtain an equivalent \galfit\ model. First, we run \galfit\ with \re\ fixed at the \sx\ value. If \galfit\ successfully converges to a realistic model (no parameters between `*' and $n < 8$), we re-run \galfit\ with all parameters except \re\ fixed at the new model values.
        Using this method, we can estimate for each object (which has not converged with the standard procedure) a \galfit\ model and its uncertainties. The models are thus consistent with the standard approach, except that they are constrained by the \sx\ input value. We used this method for two objects of the full sample (see Sect.~\ref{subsec:grb_hosts_properties}).

  \subsection{GRB hosts properties}
  \label{subsec:grb_hosts_properties}

  \subsubsection{Structural parameters}
  \label{subsubsec:grb_structural parameters}
    
    \begin{figure}
        \centering
        \includegraphics[width=\hsize]{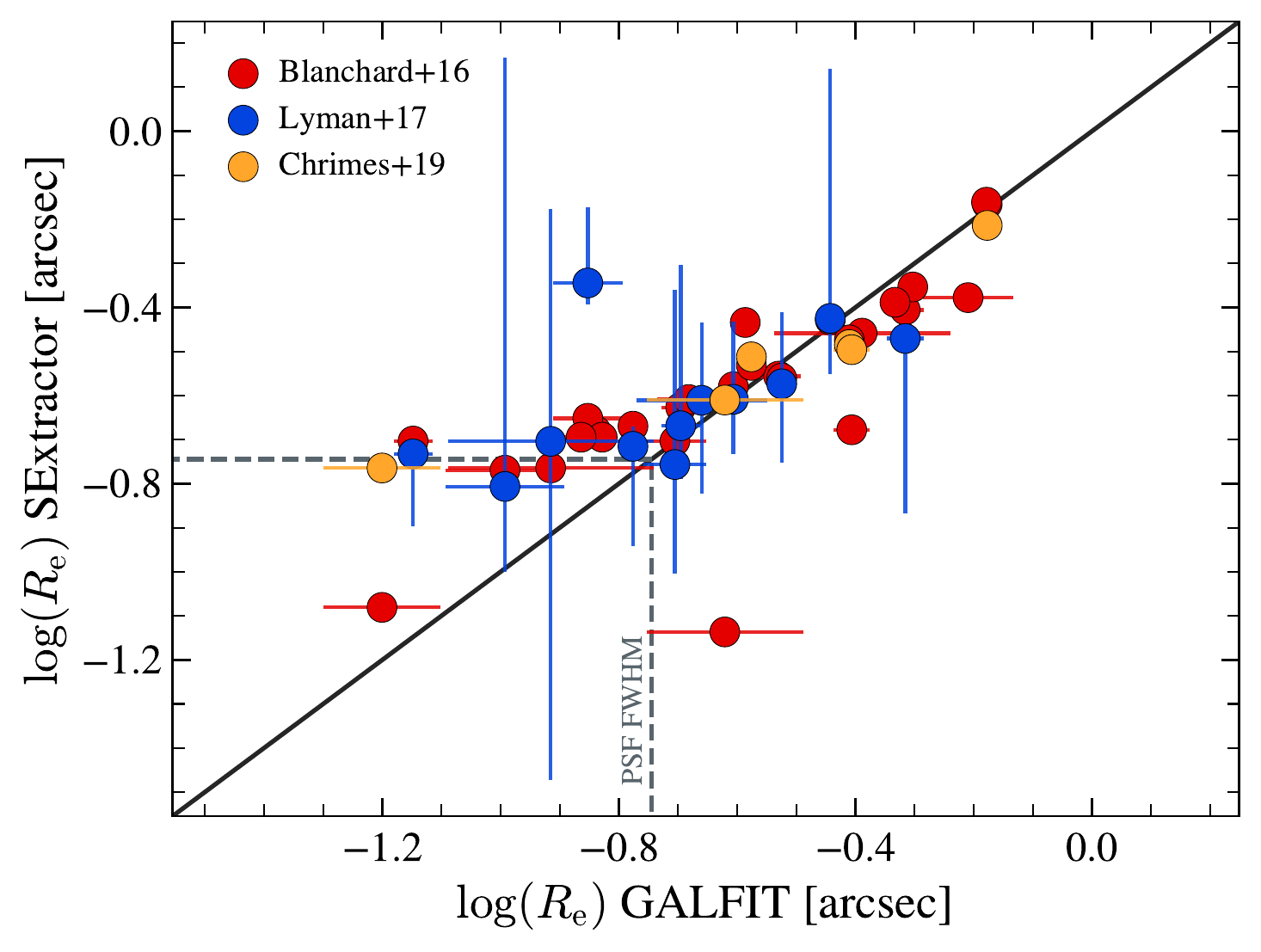}
        \caption{Half-light radius of GRB hosts estimated by \galfit\ in this work (\textit{x axis}) compared to estimates based on \sx\ extracted from the literature (\textit{y axis}). The FWHM of the PSF is visible as a gray dashed line.}
        \label{fig:re_comparison}
    \end{figure}

    \begin{table*}
     \caption{Physical and structural properties of GRB hosts at $1 < z < 3.1$.}
     \renewcommand{\arraystretch}{1.25}
     \centering
     \begin{tabular}{lcccccccccc}
      \hline \hline
        GRB & $z$ & $\log(\mathrm{M}_{\ast} / \mathrm{M}_{\odot})$ & $\mathrm{SFR} \, (\mathrm{M}_{\odot}/\mathrm{yr})$ & AB Mag & $\mathrm{R}_e$ (arcsec) & $n$ & $q$ & Ref. \\
      \hline
        050315                    & 1.95                                      & 9.77                     & > 7.57                   & 23.67 $\pm$ 0.08 & 0.20 $\pm$ 0.02 & 1.86 $\pm$ 0.52 & 0.37 $\pm$ 0.04 & 9, 1 \tstrut \\
        \textbf{050408}           & 1.24                                      & 9.37$^{+0.21}_{-0.24}$   & \dots                    & 23.48 $\pm$ 0.07 & 0.30 $\pm$ 0.03 & 1.01 $\pm$ 0.41 & 0.19 $\pm$ 0.06 & 1 \\
        060502A \tablefootmark{c} & 1.51                                      & \dots                    & \dots                    & \dots            & \dots           & \dots           & \dots           & \\
        \textbf{060719}           & 1.53                                      & 9.84                     & 7.1$^{+18.9}_{-3.9}$     & 23.24 $\pm$ 0.08 & 0.21 $\pm$ 0.03 & 4.12 $\pm$ 0.99 & 0.37 $\pm$ 0.06 & 9, 10 \\
        \textbf{060814}           & 1.92                                      & 10.43$^{+0.12}_{-0.12}$  & 56.0$^{+9}_{-9}$         & 23.04 $\pm$ 0.03 & 0.29 $\pm$ 0.01 & 1.61 $\pm$ 0.18 & 0.36 $\pm$ 0.02 & 2, 2 \\
        061007                    & 1.26                                      & 8.9$^{+0.4}_{-0.5}$      & 4.4$^{+6.2}_{-2.1}$      & 23.70 $\pm$ 0.05 & 0.36 $\pm$ 0.02 & 0.60 $\pm$ 0.10 & 0.50 $\pm$ 0.04 & 4, 7 \\
        070125 \tablefootmark{b}  & 1.55                                      & \dots                    & \dots                    & \dots            & \dots           & \dots           & \dots           & \\
        \textbf{070208}           & 1.17                                      & 9.87$^{+0.26}_{-0.19}$   & \dots                    & 22.35 $\pm$ 0.01 & 0.26 $\pm$ 0.01 & 0.36 $\pm$ 0.03 & 0.60 $\pm$ 0.01 & 1 \\
        \textbf{070306}           & 1.50                                      & 10.48$^{+0.06}_{-0.06}$  & 38.0$^{+2}_{-2}$         & 21.71 $\pm$ 0.04 & 0.14 $\pm$ 0.01 & 4.50 $\pm$ 0.73 & 0.35 $\pm$ 0.01 & 2, 10 \\
        071122                    & 1.14                                      & 9.75$^{+0.26}_{-0.22}$   & \dots                    & 22.56 $\pm$ 0.06 & 0.48 $\pm$ 0.04 & 1.45 $\pm$ 0.17 & 0.47 $\pm$ 0.02 & 1 \\
        080319C                   & 1.95                                      & 8.82$^{+0.37}_{-0.52}$   & \dots                    & 25.35 $\pm$ 0.09 & 0.09 $\pm$ 0.03 & 0.24 $\pm$ 1.11 & 0.64 $\pm$ 0.28 & 1 \\
        \textbf{080325}           & 1.78                                      & 10.75$^{+0.07}_{-0.07}$  & 66.09$^{+24.3}_{-24.3}$  & 22.50 $\pm$ 0.02 & 0.50 $\pm$ 0.01 & 0.31 $\pm$ 0.03 & 0.61 $\pm$ 0.02 & 5, 11 \\
        080520                    & 1.55                                      & 9.2$^{+0.27}_{-0.31}$    & \dots                    & 24.06 $\pm$ 0.12 & 0.14 $\pm$ 0.02 & 2.14 $\pm$ 0.85 & 0.72 $\pm$ 0.12 & 1 \\
        \textbf{080603A}          & 1.69                                      & 10.04$^{+0.44}_{-0.23}$  & \dots                    & 22.75 $\pm$ 0.04 & 0.15 $\pm$ 0.01 & 1.69 $\pm$ 0.26 & 0.75 $\pm$ 0.05 & 1 \\
        \textbf{080605}           & 1.64                                      & 10.09$^{+0.15}_{-0.15}$  & 44.9$^{+22.9}_{-22.9}$   & 22.34 $\pm$ 0.04 & 0.07 $\pm$ 0.01 & 4.48 $\pm$ 1.51 & 0.33 $\pm$ 0.07 & 5, 10 \\
        080707                    & 1.23                                      & 9.68$^{+0.27}_{-0.24}$   & \dots                    & 22.87 $\pm$ 0.05 & 0.25 $\pm$ 0.01 & 2.04 $\pm$ 0.24 & 0.37 $\pm$ 0.02 & 1 \\
        080805                    & 1.51                                      & 9.53$^{+0.22}_{-0.22}$   & 20.6$^{+12}_{-12}$       & 23.09 $\pm$ 0.04 & 0.30 $\pm$ 0.02 & 1.77 $\pm$ 0.23 & 0.31 $\pm$ 0.04 & 5, 10 \\
        080928 \tablefootmark{b}  & 1.69                                      & \dots                    & \dots                    & \dots            & \dots           & \dots           & \dots           & \\
        081008 \tablefootmark{b}  & 1.97                                      & \dots                    & \dots                    & \dots            & \dots           & \dots           & \dots           & \\
        \textbf{090113}           & 1.75                                      & 9.89                     & 17.9$^{+10.1}_{-4.8}$    & 22.87 $\pm$ 0.02 & 0.27 $\pm$ 0.01 & 1.11 $\pm$ 0.08 & 0.74 $\pm$ 0.02 & 8, 10 \\
        \textbf{090407}           & 1.45                                      & 10.02$^{+0.11}_{-0.11}$  & 14.06$^{+4.87}_{-4.87}$  & 22.92 $\pm$ 0.04 & 0.39 $\pm$ 0.02 & 1.16 $\pm$ 0.15 & 0.31 $\pm$ 0.02 & 5, 10 \\
        090418A                   & 1.61                                      & 9.61                     & \dots                    & 23.58 $\pm$ 0.04 & 0.17 $\pm$ 0.01 & 1.27 $\pm$ 0.26 & 0.43 $\pm$ 0.04 & 9 \\
        091208B \tablefootmark{b} & 1.06                                      & \dots                    & \dots                    & \dots            & \dots           & \dots           & \dots           & \\
        \textbf{100615A}          & 1.40                                      & 8.6$^{+0.2}_{-0.2}$      & 8.6$^{+13.9}_{-4.4}$     & 23.90 $\pm$ 0.04 & 0.06 $\pm$ 0.01 & 3.50 $\pm$ 1.65 & 0.36 $\pm$ 0.12 & 4, 10 \\
        \textbf{120119A}          & 1.73                                      & 9.58$^{+0.14}_{-0.14}$   & 25.5$^{+14.1}_{-14.1}$   & 23.24 $\pm$ 0.12 & 0.14 $\pm$ 0.02 & 5.28 $\pm$ 2.15 & 0.49 $\pm$ 0.06 & 5, 10 \\
        \textbf{140331A}          & 1.00 \tablefootmark{a} $^{+0.11}_{-0.04}$ & 11.22$^{+0.11}_{-0.17}$  & 5.3$^{+4.3}_{-2.4}$      & \dots            & \dots           & \dots           & \dots           & 6, 6 \\
        150314A                   & 1.76                                      & 10.01$^{+0.45}_{-0.26}$  & \dots                    & 23.00 $\pm$ 0.07 & 0.27 $\pm$ 0.03 & 3.05 $\pm$ 0.48 & 0.73 $\pm$ 0.04 & 1 \\
        \textbf{160509A}          & 1.17                                      & 9.8$^{+0.26}_{-0.22}$    & \dots                    & 22.53 $\pm$ 0.03 & 0.26 $\pm$ 0.01 & 1.98 $\pm$ 0.13 & 0.36 $\pm$ 0.03 & 1 \bstrut \\
     \hline
        \textbf{050401}           & 2.90                                      & 9.61                     & > 3.17                   & 25.03 $\pm$ 0.09 & 0.10 $\pm$ 0.02 & 2.45 $\pm$ 1.93 & 0.30 $\pm$ 0.20 & 9, 1 \tstrut \\
        050406X \tablefootmark{c} & 2.44                                      & \dots                    & > 1.69                   & \dots            & \dots           & \dots           & \dots           & 1 \\
        060124                    & 2.30                                      & 8.7$^{+0.47}_{-0.54}$    & \dots                    & 25.83 $\pm$ 0.21 & 0.12 $\pm$ 0.05 & 0.39 $\pm$ 0.99 & 0.99 $\pm$ 0.43 & 1 \\
        \textbf{070521}           & 2.09                                      & 10.65$^{+0.21}_{-0.002}$ & 49.85$^{+72.33}_{-2.86}$ & 22.93 $\pm$ 0.18 & 0.22 $\pm$ 0.06 & 5.92 $\pm$ 2.08 & 0.51 $\pm$ 0.06 & 3, 3 \\
        \textbf{070802}           & 2.45                                      & 9.57$^{+0.19}_{-0.19}$   & 32.2$^{+17.8}_{-17.8}$   & 23.74 $\pm$ 0.28 & 0.41 $\pm$ 0.14 & 3.76 $\pm$ 1.48 & 0.63 $\pm$ 0.09 & 5, 10 \\
        \textbf{071021}           & 2.45                                      & 11.08$^{+0.05}_{-0.05}$  & 90.0$^{+5}_{-5}$         & 23.20 $\pm$ 0.05 & 0.30 $\pm$ 0.02 & 1.68 $\pm$ 0.26 & 0.40 $\pm$ 0.03 & 2, 2 \\
        071031 \tablefootmark{b}  & 2.69                                      & \dots                    & \dots                    & \dots            & \dots           & \dots           & \dots           & \\
        \textbf{080207}           & 2.09                                      & 11.3$^{+0.02}_{-0.02}$   & 250.0$^{+13}_{-13}$      & 23.38 $\pm$ 0.84 & 0.67 $\pm$ 0.02 & 0.33 $\pm$ 0.05 & 0.76 $\pm$ 0.03 & 2, 2 \\
        080603B \tablefootmark{b} & 2.69                                      & \dots                    & \dots                    & \dots            & \dots           & \dots           & \dots           & \\
        \textbf{080607}           & 3.04                                      & 10.44$^{+0.13}_{-0.13}$  & 35.2$^{+13.9}_{-13.9}$   & 24.01 $\pm$ 0.19 & 0.62 $\pm$ 0.11 & 1.99 $\pm$ 0.37 & 0.56 $\pm$ 0.05 & 5, 5 \\
        081121                    & 2.51                                      & 9.24                     & \dots                    & 24.70 $\pm$ 0.09 & 0.20 $\pm$ 0.02 & 1.63 $\pm$ 0.60 & 0.26 $\pm$ 0.07 & 9 \\
        \textbf{081221}           & 2.26                                      & 10.58$^{+0.02}_{-0.02}$  & 35.0$^{+2}_{-2}$         & 23.23 $\pm$ 0.03 & 0.46 $\pm$ 0.01 & 0.29 $\pm$ 0.05 & 0.37 $\pm$ 0.02 & 2, 2 \\
        \textbf{090404}           & 3.00 \tablefootmark{a} $^{+0.83}_{-1.82}$ & 11.1                     & 381.0                    & 23.74 $\pm$ 0.06 & 0.66 $\pm$ 0.03 & 0.68 $\pm$ 0.08 & 0.33 $\pm$ 0.02 & 13, 13 \\
        090426S                   & 2.61                                      & 9.0$^{+0.46}_{-0.5}$     & 14.4$^{+2}_{-2}$         & 25.53 $\pm$ 0.14 & 0.05 $\pm$ 0.03 & 2.25 $\pm$ 4.08 & 0.14 $\pm$ 0.57 & 1, 14 \\
        \textbf{110709B}          & 2.09                                      & 9.2                      & \dots                    & 24.58 $\pm$ 0.21 & 0.24 $\pm$ 0.07 & 3.87 $\pm$ 1.85 & 0.60 $\pm$ 0.13 & 9 \\
        \textbf{111215A}          & 2.06 \tablefootmark{a} $^{+0.10}_{-0.16}$ & 10.5$^{+0.1}_{-0.2}$     & 34.0$^{+33}_{-13}$       & 22.41 $\pm$ 0.05 & 0.39 $\pm$ 0.03 & 1.92 $\pm$ 0.23 & 0.53 $\pm$ 0.02 & 12, 12 \bstrut \\
     \hline
     \label{tab:grb_sample}
    \end{tabular}
    \tablefoot{\tablefoottext{a}{Photometric redshift}; \tablefoottext{b}{No host detected}; \tablefoottext{c}{\galfit\ has not converged}. Names in bold are dark GRBs.}
    \tablebib{(1)~This work; (2)~\citet{hsiao2020a}; (3)~\citet{hashimoto2019a}; (4)~\citet{palmerio2019a}; (5)~\citet{corre2018a}; (6)~\citet{chrimes2018a}; (7)~\citet{vergani2017a}; (8)~\citet{kruhler2017b}; (9)~\citet{perley2016b}; (10)~\citet{kruhler2015a}; (11)~\citet{hashimoto2015a}; (12)~\citet{vanderhorst2015a}; (13)~\citet{hunt2014a}; (14)~\citet{levesque2010a}.}
    \end{table*}

    The structural parameters and their uncertainties are presented in Table \ref{tab:grb_sample}. We provide for each host galaxy the $F160W$ AB magnitude, the half-light radius, the Sérsic index and the axis ratio returned by \galfit. A total of 35/37 GRB host galaxies converge successfully to realistic parameters. The models and the residuals maps are visible in Appendix~\ref{app:galfit_models} (Figs.~\ref{fig:galfit_models_1}-\ref{fig:galfit_models_3}). \\
    We use a specific treatment for the host galaxy of GRB~080319C. \cite{perley2009a} and \cite{lyman2017a} reported that a bright foreground galaxy is probably superimposed on the true host galaxy. The redshift of this object was determined from absorption lines in the GRB afterglow. No spectroscopic observations were performed to confirm the association with the host galaxy. The results of \galfit\ using a single Sérsic model show a residual source located in the South-East of the main object. This source is consistent with the GRB position and supports the hypothesis that an object is overlapping the true host. We use two Sérsic components to model and mitigate the contamination of the superimposed object. We then add a single Sérsic component to the overall \galfit\ model to fit the residual source near the GRB location.
    The host galaxies of GRB~070802 and GRB~090404 do not converge to realistic parameters ($n > 8$) with the standard approach. We note that bright sources are close to the host galaxy and likely contaminate it. For these objects, we use the procedure described in Sect.~\ref{subsec:Galaxy_structural_parameters} where the \re\ is maintained at the \sx\ value.
    Finally, the host galaxies of GRB~060502A and 050406X do not converge even when keeping the \re\ fixed at the first guess of \sx. These objects are very faint sources with a low S/N and might diverge numerically easily due to contamination by neighboring objects.

    \cite{blanchard2016a}, \cite{lyman2017a} and \cite{chrimes2019a} reported the measurement of half-light radii using \sx\ for GRB hosts mainly observed by the WFC3 in the $F160W$ filter. A fraction of our GRB hosts matches their objects. The comparison between their \sx\ and our \galfit\ values is shown in Fig.~\ref{fig:re_comparison}. We find good agreement for objects with a \re\ greater than the full width at half maximum (FWHM) of the PSF. For objects with a half-light radius derived by \sx\ and close to the FWHM, we find that \galfit\ returns smaller \re. We expect this behavior because \galfit\ convolves its models with the PSF function. It can therefore capture smaller structures of the galaxy.

    \subsubsection{Stellar mass}
    \label{subsubsec:stellar_mass}
\begin{figure}[t!]
        \centering
        \includegraphics[width=\hsize]{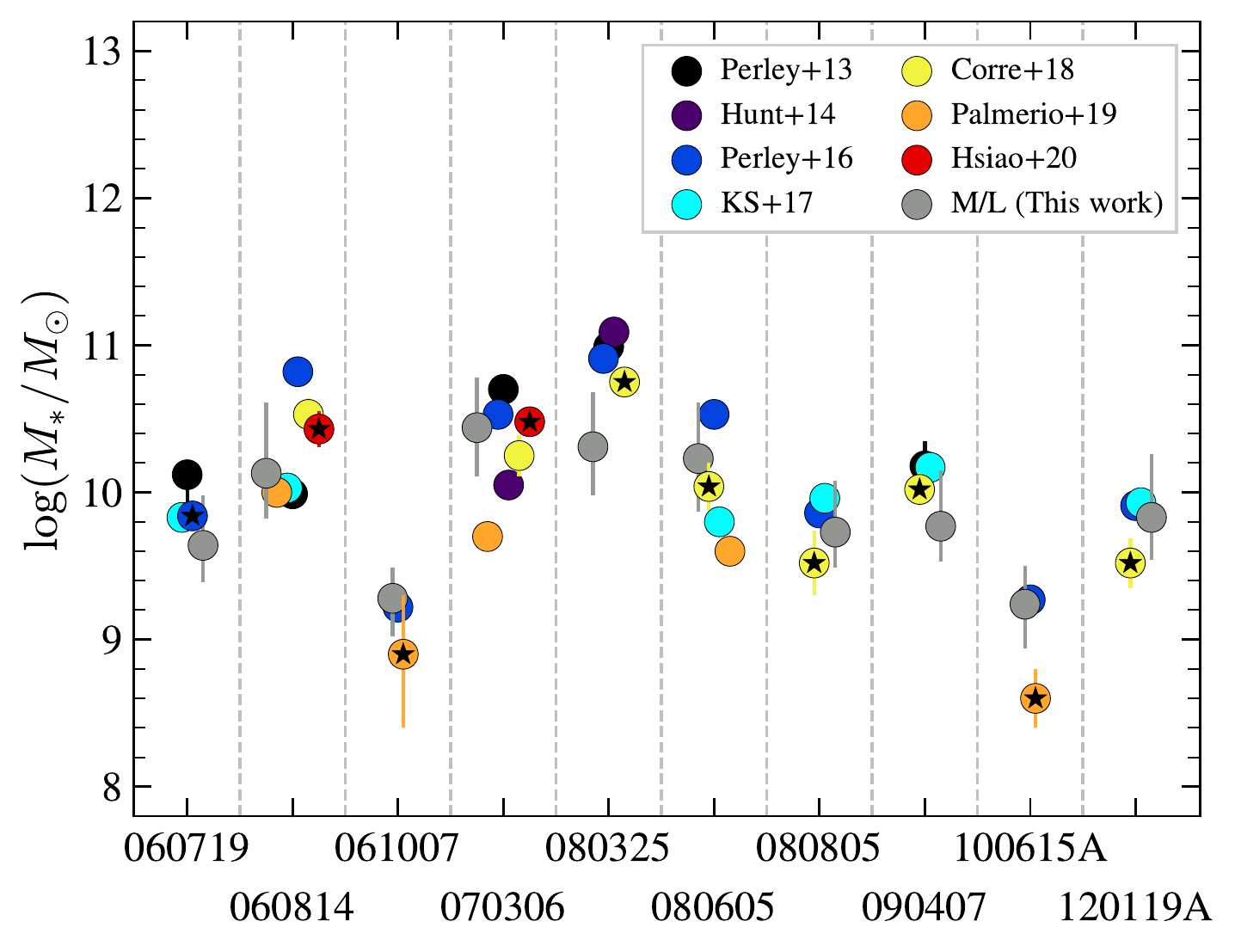}
        \includegraphics[width=\hsize]{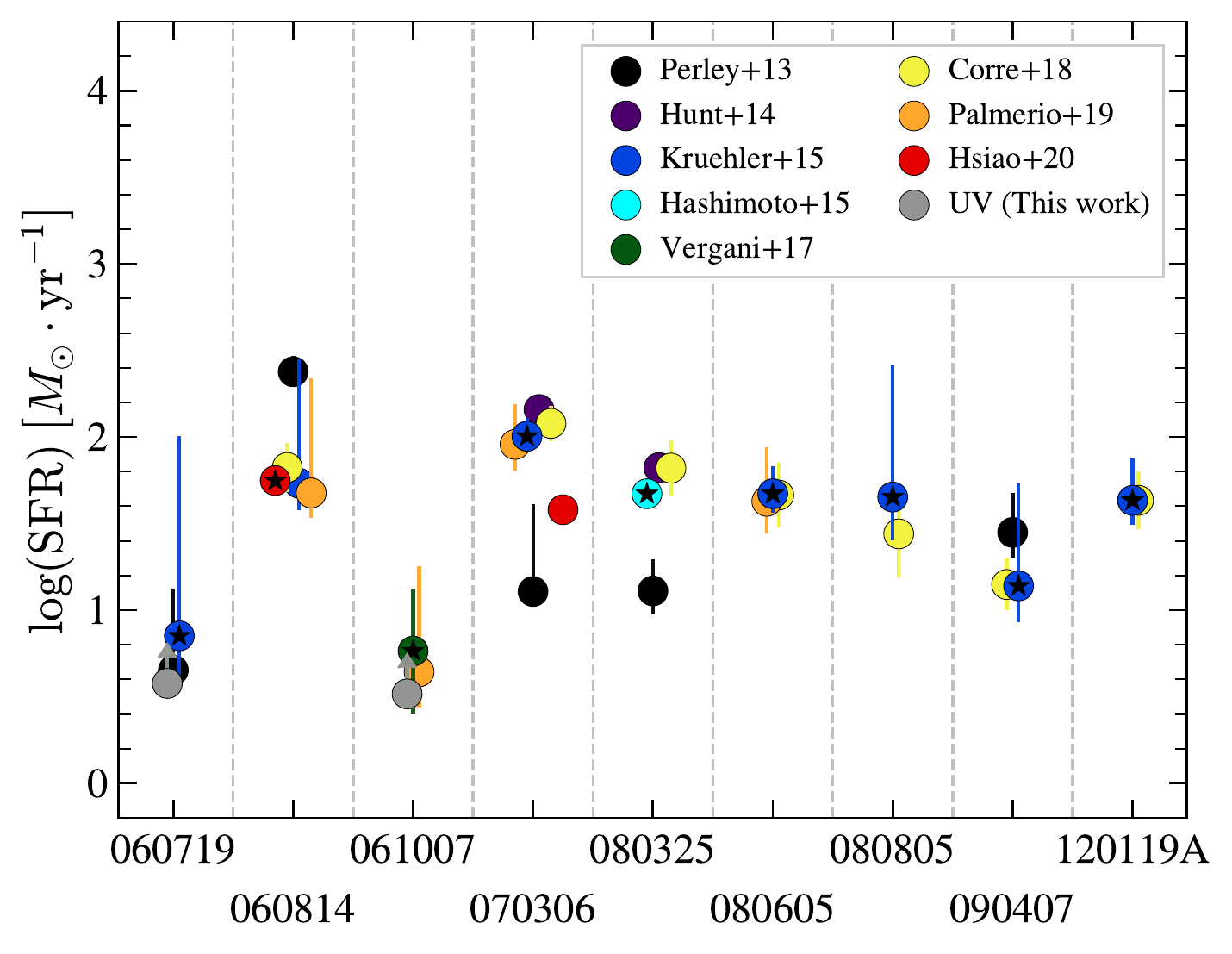}
          \caption{Compilation of stellar mass (\textit{top panel}) and SFR (\textit{bottom panel}) estimates for GRB hosts at \zmed. Each circle corresponds to an estimate from the literature or determined as described in Sect.~\ref{subsec:grb_hosts_properties}. The circles marked with a star represent the estimates used in our analysis.}
          \label{fig:SED_dispersion}
      \end{figure}

        The stellar masses of GRB host galaxies used in this work were mostly gathered from the literature. For some objects, we find multiple estimates where most of them were obtained with SED fitting using several photometric points. Only stellar masses from \cite{perley2016b} were derived using a single photometric point (\textit{Spitzer}/IRAC 3.6 $\mu$m band) and a method based on a mass-redshift grid of galaxy SED models. We compare all these estimates in Fig.~\ref{fig:SED_dispersion} (\textit{top panel}) and we note a significant discrepancy in many cases (up to $\sim 0.9$~dex). SED fitting codes based on an energy balance principle (e.g., CIGALE, \citealt{noll2009a, boquien2019a}) can model the stellar luminosity absorbed by dust and its re-emitted luminosity in the IR. If a far-infrared (FIR) band is used to constrain the models, a more realistic attenuation value can be derived and thus we expect a more accurate stellar mass. For this reason, we select preferentially the estimates in the following order: (1) SED fitting with optical/near-IR (NIR) and FIR measurements using energy balance code, (2) SED fitting with optical/NIR measurements, (3) a mass-to-light ratio.

        For GRB hosts with no stellar mass reported in the literature and not enough photometric points to determine a stellar mass from a SED fitting, we derive our own estimate based on a mass-to-light (M/L) ratio \citep{bell2001a} applied to the $F160W$ magnitude determined by \galfit. We use the COSMOS2015 \citep{laigle2016a} catalog to find a relation between stellar mass and NIR luminosity at a given redshift. The COSMOS2015 survey covers a larger area ($\sim 2~\mathrm{degree}^2$) than the 3D-HST/CANDELS survey and gives access to a larger number of galaxies ($> 500\,000$ objects). The catalog provides a total of 16 photometric bands from the ultraviolet to the mid-infrared, including the H band ($\lambda_{\rm{mean}} \sim 1.64\ \mu$m) of the \textit{VISTA} infrared camera. Given that magnitudes of GRB hosts are obtained with WFC3/$F160W$ filter ($\lambda_{\rm{mean}} \sim 1.54\ \mu$m), we apply a color correction on each GRB host magnitude. To estimate this value, we match the objects that were observed in the 3D-HST/COSMOS field and the COSMOS2015 catalog. We measure a mean difference of $0.08$ mag between the two filters. Finally, we correct the GRB host galaxy magnitudes for the Galactic extinction using \cite{schlafly2011a} measurements. \\
        To determine the M/L ratio, we first select all star-forming galaxies \texttt{(CLASS=1)} from the COSMOS2015 catalog at $z_{\mathrm{target}}\pm 0.1$. We then fit a linear relation between $H_{VISTA}$ magnitudes and stellar masses of the identified objects. We finally use the linear model and the GRB host magnitudes earlier corrected for color excess and Galactic extinction to obtain the stellar mass.
        We note that our M/L ratio is based on magnitude in COSMOS2015 determined from aperture photometry using \sx\ while our magnitude is determined by \galfit. \cite{skelton2014a} showed that for the 3D-HST catalog the median difference is lower than 0.04 mag in the range 21 < $H_{F160W}$ < 24 between \sx\ and \galfit\ measurements. We do not correct this effect, which would have only a minimal consequence on the estimated stellar mass. In addition, we compare the stellar masses derived using the COSMOS2015 catalog with those calculated from the M/L ratios of the 3D-HST catalog. We find good agreement between the two estimates for the entire sample of GRB hosts. In our analysis, we use the estimates from the COSMOS2015 catalog which are based on a larger statistic. The uncertainties are derived by propagating the uncertainty of the \galfit\ magnitude models\footnote{More particularly, we use the largest uncertainty values returned by \galfit\ or the MC approach, as described in the Sect.~\ref{subsub:uncertainties}.}. We select all galaxies inside $z_{\mathrm{target}}\pm 0.1$ and with a $\mathrm{mag}_{\mathrm{target}}\pm \delta \mathrm{mag}_{\mathrm{target}}$. We then compute the $1\sigma$ error by taking the $16^{\mathrm{th}}-$percentile and $84^{\mathrm{th}}-$percentile of the resulting galaxy distribution.

\begin{figure}
        \centering
        \includegraphics[width=\hsize]{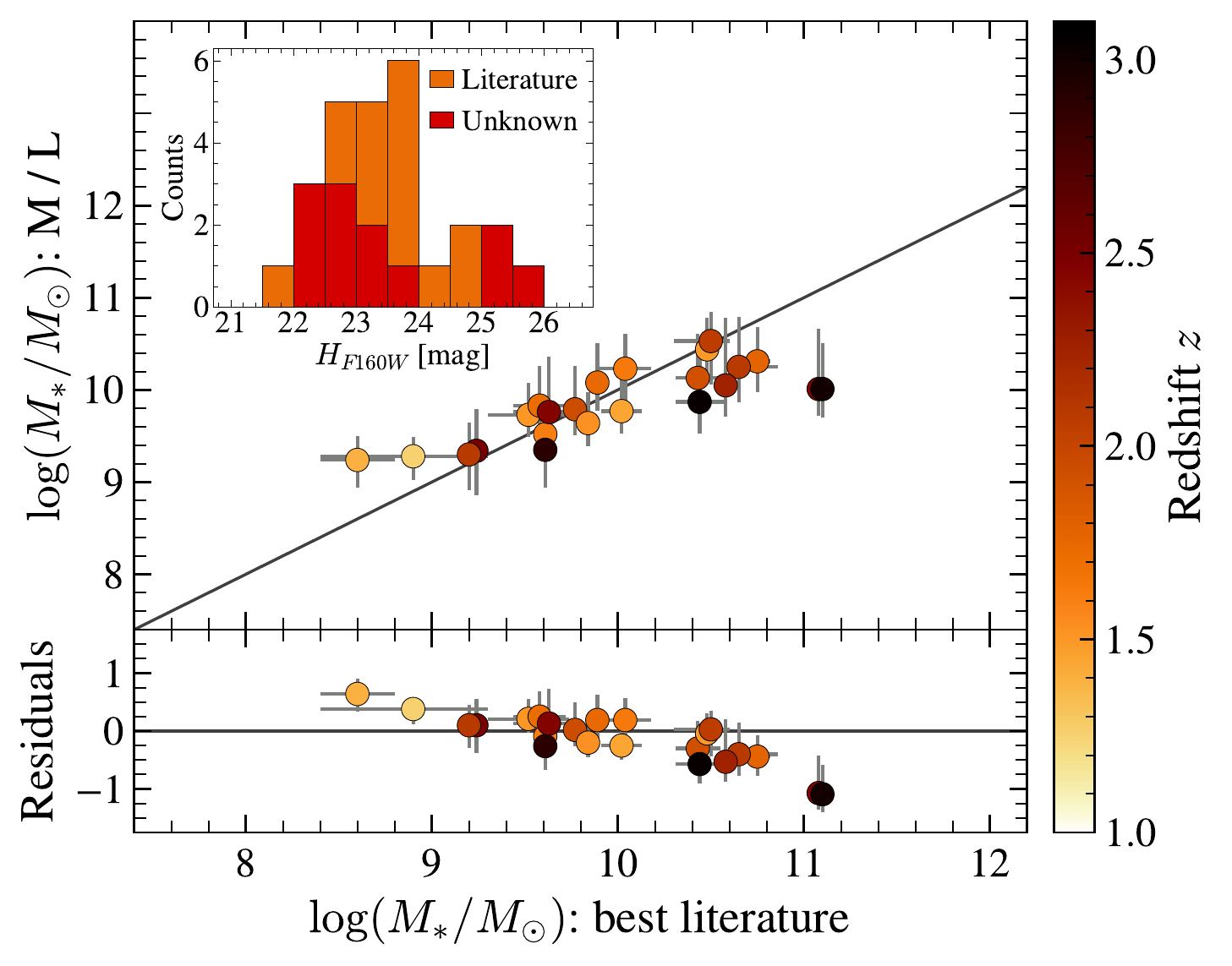}
        \caption{Stellar masses of GRB hosts estimated from mass-to-light ratios, compared with the best estimates selected from the literature (Sect.~\ref{subsec:grb_hosts_properties}). The residuals are shown in the \textit{bottom panel}. Each GRB host is color-coded according to its redshift. The inset represents their associated $F160W$ magnitude distribution (orange histogram) and the distribution of the GRB hosts for which no stellar mass measurement exists in the literature (red histogram).}
        \label{fig:stellar_mass_from_Hmag}
    \end{figure}

        As a sanity check, we also apply this M/L procedure to the hosts with stellar masses determined in the literature and selected according to the requirements described above.
        The comparison between the two is visible in Fig.~\ref{fig:stellar_mass_from_Hmag}, where we color code GRB hosts according to their redshift.
        We find an overall agreement between the two estimates, but we also note a linear trend evolving with stellar mass, where low (high) stellar mass galaxies tend to be overestimated (underestimated). In particular, two GRB hosts have mass estimates differing by more than 0.8~dex (GRB~071021 and GRB~090404). These sources are located at $z > 2.4$ where the WFC3/$F160W$ filter probes bluer wavelengths, more subject to dust extinction. As GRB host magnitudes are not corrected for galaxy attenuation, they might lead to underestimate the stellar mass derived from a M/L ratio.
        In addition, we use a single Sérsic profile to model each GRB host galaxy. This enabled us to catch most of the flux for the majority of objects, but in some cases (e.g., GRB~080207, GRB~111215A) more components would have been required to improve the fit and the magnitude estimate.
        This might have contributed to underestimate their total flux and therefore their stellar mass. With this caution in mind, we note however that the few GRB hosts with mass estimates relying on this M/L approach (see Table~\ref{tab:grb_sample}) have redshifts and stellar masses in the range where Fig.~\ref{fig:stellar_mass_from_Hmag} reveals consistent results with the more conventional SED fitting method. This supports therefore the reliability of our measurements, which should not introduce any additional systematics given their otherwise large statistical uncertainties.

    \subsubsection{Star formation rate}
    \label{subsubsec:star_formation_rate}
        The star formation rate of GRB host galaxies are also gathered from the literature. In a similar way to stellar masses, we show in Fig.~\ref{fig:SED_dispersion} (\textit{bottom panel}) the dispersion of \sfr\ measurements obtained for a same sources. We yet observe a better agreement between \sfr s than previously found for \mass. 
        We note in several cases that the SED fitting solution used to estimate these star formation rates \citep[e.g.,][]{corre2018a, palmerio2019a} was constrained using \sfr\ measurements determined from emission lines fluxes published in other works \citep[e.g.,][]{kruhler2015a}.
        This contributes to reduce the dispersion between estimates observed in Fig.~\ref{fig:SED_dispersion}. In general, it is expected that estimates including observations in the FIR give a more reliable SFR because the thermal emission of cold dust heated by O/B stars is more accurately modeled.
        We therefore preferentially select the \sfr\ estimated by (1) SED fitting with optical/NIR and FIR measurements using energy balance code, (2) Dust corrected \halpha\ luminosity (3) SED fitting with optical/NIR measurements.
        Only for GRB~070306, we consider the \sfr\ based on \halpha\ luminosity instead of the SED fitting with FIR observations. Indeed, the SED of the galaxy in \cite{hsiao2020a} does not match correctly the \Herschel\ observations. The model seems to underestimate the IR luminosity and thus the total \sfr\ of the galaxy, as also suggested by the higher SFR estimate obtained from \halpha.
        At \zmed\, we have a majority (10/11) of SFRs from \halpha\ and one estimate based on SED fitting including a FIR measurement (ALMA detection). The tracers are more diversified at \zhigh\ with 5 over 9 objects from SED fitting with ALMA detection, one from \halpha, two from SED fitting with optical/NIR measurements and one from the rest-frame UV continuum emission.

        If no \sfr\ is found in the literature, we derive a lower limit value based on the $R$-band magnitude of \cite{hjorth2012a}. This filter probes the UV rest-frame of the galaxy at $z > 1$. The UV light is mainly radiated by young and short-lived stars. It is another indicator of recent \sfr\ in the galaxy. However, the UV radiation is subject to dust reddening caused by the Galactic center or the galaxy itself. We correct UV magnitudes from the Galactic extinction using \cite{schlafly2011a} measurements. We then derive a \sfruv\ using the relation from \cite{kennicutt1998a}. Nevertheless, we only consider these values as lower limits of the \sfr\ since the UV luminosities are not corrected for the dust attenuation in the host itself.

  \subsection{Stellar mass and Star formation surface densities}
    The stellar mass and star formation surface densities are derived using Eqs.~(\ref{eq:sigmass}) and (\ref{eq:sigsfr}). The \mass\ and SFR are divided by the galaxy projected area defined by the half-light radius. As it contains half of the total light of the galaxy, a correction factor of 1/2 is applied to the \mass\ and \sfr\ assuming that the matter is uniformly distributed inside the galaxy.

      \begin{equation} \label{eq:sigmass}
        \sigmass = \log \left( \frac{\mathrm{M}/2}{\pi R_e^2} \right), \, \text{where M is the stellar mass in } \Msun.
      \end{equation}

      \begin{equation} \label{eq:sigsfr}
        \sigsfr = \log \left( \frac{\sfr/2}{\pi R_e^2} \right)
      \end{equation}

    We derive the \sigmass\ and \sigsfr\ errors by propagating the uncertainties of the stellar masses, the star formation rates and the half-light radii.

  \subsection{Statistical tests}
  \label{subsec:statistical_tests}
    We apply the Kolmogorov-Smirnov (\ks) test to compare the CDFs of GRB hosts and field galaxies. We consider a similar Bayesian inference framework as described by \cite{palmerio2019a}. This approach considers that each parameter value (\re, \sigmass, \sigsfr) is described by an asymmetric Gaussian. The center of the distribution is given by the value in Table \ref{tab:grb_sample} and the asymmetrical standard deviation is given by errors associated with that value.
    We sample each Gaussian probability density function (PDF) with $10\,000 \: (N_{real})$ MC realizations. We then build $N_{real}$ different CDFs for the GRB hosts and 3D-HST samples. Finally, we compute $N_{real}$ MC realizations of the two sided \ks\ test from the previous samples of CDFs. We thus obtain a distribution function of D-statistic and \pvalue s that provide confidence bounds on the \ks\ test.

\section{Results}
\label{sec:Results}

  \subsection{\reMass\ relation}
  \label{subsec:Mass-size_relation}

    \begin{figure*}[h!]
        \begin{subfigure}[b]{0.5\hsize}
            \centering
            \includegraphics[width=\hsize]{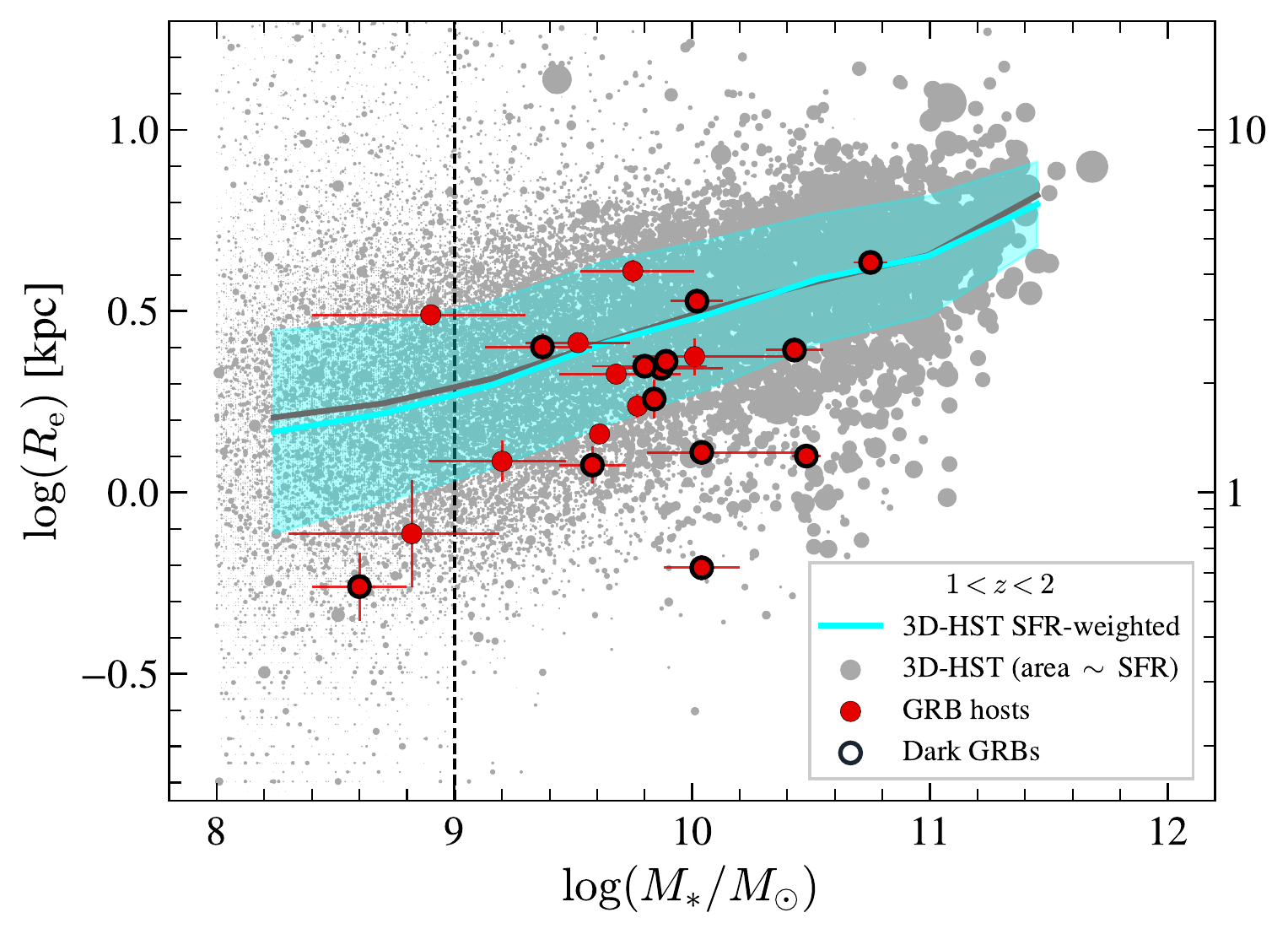}
            \includegraphics[width=0.85\hsize]{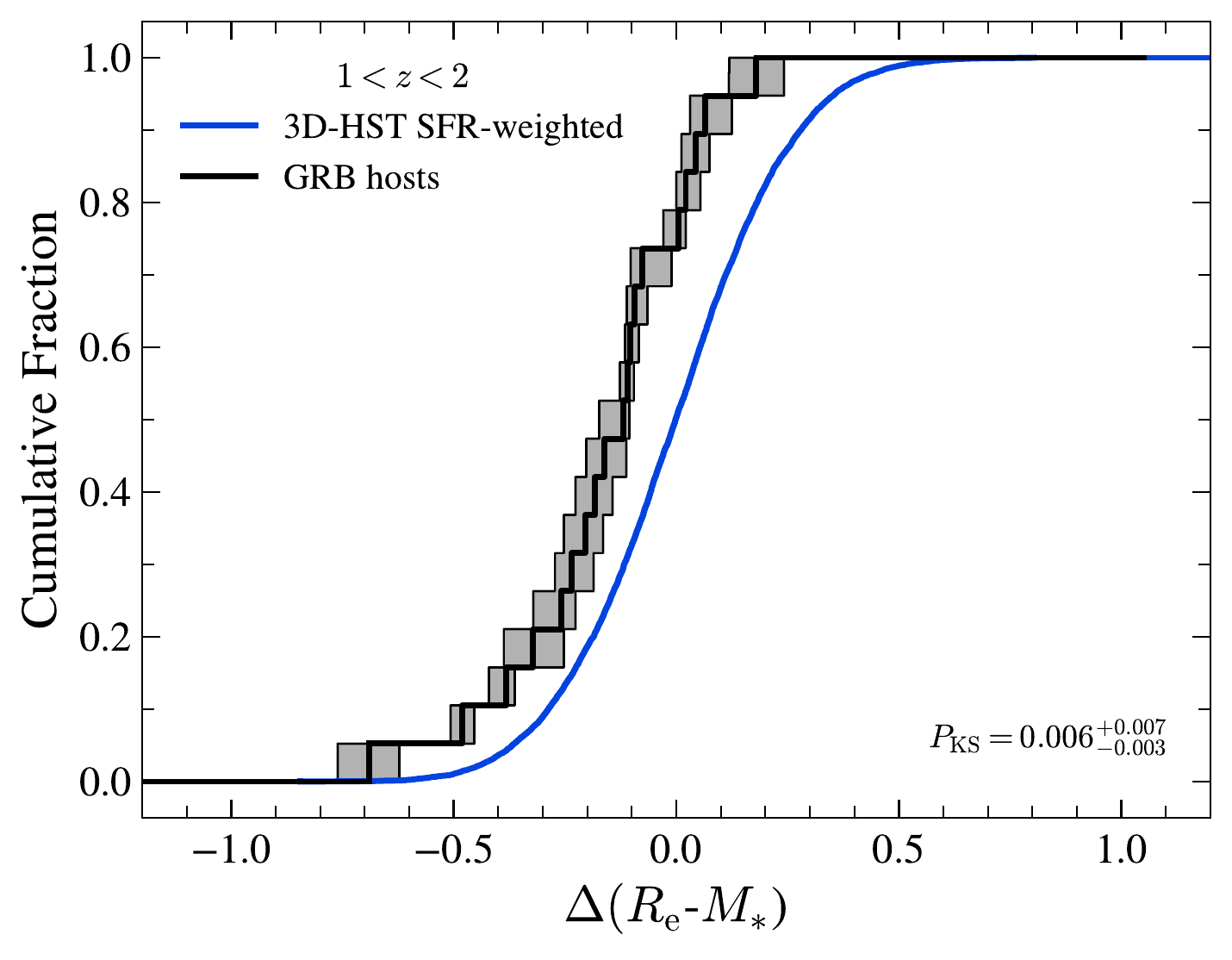}
        \end{subfigure}
        \hfill
        \begin{subfigure}[b]{0.5\hsize}
            \centering
            \includegraphics[width=\hsize]{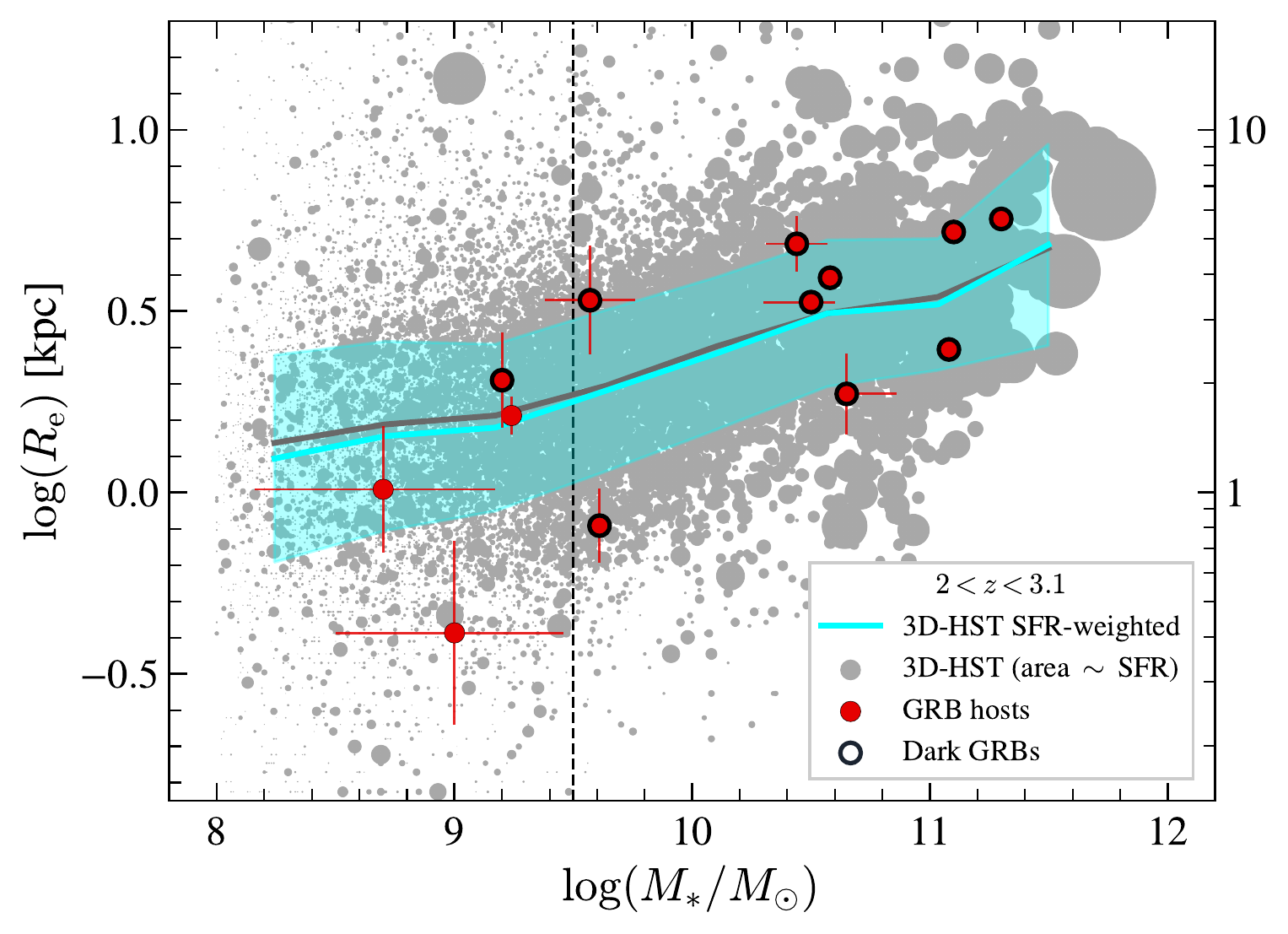}
            \includegraphics[width=0.85\hsize]{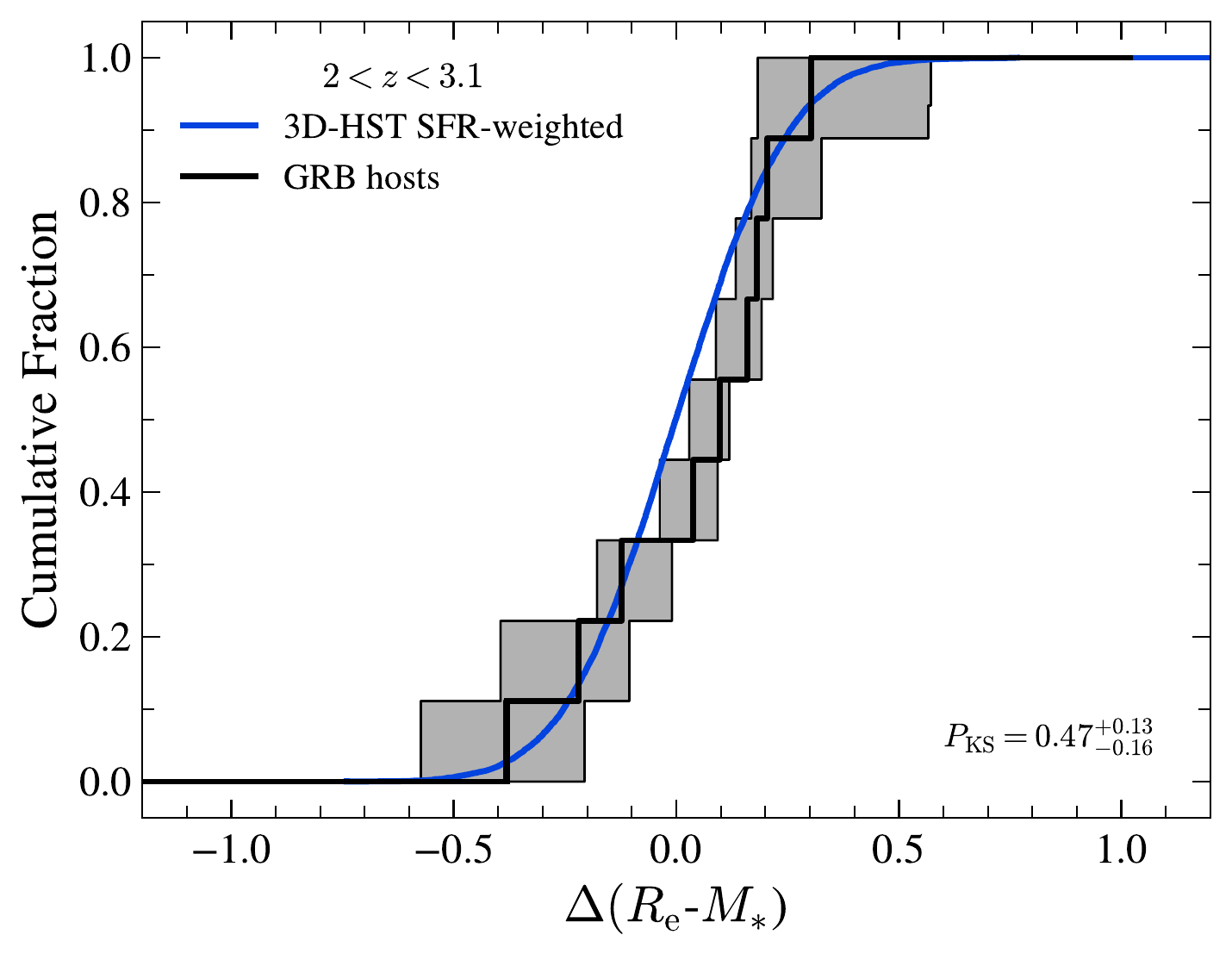}
        \end{subfigure}
        \caption{\textit{Top panels}: Half-light radius against stellar mass for GRB hosts and star-forming galaxies at \zmed\ (\textit{left panel}) and at \zhigh\ (\textit{right panel}). The GRB host galaxies are displayed as red circles and the 3D-HST star-forming galaxies are shown as gray circles with an area proportional to their SFR. The median of the star-forming population is shown as a dashed gray line. The dashed cyan line represents the expected median of a GRB hosts population that does not suffer from bias to trace the SFR (gray circles weighted by their SFR). The 1$\sigma$ uncertainty of the cyan median is given as a shaded cyan region. The vertical dashed black line is the mass-completeness limit of the 3D-HST survey. \\
        \textit{Bottom panels}: Cumulative distribution function of \deltare. The \deltare\ represents the distance between GRB hosts and the SFR-weighted \reMass\ relation of the \textit{top panels}. The blue curve is a Gaussian CDF with a mean of 0 and a standard deviation defined by the $1\sigma$ errors of the shaded cyan area at the GRB hosts positions. The \pvalue\ returned by the two-sided \ks\ test is provided in the right bottom part of panels.}
        \label{fig:M-size_relation}
    \end{figure*}

    In Fig.~\ref{fig:M-size_relation} (\textit{top panel}), we show the half-light radii (\re) against stellar masses for the GRB hosts and 3D-HST star-forming galaxies. As mentioned previously, it is commonly accepted that long GRBs are related to recent star formation activity in their host. If GRB hosts trace the star-forming sources with no bias, then galaxies with higher SFR should have a higher probability to produce a GRB. Based on this assumption, we weight the control sample by its SFR to mimic a population of galaxies that should host GRBs with no environmental dependence. The resulting SFR-weighted \reMass\ relation (cyan curve) is close to the median relation characterizing the field galaxies (gray curve). This is expected because for a given stellar mass, the radius does not depend much on the SFR, as can be seen from the sizes of the gray circles in Fig.~\ref{fig:M-size_relation}. The 1$\sigma$ uncertainty associated with the SFR-weighted \reMass\ relation (cyan region) is derived using the median absolute deviation (MAD) estimator. \\
    Figure~\ref{fig:M-size_relation} (\textit{top left panel}) clearly shows that GRB hosts are markedly different from the general population. Indeed, we note a larger number of GRB hosts below the SFR-weighted relation at \zmed. If GRB hosts were truly representative of the overall population of star-forming sources, we would expect approximately equal numbers above and below the SFR-weighted relation.
    For GRB hosts at \zhigh, we observe however a more uniform distribution of sizes with respect to field galaxies, and a better agreement with the SFR-weighted relation.
    The predominance of dark GRB hosts in the samples and how these results are representative of the whole GRB host population will be further discussed in Sect.~\ref{sub:hosts_of_grbs_with_dark_vs_optically_bright_afterglows}. \\
    Because GRB hosts commonly occur in faint and low stellar mass galaxies \citep{fruchter2006a}, and given the positive trend of the \reMass\ relation (massive galaxies are also larger), a straight comparison between the distribution of their sizes and that of the overall population of star-forming sources, irrespective of their stellar mass, would necessarily be biased. 
    To quantify if GRB host galaxies are not smaller only because they explode in faint galaxies, we measure for each GRB host the distance between its position in Fig.~\ref{fig:M-size_relation} (\textit{top panel}) and the SFR-weighted \reMass\ relation at the same stellar mass, defined by Eq.~(\ref{eq:deltaRe}) and denoted as \deltare\ hereafter. If GRB hosts are representative of field star-forming sources, we should expect that the CDF of \deltare\ to be distributed around zero. To test this assumption, we apply a \ks\ test as described in Sect.~\ref{subsec:statistical_tests}, only considering GRB hosts and field galaxies above the 3D-HST mass-completeness limits quoted earlier. We compare the \deltare\ CDF to a Gaussian CDF centered on zero and with a standard deviation given by the median value of the 1$\sigma$ uncertainties associated with the SFR-weighted relation at each GRB host position. The CDFs and their associated uncertainties are shown in Fig.~\ref{fig:M-size_relation} (\textit{bottom panel}). For GRB hosts at \zmed, the two sided \ks\ test returns a probability of $\pks = 0.006$. We can rule out the null hypothesis that the two samples are drawn from the same underlying population. At \zhigh, we obtain a \ks\ probability of $\pks = 0.47$. In this case, the null hypothesis of the \ks\ test cannot be rejected and we cannot rule out the possibility that both samples are drawn from the same distribution.

    \begin{equation} 
    \label{eq:deltaRe}
        \deltare = \log{(R_{\rm{e,\, GRBH}})} - \log{(R_{\rm{e,\, 3D\text{-}HST\, SFR\text{-}weighted,\, median}})}
    \end{equation}

  \subsection{\sigmassMass and \sigsfrMass\ relations}
  \label{subsec:Mass-sigma_relation}
    \begin{figure*}[h!]
        \begin{subfigure}[b]{0.5\hsize}
            \centering
            \includegraphics[width=\hsize]{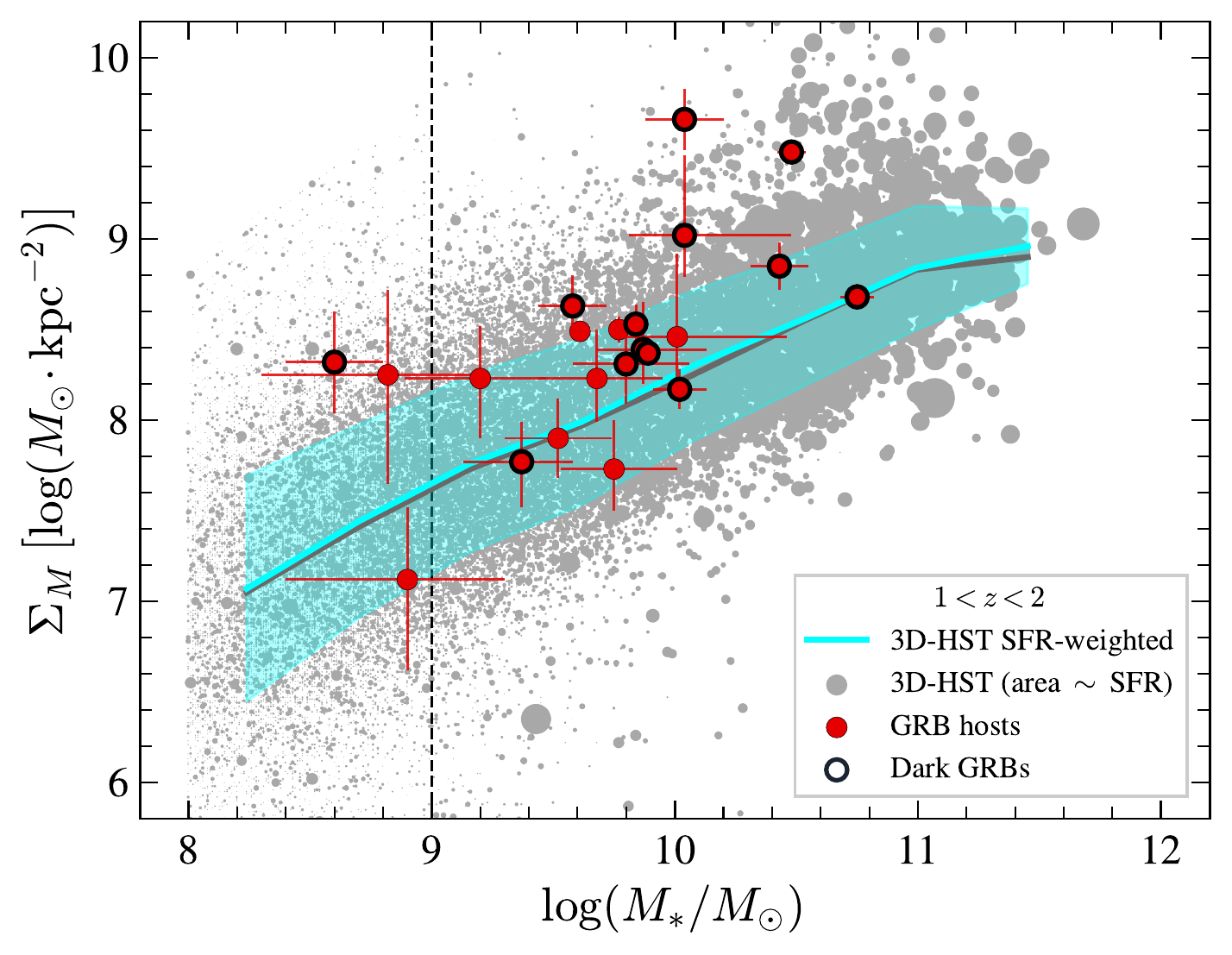}
            \includegraphics[width=0.85\hsize]{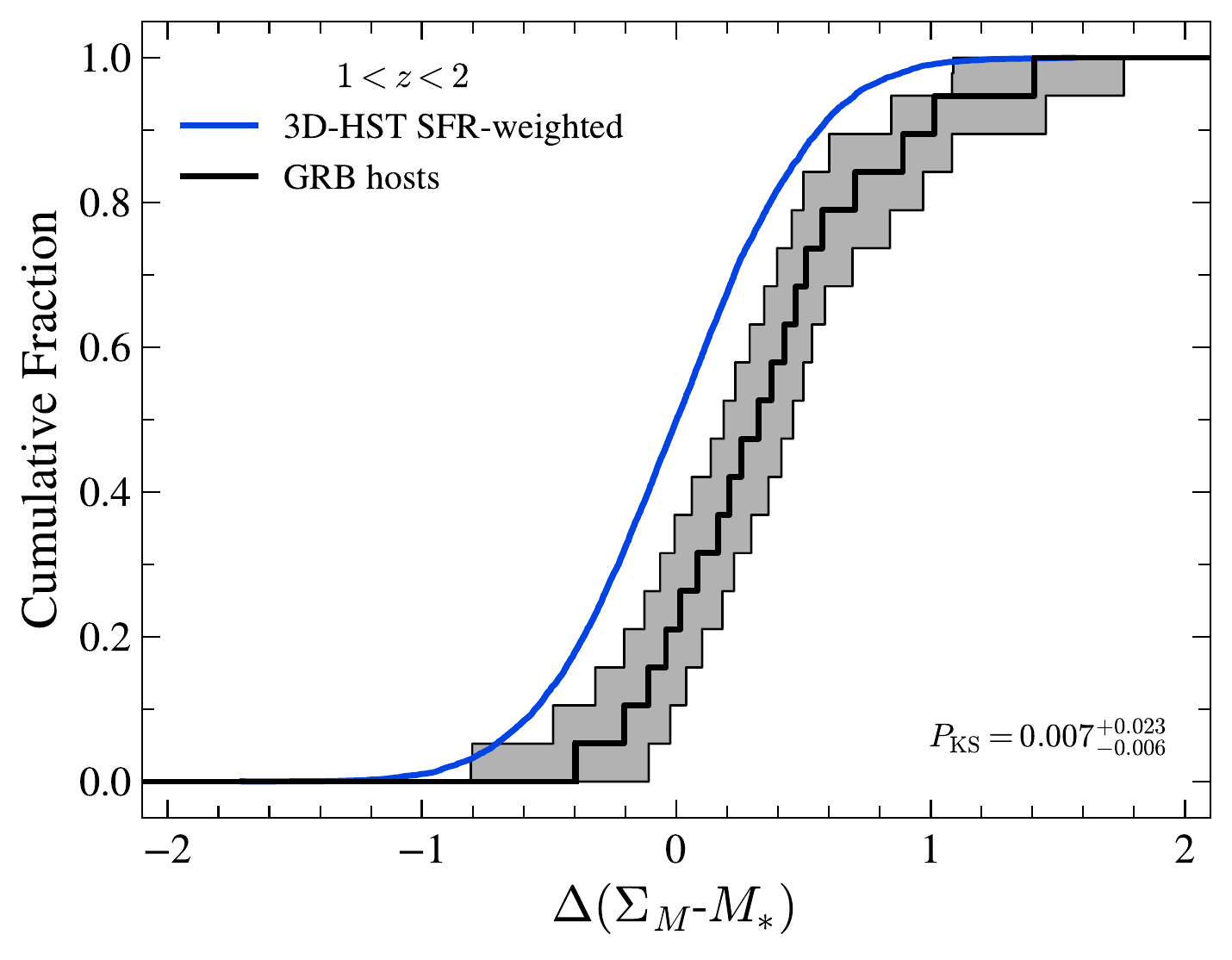}
        \end{subfigure}
        \hfill
        \begin{subfigure}[b]{0.5\hsize}
            \centering
            \includegraphics[width=\hsize]{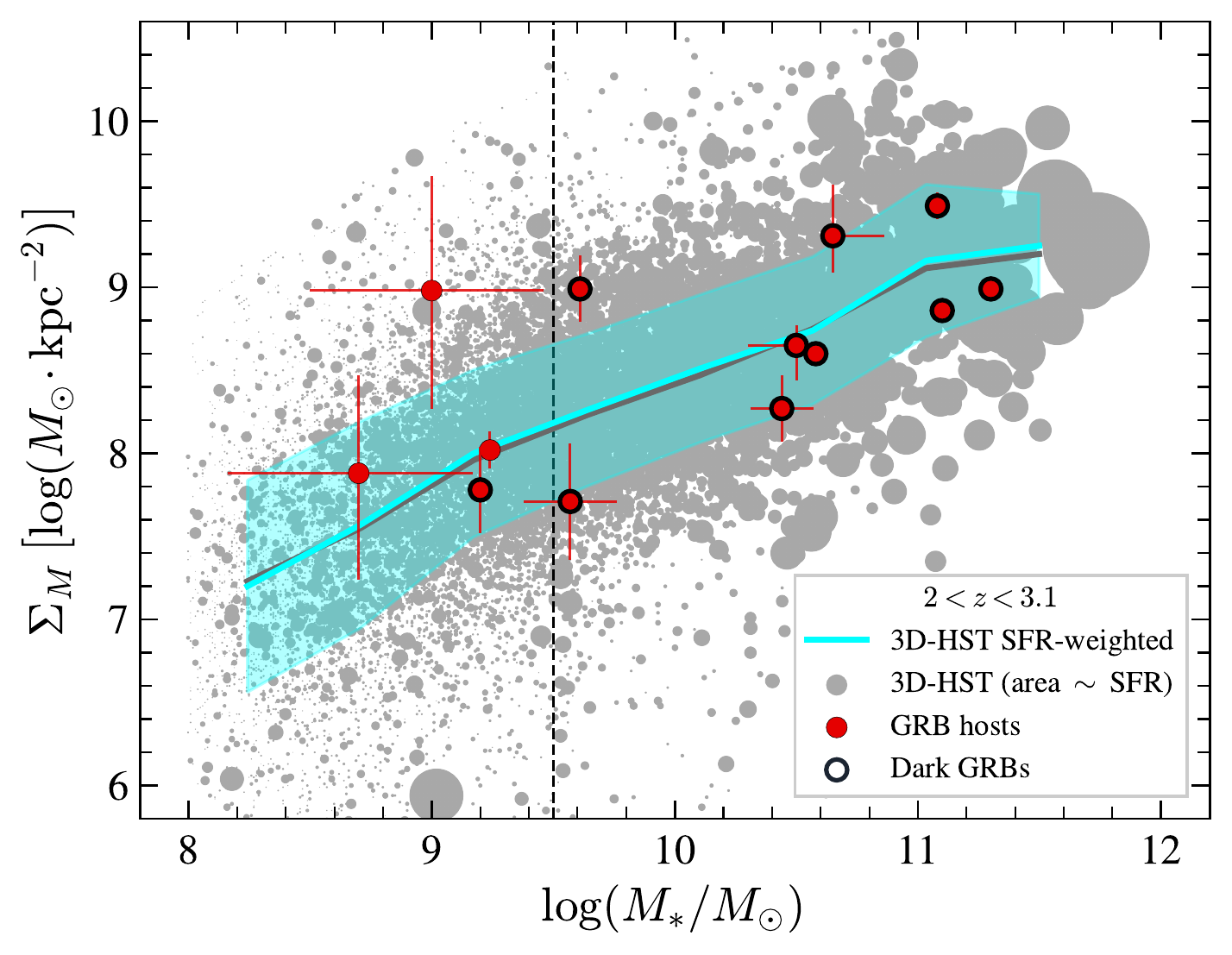}
            \includegraphics[width=0.85\hsize]{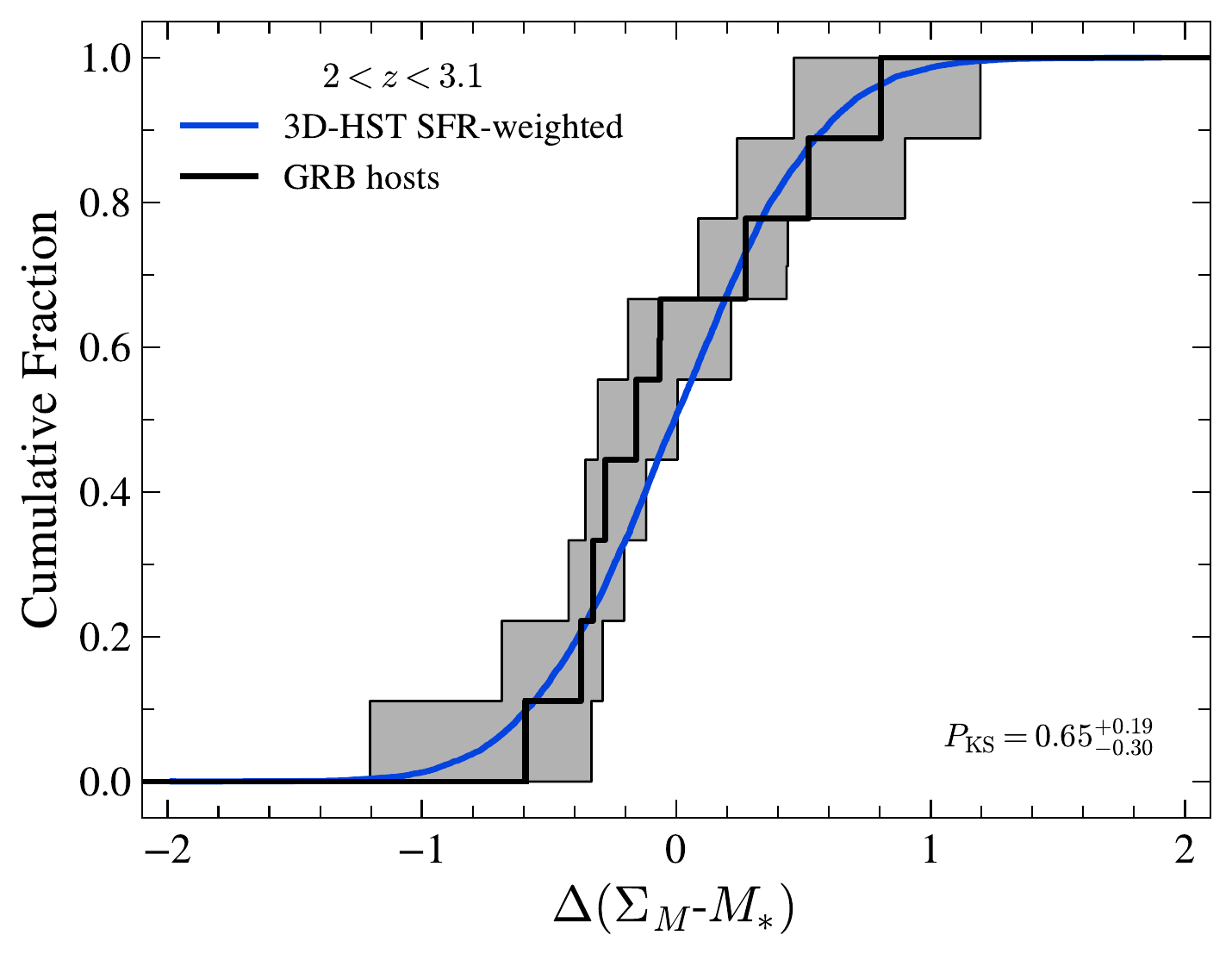}
        \end{subfigure}
        \caption{As in Fig.~\ref{fig:M-size_relation} but showing the stellar mass surface density (\sigmass) against stellar mass.}
        \label{fig:M-sigma_mass_relation}
    \end{figure*}
 
     \begin{figure*}[h!]
        \begin{subfigure}[b]{0.5\hsize}
            \centering
            \includegraphics[width=\hsize]{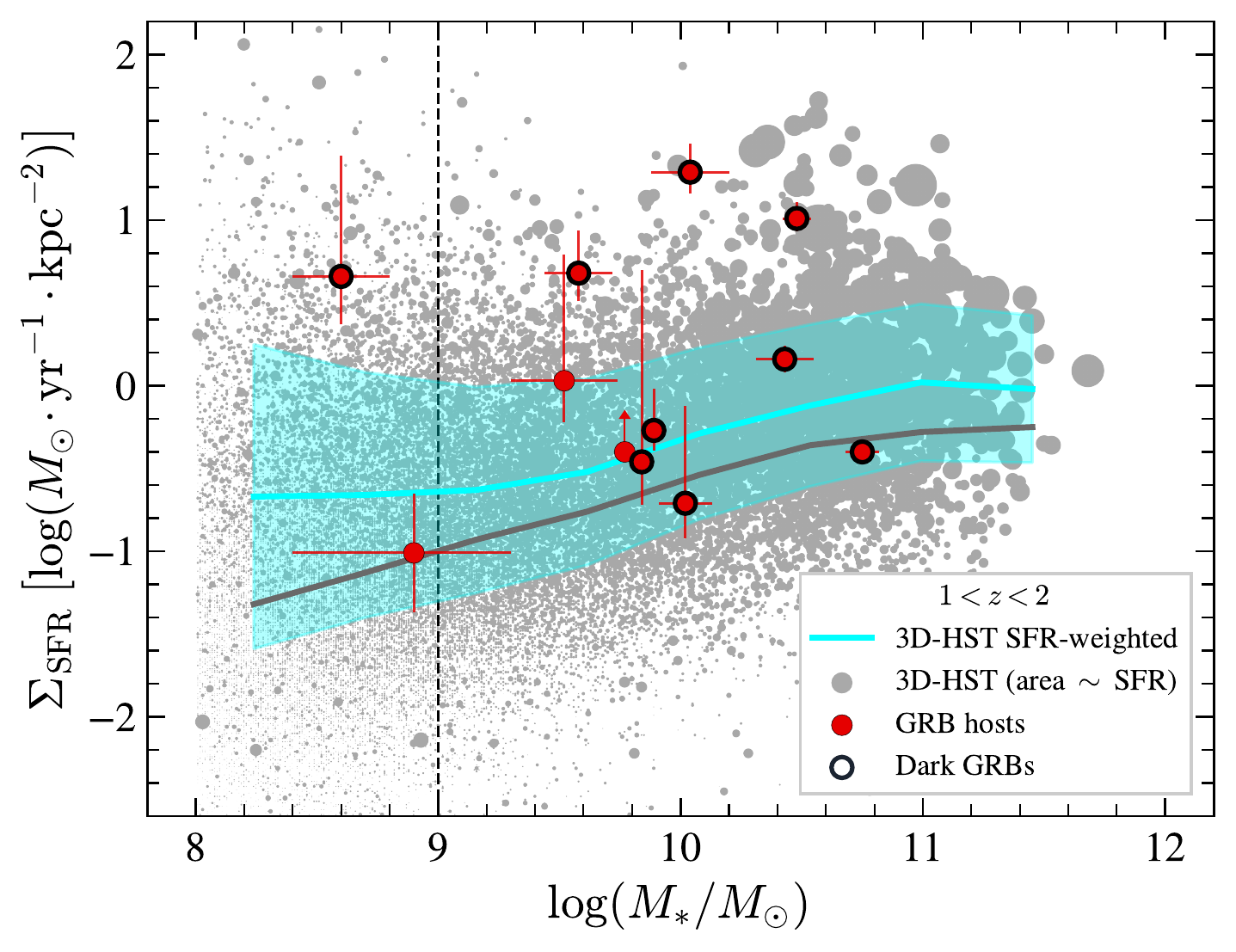}
            \includegraphics[width=0.85\hsize]{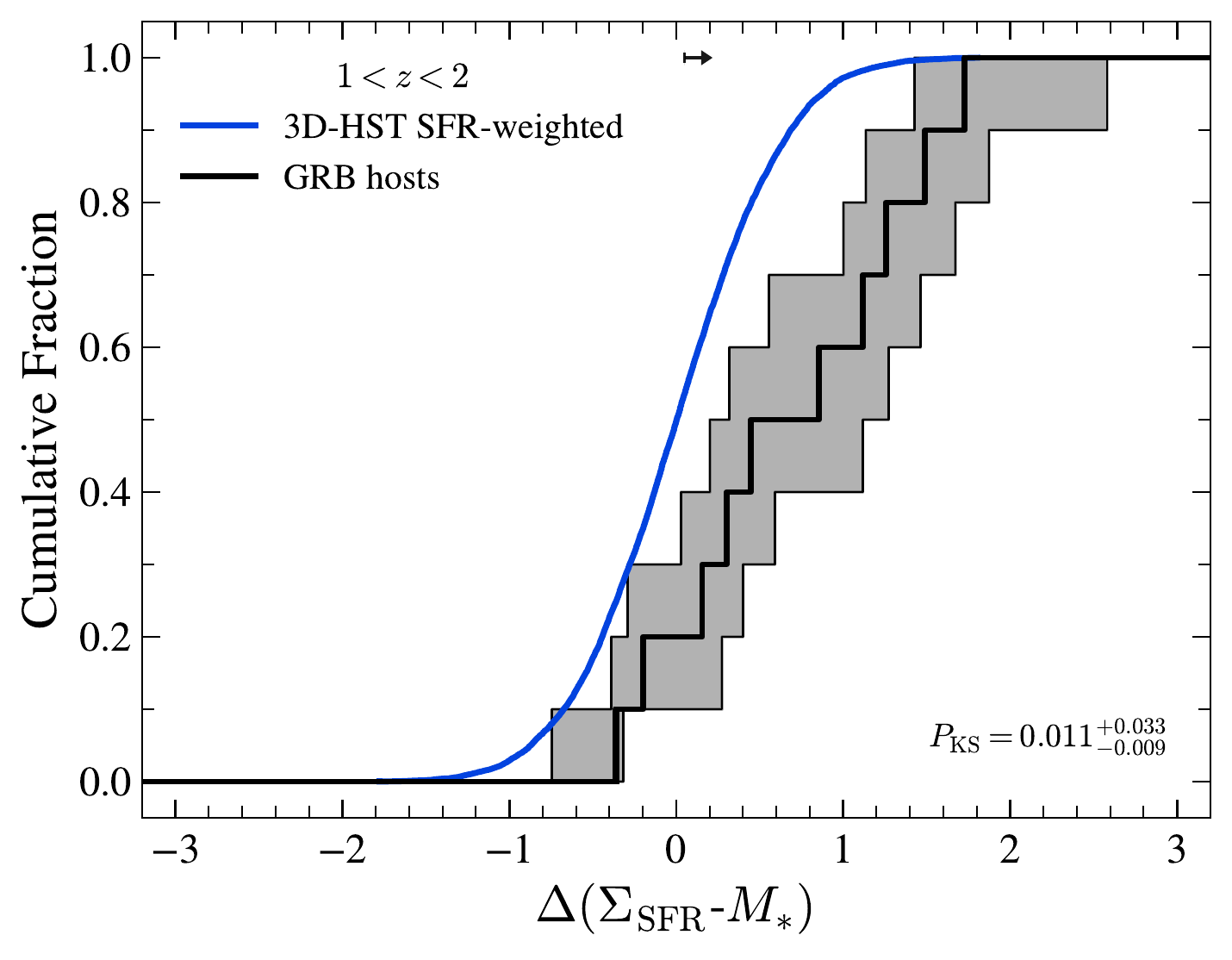}
        \end{subfigure}
        \hfill
        \begin{subfigure}[b]{0.5\hsize}
            \centering
            \includegraphics[width=\hsize]{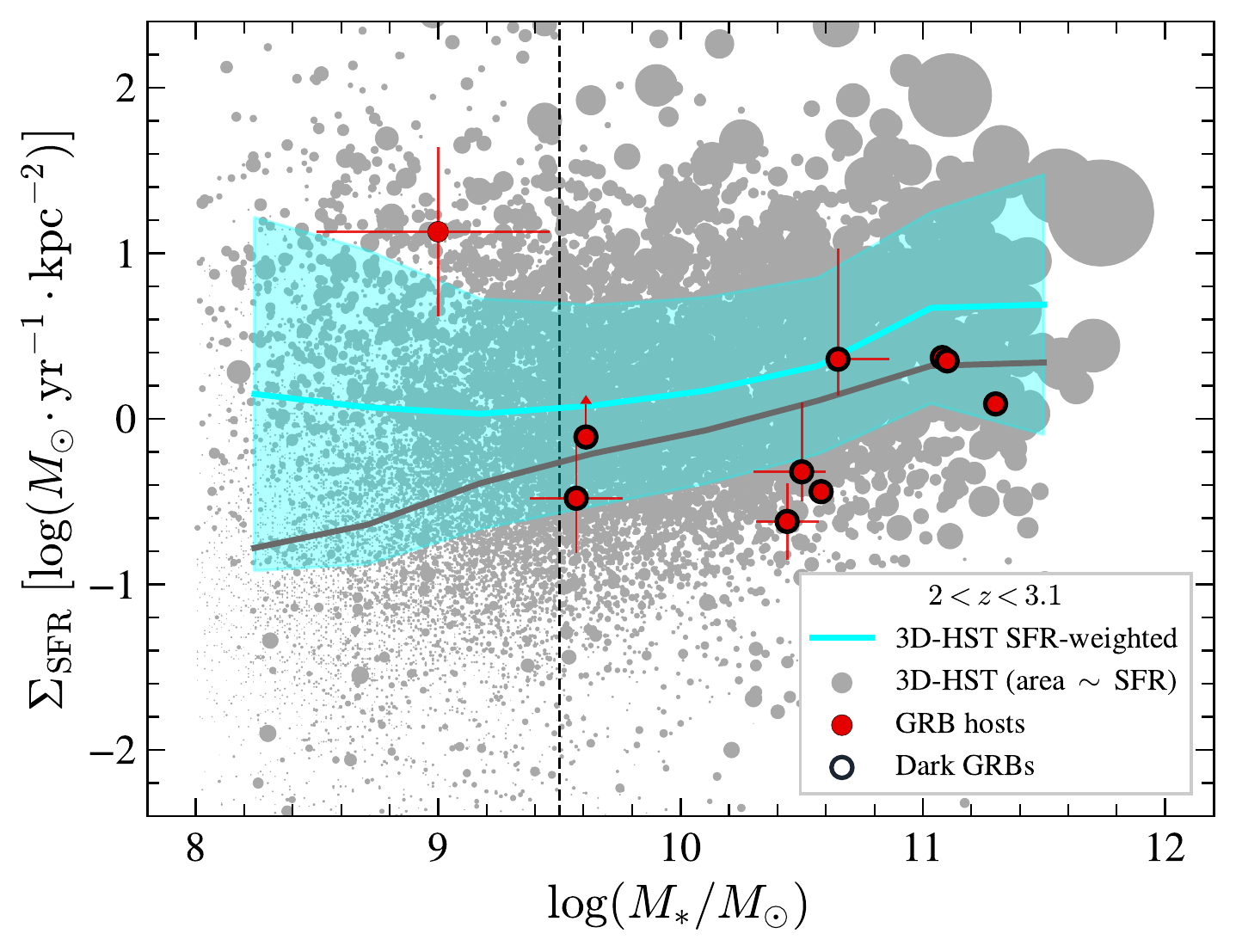}
            \includegraphics[width=0.85\hsize]{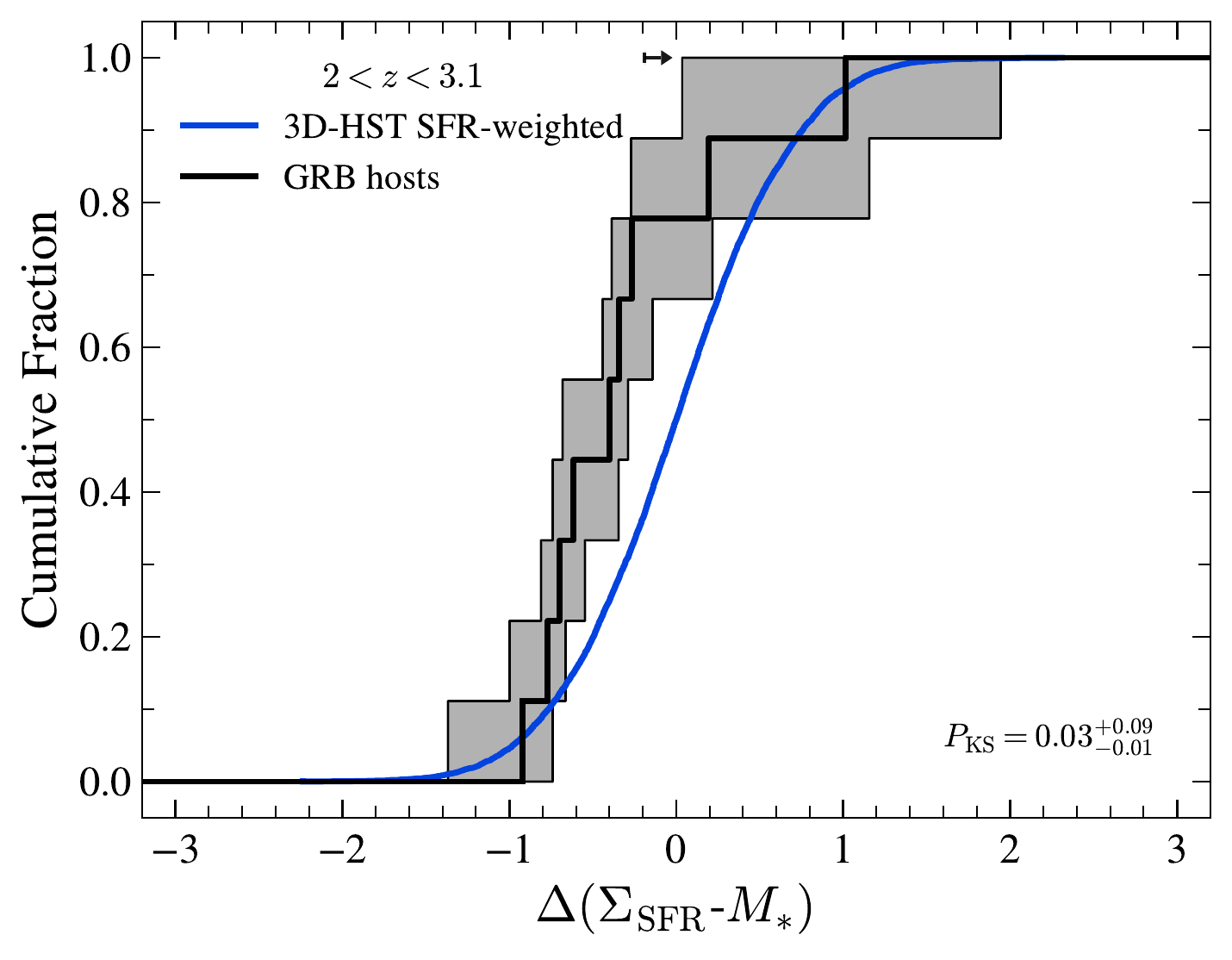}
        \end{subfigure}
        \caption{As in Fig.~\ref{fig:M-size_relation} but showing the star formation surface density (\sigsfr) against stellar mass. Lower limits are represented as arrows at the top of the CDF plots.}
        \label{fig:M-sigma_sfr_relation}
    \end{figure*}

    Figures~\ref{fig:M-sigma_mass_relation} and \ref{fig:M-sigma_sfr_relation} (\textit{top panels}) show stellar mass and star formation surface densities against stellar mass for GRB hosts and field galaxies. At \zmed, we note for the \sigmassMass\ and \sigsfrMass\ planes (\textit{top left panels}) that the GRB sample is clearly different from the general population, with GRB hosts being placed in a region with higher density values.
    Regarding the stellar mass densities, we find indeed a larger number of GRB hosts above the SFR-weighted \sigmassMass\ relation at \zmed, while a more homogeneous distribution of host galaxies is found at \zhigh\ with respect to the field.
    This is probably a direct consequence of the trend discussed earlier regarding the size distribution of GRB host galaxies, as the hosts at \zmed\ appear to be smaller (and therefore denser) than typical star-forming sources at comparable stellar masses (see Fig.~\ref{fig:M-size_relation}).
    Similarly, we see that GRB hosts at \zmed\ exhibit star formation surface densities often higher than typically measured in the field, as a larger number of GRB hosts lie above the SFR-weighted \sigmassMass\ relation. At \zhigh, our results are nonetheless more intriguing, as we observe an opposite trend where the majority of GRB hosts are below the SFR-weighted \sigsfrMass\ relation.

    In a similar way to the approach followed to compare the size distribution, we measure the distance between the GRB hosts and the SFR-weighted \sigmassMass\ and \sigsfrMass\ relations, defined by Eqs.~(\ref{eq:deltaSigMass}) and (\ref{eq:deltaSigSfr}) and denoted as \deltasigmass\ and \deltasigsfr, respectively. We then compute their CDFs using the prescriptions given in Sect.~\ref{subsec:statistical_tests}. The CDFs are shown in the \textit{bottom panels} of Figs.~\ref{fig:M-sigma_mass_relation} and \ref{fig:M-sigma_sfr_relation}. We compare these CDFs to a CDF centered on zero defined in a similar manner as for the \deltare\ parameter. At \zmed, for both parameters we observe a fraction $> 70\%$ of GRB hosts with a distance $>0$ to their SFR-weighted relation. We obtain a \pvalue\ of $\pks = 0.007$ and $\pks = 0.011$ for the \deltasigmass\ and \deltasigsfr, respectively. We can rule out the null hypothesis in both cases. In other words, the \ks\ test suggests that the GRB host sample is a distinct population from the general star-forming galaxy population.
    At \zhigh, the median of \ks\ realizations for \deltasigsfr\ rejects the null hypothesis at the 3\% significance level but can also be reconciled with the null hypothesis given the \ks\ uncertainty. Finally at \zhigh\ for \deltasigmass, the \ks\ test confirms that GRB host galaxies have stellar mass surface densities consistent with those typically found among star-forming sources with similar stellar mass.

    \begin{equation} 
    \label{eq:deltaSigMass}
        \deltasigmass = \Sigma_{\rm{M},\, GRBH}- \Sigma_{\rm{M},\, 3D\text{-}HST\, SFR\text{-}weighted,\, median}
    \end{equation}
    \begin{equation} 
    \label{eq:deltaSigSfr}
        \deltasigsfr = \Sigma_{\rm{SFR},\, GRBH} - \Sigma_{\rm{SFR},\, 3D\text{-}HST\, SFR\text{-}weighted,\,median}
    \end{equation}

    \subsection{Hosts of GRBs with dark vs optically-bright afterglows} 
    \label{sub:hosts_of_grbs_with_dark_vs_optically_bright_afterglows}
    We further investigate whether the deviations found may be due to a predominance of dark GRB hosts in the sample and how these results may be extended to the whole GRB host population.
    Previous studies on the nature of dark bursts and the properties of their host galaxy found a population of galaxies more massive, with a typical stellar mass of about $ 10^{10}\ \Msun$, more luminous and with redder colors than optically-bright GRB hosts \citep{kruhler2011b, rossi2012a, svensson2012a, perley2013a, hunt2014a, perley2016b}.
    Only a few cases of dark GRBs with low-mass host galaxies have been reported \citep[e.g., GRB~080605 and GRB~100621A,][]{kruhler2011b}. The apparent relation between dark GRBs and massive galaxies suggests that the GRB obscuration is mainly due to the dust within the host rather than a dense local environment (clumps) surrounding the GRB. Although a more complex situation with a combination of the two is more likely to be realistic, e.g., \cite{corre2018a} showed that for half of their sample a very clumpy local dust distribution near the burst is necessary to reproduce the galaxy attenuation curves.
    Furthermore, \cite{chrimes2019a} analyzed a sample of 21 dark GRBs observed with the HST in F606W and F160W filters. They found that dark GRB host galaxies are physically larger but have a morphology (i.e., Spirals/Irregulars) similar to those of optically-bright GRB hosts. They reported no particular evidence of differences in concentration, asymmetry or ellipticity between the two populations. These findings support the view that dark and optically-bright GRB hosts share common morphological properties, except, not surprisingly, for the galaxy size \citep[galaxies with higher stellar mass are also larger,][]{vanderwel2014a}.

    At \zmed\ and for $9 \leq \lmass \leq 10.2$, the two host subpopulations of dark and optically-bright GRB afterglows exhibit properties that overlap each other, as it is visible in the \textit{top left panel} of Fig.~\ref{fig:M-size_relation}.
    For these sources, we first investigate the distributions of dark versus optically-bright GRB hosts in the \reMass\ plane. To prevent the effect of the positive mass-size trend, we use the \deltare\ and calculate a median value for each population. We find a median offset of -0.113, -0.105 and -0.109~dex (i.e., $\sim$ 22\% smaller) for dark hosts, optically-bright hosts and the whole subsample, respectively. This indicates that at this redshift range, the size distributions of the two population are consistent with each other, and that the tendency for GRB host galaxies to be smaller than the field is not driven by the large number of dark GRBs within our initial selection. In addition, we note that the median Sérsic index and axis ratio are also similar for both populations, which supports that the dark and optically-bright GRB hosts share similar morphological properties, at least for this stellar mass range. As suggested by the \reMass\ plane, we also find that the two populations are consistent in terms of \deltasigmass. In the case of \deltasigsfr, the lower statistic makes the comparison more challenging. We note, however, that the two remaining optically-bright hosts are in favor of a consistent trend between the two populations.

    At \zhigh, our sample is mainly composed of dark GRB hosts with $\lmass > 10.5$. Therefore, it is not possible to extend the comparison between the two subpopulations of host galaxies as discussed above, and additional HST observations of the hosts of bright afterglows would be needed to securely conclude on the properties of the whole GRB host population in this redshift range. Because the radii of dark GRB hosts appear more consistent with the size of field galaxies at \zhigh, we argue that a different behavior for the size of the hosts of optically-bright GRB afterglows would be difficult to interpret. In this case indeed, one would have to explain either why the latter remain more compact than the host of dark GRBs despite their lower obscuration, or why they become on the other hand much larger than field sources. However, in the absence of clear observational constraints on their physical size, we acknowledge that caution should be considered regarding the interpretation of our results for this redshift range.

  \subsection{Evolution across $z$}
  \label{sub:evolution_across_z}
    \begin{figure}[t]
        \centering
        \includegraphics[width=\hsize]{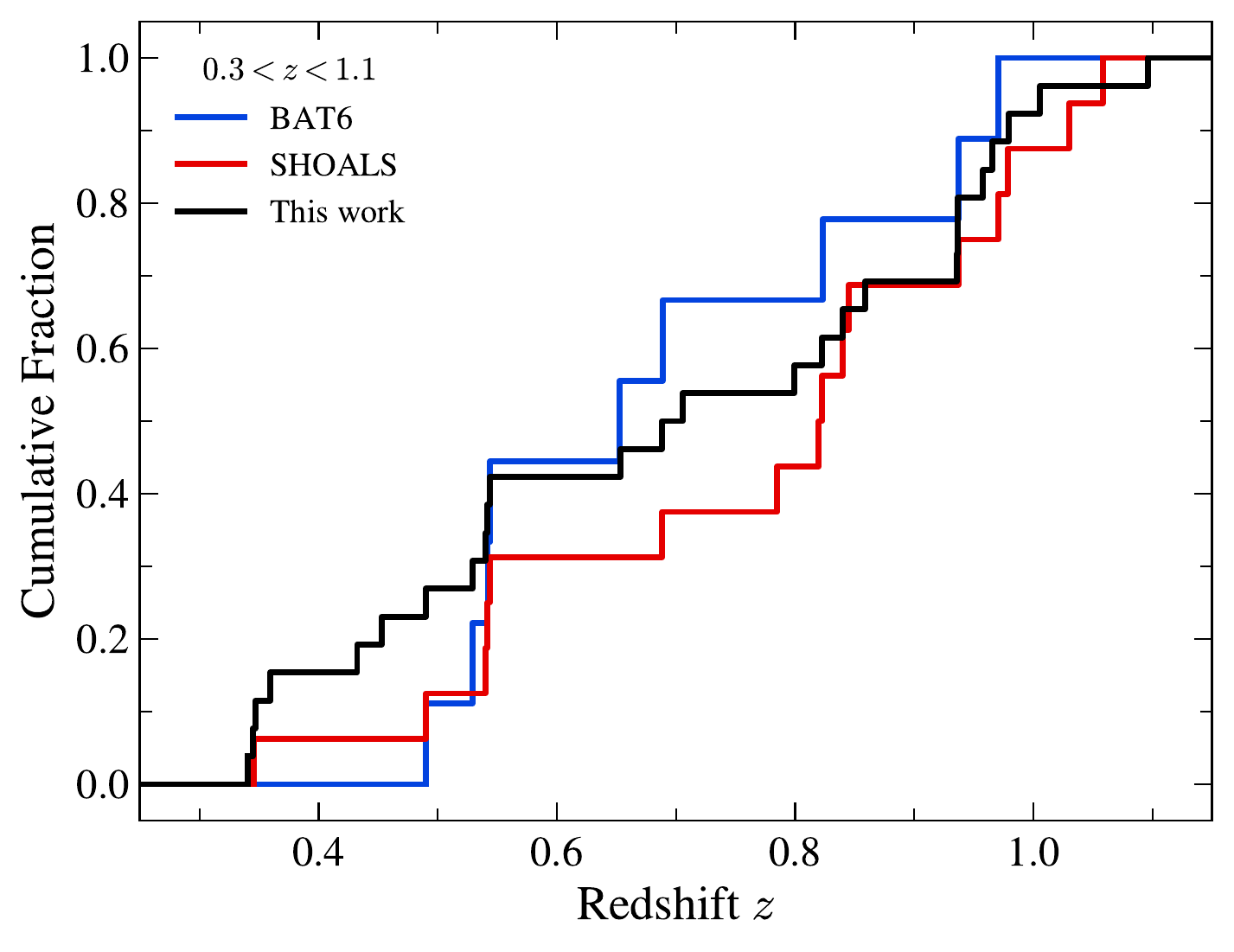}
        \includegraphics[width=\hsize]{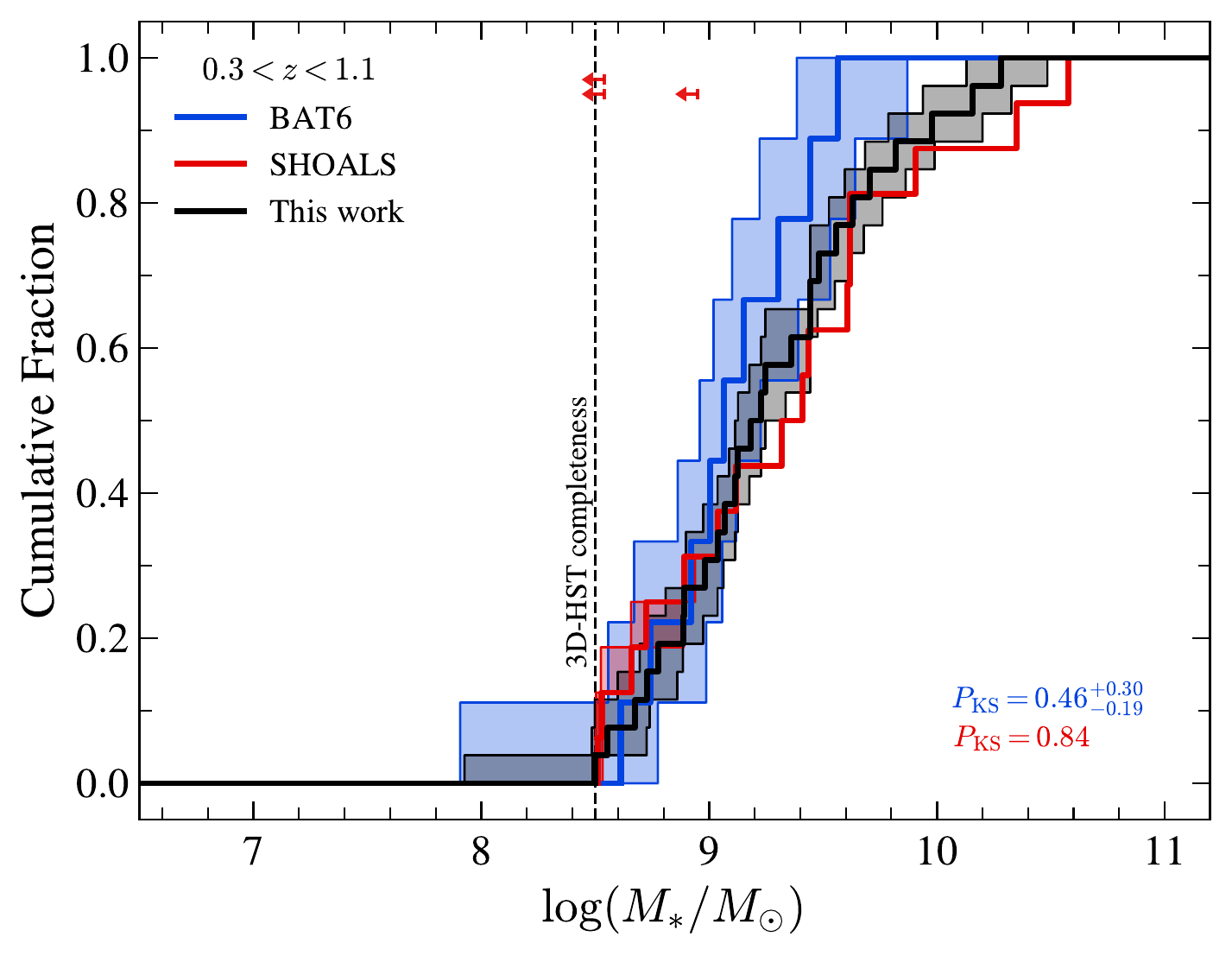}
        \caption{Cumulative distributions as in Fig.~\ref{fig:cumul_unbiased_sample_z} (\textit{top panel}) and Fig.~\ref{fig:cumul_unbiased_sample_mass} (\textit{bottom panel}) but showing the GRB host sample combined from \cite{kelly2014a} and \cite{blanchard2016a} at \zlow.}
        \label{fig:cumul_unbiased_sample_zlow}
    \end{figure}

    The results previously described in Sects.~\ref{subsec:Mass-size_relation} and \ref{subsec:Mass-sigma_relation} reveal that a deviation between GRB hosts and field star-forming galaxies on the \reMass, \sigmassMass\ and \sigsfrMass\ planes exists at \zmed. We further investigate how this deviation compares with the trend previously discussed in the literature for GRB hosts at $z < 1$. \\
    At \zlow, we use a GRB host sample extracted jointly from \cite{kelly2014a} and \cite{blanchard2016a}.
    Their respective selections are not drawn from unbiased GRB samples. We use a similar method to the one described in Sect.~\ref{subsec:Completeness of the samples} and evaluate how the CDFs in redshift and stellar mass are distributed compared to the hosts of unbiased GRB samples (\BATsix\ and \SHOALS). The results are visible in Fig.~\ref{fig:cumul_unbiased_sample_zlow}. We find that the CDFs are consistent with each other for both parameters. This confirms that our sample at \zlow\ probes a similar range of stellar masses as the host galaxies of unbiased GRB samples and that our sample does not suffer from an important bias toward low or high mass galaxies.

    In addition, to enable a comparison as fair as possible with our previous work at $z > 1$ and limit the potential systematic bias between the different studies, we only consider the size measurements reported by these authors. We then recompute the \sigmass\ and \sigsfr\ using \mass\ and SFR extracted from literature in a way similar to the one described in Sect.~\ref{subsec:Galaxy_structural_parameters}.
    From \cite{kelly2014a}, we select the $r_{50}$ determined from the \SDSS\ photo pipeline. A part of their sample matches the sample of \cite{wainwright2007a} who used \galfit\ and a single Sérsic profile to measure galaxy sizes. The comparison of estimates shows good agreement between the two methods. This confirms that using sizes from the \SDSS\ photo pipeline should not introduce a significant bias compared to our method.
    From \cite{blanchard2016a}, we select \re\ determined by \sx. In Fig.~\ref{fig:re_comparison}, we show that the majority of size measurements are consistent between \galfit\ and \sx\ except when the galaxy size becomes close to the PSF size. Given that only 2/18 objects at $z < 1$ are smaller than the PSF size, this effect should not significantly affect the results. Finally, the majority of the objects (13/18) were observed with the $F160W$ filter of the WFC3/IR camera. The others were targeted with a filter close to the R-band, which also mostly probes the bulk of the stellar component for these sources at low redshift. Hence, it should not introduce any additional bias to our study.

    In \cite{kelly2014a}, GRB hosts are compared to a sample of star-forming galaxies from the Sloan Digital Sky Survey (\SDSS) DR10 catalog at $z < 0.2$. Because a non negligible fraction of our combined GRB host low-redshift sample is located at $z > 0.2$, we perform a new comparison by considering a control sample from the 3D-HST survey at similar redshift, applying the same analysis method as performed at \zmed\ and \zhigh. Although 3D-HST may be more suited to galaxies at larger distances (i.e., $z > 1$), we note that it still remains the optimal HST survey combining estimates of redshift, size, stellar mass and star formation rate for sources at low to intermediate redshifts. It represents therefore the best available control sample to study the densities of GRB hosts at \zlow. Our combined sample of GRB host galaxies at \zlow\ yields similar results to those obtained by \cite{kelly2014a}. We find indeed that the majority of GRB hosts are located above the SFR-weighted \sigmassMass\ and \sigsfrMass\ relations of field galaxies, while they fall below the \reMass\ relation driven by the control sample of star-forming sources. However, the apparent deviations that we measure are much less pronounced than those derived by \cite{kelly2014a}, which may be due to the different control sample used in their analysis.
    We compute the CDFs for each parameter and perform \ks\ tests. The results reveal that we can reject the null hypothesis with a significance level of $\lesssim$ 5\% for all parameters ($\pks = 0.01$, $\pks = 0.04$ and $\pks = 0.006$ for \deltare, \deltasigmass\ and \deltasigmass, respectively).

\begin{figure}
        \centering
        \includegraphics[width=\hsize]{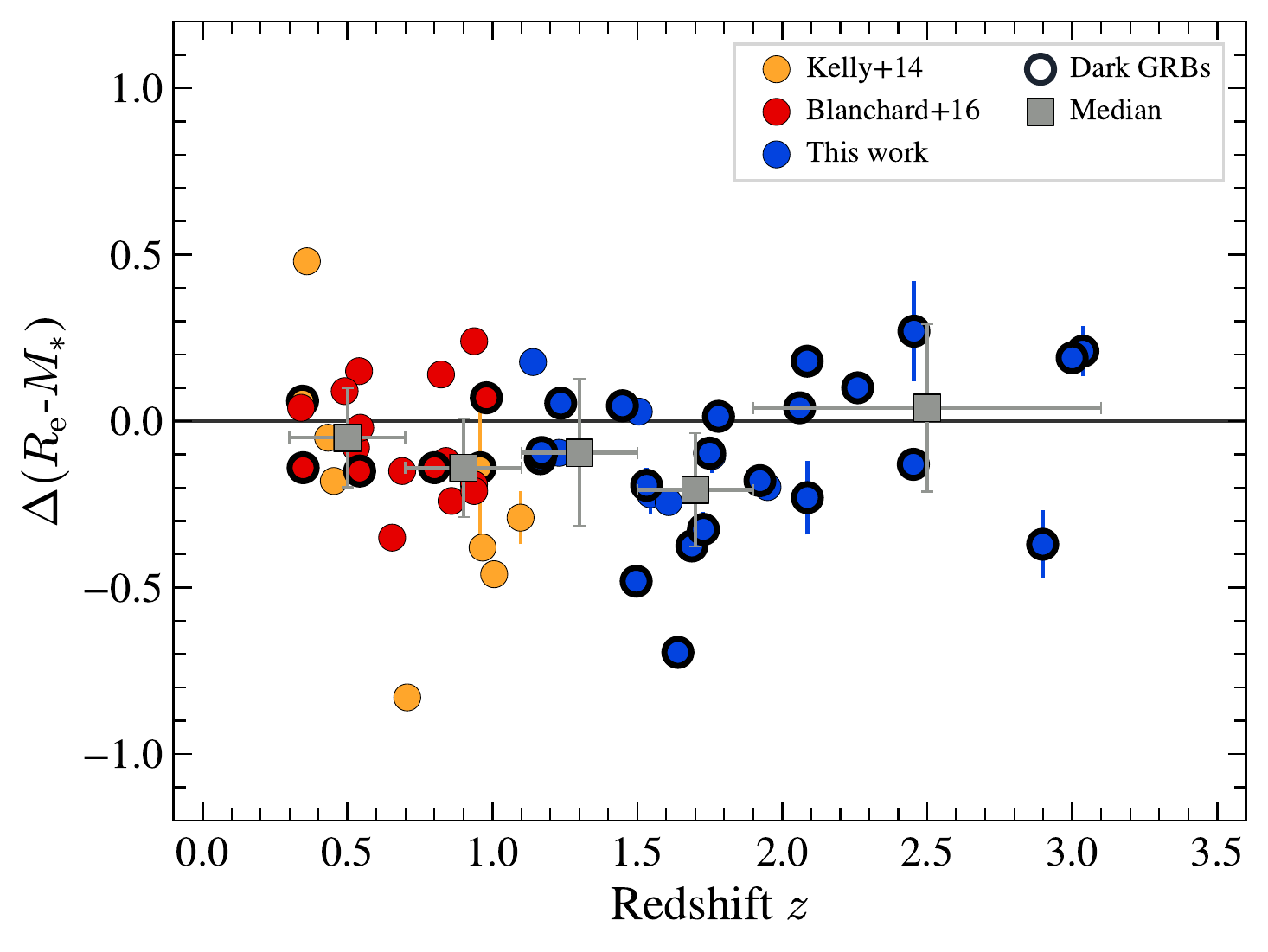}
        \caption{\deltare\ as a function of the redshift for GRB host galaxies. The orange and red circles are GRB hosts at $z \lesssim 1$ from \cite{kelly2014a} and \cite{blanchard2016a}, respectively. The blue circles are the GRB hosts considered in this study. Dark GRBs are highlighted by a thick black circle. The gray squares are the median of the \deltare\ for each redshift bin. The associated error bars are the standard deviation using the MAD estimator. The black line at $y=0$ represents the expected median values for a GRB hosts population that do not suffer from environment bias.}
        \label{fig:re_offset_redshift}
    \end{figure}

\begin{figure}
        \centering
            \includegraphics[width=\hsize]{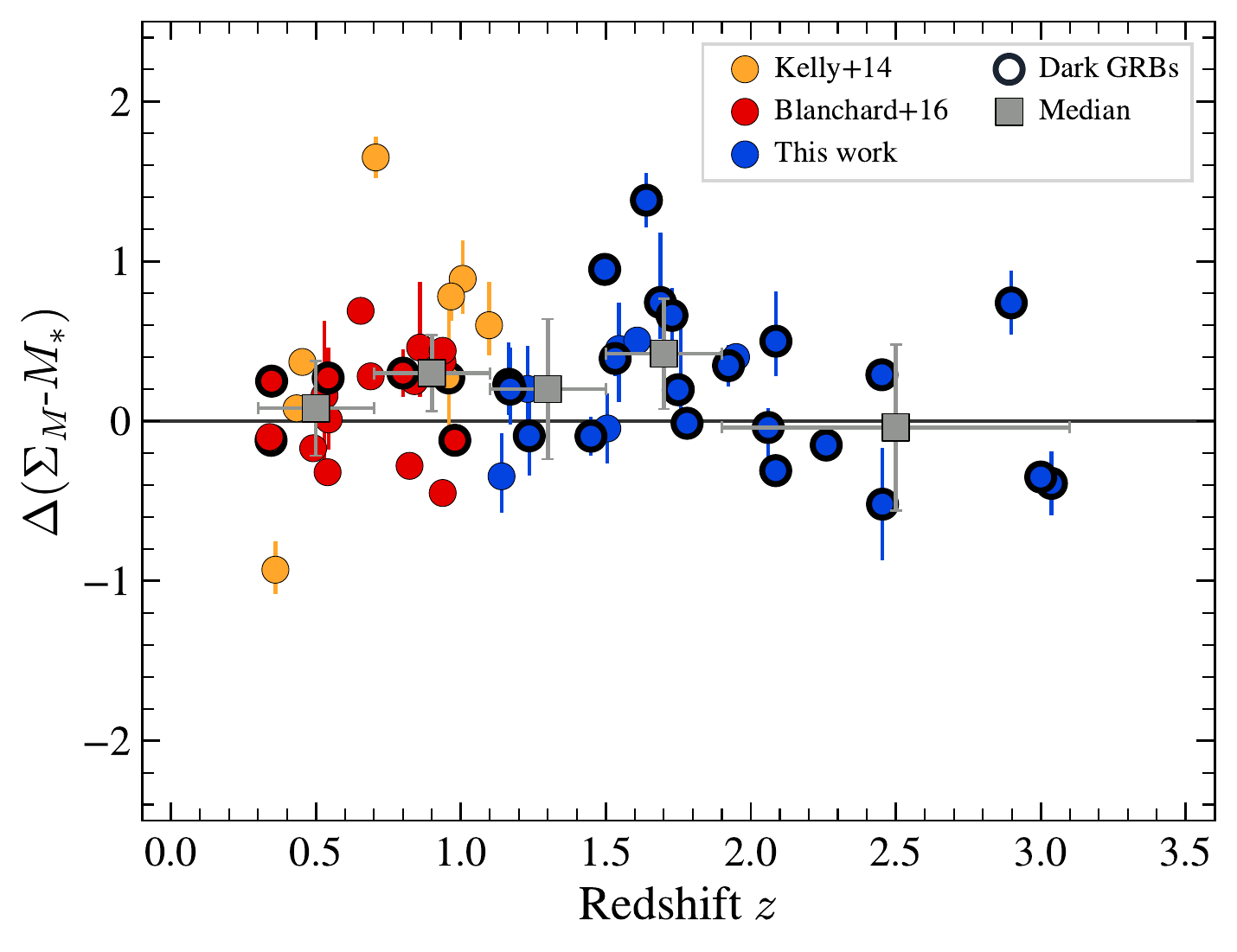}
        \caption{As in Fig.~\ref{fig:re_offset_redshift} but showing the \deltasigmass\ against redshift.}
        \label{fig:sigmass_offset_redshift}
    \end{figure}

\begin{figure}
        \centering
        \includegraphics[width=\hsize]{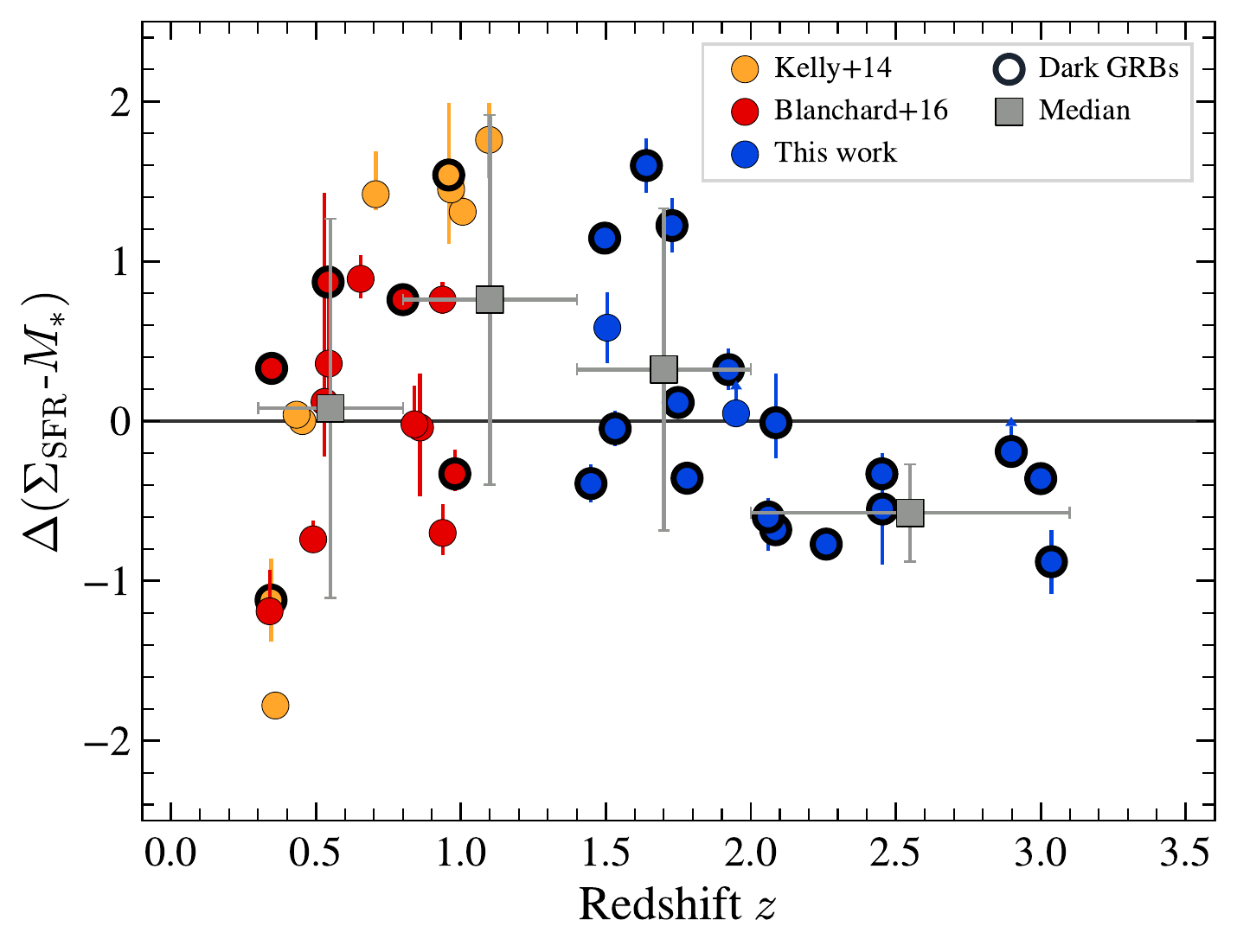}
        \caption{As in Fig.~\ref{fig:re_offset_redshift} but showing the \deltasigsfr\ against redshift.}
        \label{fig:sigsfr_offset_redshift}
    \end{figure}

    In Fig.~\ref{fig:re_offset_redshift}, we show the evolution of \deltare\ at $0.3 < z < 3.1$. We divide the redshift range in five bins and determine the corresponding median value for each redshift bin. At $z \lesssim 2$, we observe that the median \deltare\ is systematically negative, with an offset that appears statistically significant despite the relatively large scatter of the individual estimates. This shows again that the overall population of GRB host galaxies at $z \lesssim 2$ tend to exhibit smaller sizes than typical star-forming sources with comparable stellar masses.
    As already noticed in Sect.~\ref{subsec:Mass-size_relation} from our sample at $2 < z < 3.1$, the median value at $z \gtrsim 2$ is however closer to a null deviation, meaning that GRB hosts at higher redshifts tend to be more homogeneously spread around the SFR-weighted \reMass\ relation.
    Similarly, we find an opposite behavior for \deltasigmass\ (Fig.~\ref{fig:sigmass_offset_redshift}), which reveals a statistically significant positive deviation up to $z \sim 2$, and a median value consistent with zero at higher redshift.
    Finally, the \deltasigsfr\ is shown in Fig.~\ref{fig:sigsfr_offset_redshift}. We reduce the number of redshift bins due to the smaller number of GRB hosts with a SFR value. We find that GRB host galaxies may show a slight preference toward high star formation density at low to intermediate redshifts, but all median values are also consistent with a null deviation given the large associated uncertainties.

\section{Discussion}
\label{sec:Discussion}
    The non negligible scatter observed in Figs.~\ref{fig:re_offset_redshift}, \ref{fig:sigmass_offset_redshift} and \ref{fig:sigsfr_offset_redshift} unfortunately prevents deriving an unambiguous interpretation of our data. Yet, our analysis strongly suggests that GRB host galaxies up to $z \sim 2$ tend to exhibit smaller sizes and larger densities of stellar mass and star formation than what we could expect for a SFR-weighted population of star-forming galaxies. It thus confirms and extends to higher redshifts the trend already observed for GRB hosts at low $z$, even though the offset that we find between the two populations is not as prominent as previously measured \citep[e.g.,][]{kelly2014a}.
    We stress again that our comparison was performed by properly matching the GRB host sample and the population of field sources in redshift and stellar mass. This indicates that our results cannot be explained by the combination of the more frequent occurrence of long GRBs in low-mass sources (at least up to $z \sim 2$) with the general mass-size relationship and its evolution with cosmic time. Similarly, we showed that it cannot be due to systematic uncertainties in the determination of physical parameters (e.g., mass, size) between field galaxies and GRB hosts. Finally, we believe that this effect cannot be simply due to the GRB host sample selection and in particular the larger number of GRBs with a dark afterglow compared to the overall population of long GRBs. A higher number of dark GRBs may indeed bias the host sample toward larger and more massive sources \citep{perley2013a,chrimes2020a}, but this should not affect the comparison with field galaxies at fixed stellar mass. In addition, we do not observe any apparent difference between dark and optically-bright GRB hosts at \zmed, when quantifying their size, stellar mass and SFR density offset with the field (Figs.~\ref{fig:M-size_relation}, \ref{fig:M-sigma_mass_relation} and \ref{fig:M-sigma_sfr_relation}). This is also what we find with the sample at \zlow, that is composed of a majority of optically-bright GRBs and that shows a similar trend toward compact and dense environments. 

    \subsection{The effect of the Size–Metallicity relation on GRB hosts} 
    \label{sub:the_effect_of_the_size_metallicity_relation_on_GRB_hosts}
        Previous works led to the conclusion that the production efficiency of long GRBs is mostly ruled by metallicity, with GRB formation being switched off in galaxies with metallicity higher than a threshold that is still currently debated in the literature \citep{modjaz2008a,graham2013a,kruhler2015a,vergani2015a,perley2016b,palmerio2019a}. Given the additional bias toward compact galaxies found in our analysis for GRB hosts, we further investigate the possible link between the size and metallicity of galaxies. \cite{ellison2007a} used a sample of star-forming sources from the \SDSS\ to study the possible influence of the galaxy size on the mass-metallicity (MZ) relation \citep{tremonti2004a, mannucci2010a, wuyts2014a, zahid2014a}. They observed an anti-correlation between size and metallicity at a given stellar mass. This means that for the same stellar mass, galaxies with a smaller size are also metal richer (the metallicity increases by 0.1~dex when the size is divided by a factor of 2). This result has been corroborated by \cite{brisbin2012a, harwit2015a} and it was also observed at higher redshifts by \cite{yabe2012a, yabe2014a} using a sample of star-forming galaxies up to $z \sim 1.4$. The tendency for GRB hosts to occur in denser environments could thus appear intriguing at first sight. In fact, GRB-selected galaxies appear to track the mass–metallicity relation of star-forming sources but with an offset of 0.15~dex toward lower metallicities \citep{arabsalmani2018a}. If this offset is due to the intrinsic nature of GRB hosts and not to systematic effects, the possibility that GRB hosts are more compact (and not larger) than field galaxies may indicate that the physical conditions and the environments in which long GRBs form are more complex than what has been assumed so far. Our results could also imply that the impact of metallicity and compactness separately considered is even stronger than actually seen, and their inter-correlation at a given stellar mass mitigates the global trends that we can observe among GRB hosts.
        Interestingly, based on the EAGLE cosmological numerical simulations, \cite{almeida2018a} found a similar size-metallicity anti-correlation up to $z \sim 8$ and explored its possible physical origin. It is worth to mention that their simulation reproduced the well-known observed scale relations such as the MZ relation at $z < 5$ \citep{derossi2017a} but was not designed to reproduce the relation between size and metallicity. They explored three potential explanations of this relation: (1) a recent metal-poor gas inflows that increased the size and reduced the metallicity of the galaxy (2) a more efficient star formation process in compact galaxies, the denser gas transforms more efficiently into stars resulting in a faster enrichment of the gas (3) an effectiveness of metal-rich gas outflows reduced in compact galaxies due to a deeper gravitational potential. The EAGLE simulation supports the cause (1) and discarded the causes (2) and (3). In this scenario, we may infer that long GRBs cannot be linked to young star formation triggered by recent inflow of gas at low metallicity, which would have otherwise increased the size of their hosting environment.

        To investigate in more detail this possible relationship between metallicity and stellar density in our GRB host sample, we extracted from previous studies \citep{hashimoto2015a, kruhler2015a} the gas-phase metallicity measurement determined from strong emission lines (\Zemiss). We only consider GRB hosts in \zmed, where the trend toward compact galaxies is more clearly observed. We find only a small fraction of GRB hosts (10/22) with a metallicity measurement. We also find an additional GRB host with gas-phase metallicity measurement determined using the GRB afterglow absorption lines (\Zabs) in \cite{arabsalmani2018a}. However, due to the insecure relation between \Zemiss\ and \Zabs\ especially at high-$z$ \citep{metha2020a, metha2021a}, we omit this measurement. Unfortunately, our data do not reveal any obvious trend between the metallicity of GRB hosts and the deviation of their size from field galaxies at comparable stellar mass. This may be explained by the poor statistics of the sample, and therefore, we cannot confirm or rule out a different relation for GRB hosts compared to field galaxies regarding these physical parameters.

        To further complicate the picture, we finally point out that minor interactions could also play a role in shaping the inter-dependency of size and metallicity in these different populations. Using a local sample of star-forming galaxies from the \SDSS, \cite{ellison2008a} observed that galaxies with a companion have indeed a lower metallicity for a given stellar mass and size. Detailed studies of individual GRB host revealed that GRBs can be found in interacting environments \citep{thone2011a, arabsalmani2019a}. It is also supported by previous work \citep{wainwright2007a, orum2020a} reporting that GRB hosts are often found in interacting systems with major companions ($\sim 30 \%$). This may suggest that a recent interaction of the host galaxy could also affect the conditions required to produce a long GRB, in addition to metallicity and stellar density.
 

    \subsection{A redshift-dependent bias ?} 
    \label{sub:a_redshift_dependent_bias}
        At $z \gtrsim 2$, the picture arising from our sample is different than the one observed at lower redshifts. The apparent deviation that we found at $z \lesssim 2$ between GRB hosts and field galaxies seems to disappear.
        Admittedly, the large uncertainties measured at \zhigh\ and the much lower statistics characterizing our GRB host sample at such redshifts do not allow us to conclude if this evolution is real and statistically robust, as also suggested by the significance of our \ks\ tests. 
        The SFR density of GRB hosts at \zhigh\ may even be lower than expected from the field according to our analysis (see Fig.~\ref{fig:M-sigma_sfr_relation}), although we believe this reversal is probably due to the small number of sources in our sample. 
        However, the data do suggest that the size and stellar mass density of GRB hosts at these higher redshifts are globally more representative of the overall population of star-forming galaxies in the field. On a qualitative point of view, one could argue that this evolution of the size and density of GRB hosts compared to the field is consistent with the idea that the bias, which is clearly established between the overall population of star-forming galaxies and the hosts of long GRBs at low redshifts, is progressively reduced as the redshift increases. This may thus support the hypothesis that long GRBs represent a more accurate tracer of star formation in the distant Universe than they actually do at lower redshifts. 
        On the other hand, we note that the stellar mass range probed by our sample in this redshift range ($\mass > 10^{10.5}$) is substantially larger than the one probed at \zmed\ ($\mass < 10^{10.5}$), and our GRB host selection at $z \gtrsim 2$ is also exclusively drawn from dark GRBs. While we found no apparent difference in the size and stellar densities between the hosts of optically-bright and dark GRBs at lower stellar mass and redshift, we cannot firmly exclude a possible dependence of the size deviation with stellar mass. This means that the offset observed at $z \lesssim 2$ could plausibly remain at \zhigh, if we had also included GRB hosts with lower stellar mass ($\mass < 10^{10.5}$) or more massive hosts selected with optically-bright GRBs.

    \subsection{Stellar density and GRB progenitor models} 
    \label{sub:stellar_density_and_grb_progenitor_models}
        Considering the proposed progenitor models, GRB production may be expected to depend on the density environment in addition to any metallicity bias. Several observational and theoretical studies reported that the fraction of star formation happening in young bound star clusters ($\Gamma$) may depend on the environmental properties of the host galaxy. Especially, they found that the \sigsfr\ correlates with $\Gamma$ \citep{goddard2010a, adamo2011a, kruijssen2012a, silva-villa2013a, adamo2015b}. Owing to a larger amount of stars, these clusters may more frequently produce binary systems of massive stars which are one of the candidates to form GRBs.
        However, this has to be contrasted with the results of \cite{chandar2017a} \citep[see also][]{chandar2015a, kruijssen2016a}, which showed that the relation between \sigsfr\ and $\Gamma$ presents no particular trend. The previously reported correlation would be due to a bias in the selection of galaxies leading to an estimation of $\Gamma$ mixing young and old clusters. As a consequence, young (old) clusters were systematically associated with high (low) \sigsfr\ creating an apparent correlation between $\Gamma$ and \sigsfr.
        On the other hand, several studies suggest that the IMF can evolve to top-heavy (overabundance of high mass stars) when the density of the environment increases \citep{marks2012a, haghi2020a}. If the number of massive stars increases, the probability to produce a GRB also increases. This provides a plausible explanation for the reported trend that associates GRBs with compact and dense environments.

\section{Summary and conclusions}
\label{sec:Conclusions}
    In this work, we studied a sample of long GRB host galaxies observed with the HST/WFC3 in the IR band at $1 < z < 3.1$. We compared their sizes, stellar mass and star formation rate surface densities to the ones of typical star-forming galaxies of the 3D-HST survey.
    Prior to comparison, we minimized the systematics and biases that measurement methods may introduce between samples observed under different conditions. We also verified that no systematic offset is present between the GRB hosts and the star-forming galaxies in the determination of their physical properties.
    In addition, we confronted our GRB host sample to the host galaxies of unbiased GRB samples (\BATsix\ and \SHOALS). At \zmed, we found that they are consistent with each other in terms of stellar mass and redshift distributions while at \zhigh\ we noted an offset toward more massive galaxies.
    We performed a fair comparison between the GRB hosts and the field galaxies by fixing their redshift range and stellar mass to remove any dependency that the measured properties may have on these two parameters.
    At \zmed, the results clearly showed that GRB hosts are smaller in size and have higher stellar mass and star formation rate surface densities than expected if they were truly representative of the overall population of star-forming galaxies. We also noted that the galaxy size and stellar density are consistent for the dark and optically-bright GRB host populations. 
    At \zhigh, the trend appears to evolve and GRB hosts seem to be more consistent with star-forming galaxies of the field. We even found an inversion of the tendency for the \sigsfr\ parameter, where GRB hosts have a lower star formation rate surface density than field sources. However, because of the small sample size at this redshift, we cannot rule out the possibility of a purely statistical effect. Furthermore, we cannot exclude a possible bias in our results at $z \gtrsim 2$ due to the predominance of galaxies selected from dark GRBs.
    We inserted our results in a broader context and considered at \zlow\ the size measurements from \cite{kelly2014a} and \cite{blanchard2016a}. We performed a similar analysis to the one at $1 < z < 3.1$ and found that up to $z \sim 2$, GRB hosts have a smaller size and a higher stellar mass and star formation surface densities than field galaxies.
    Finally, we investigated the possible relation between the size and metallicity bias found in the GRB host population. However, due to the limited number of metallicity measurements available in the literature for the GRB hosts in our sample, we cannot confirm or refute the anti-correlation reported for star-forming galaxies in the literature between size and metallicity at a given stellar mass. \\
    These results are part of the effort to better understand the long GRB formation and their ability to trace the CSFRH, especially at high-redshift where the trend is still poorly constrained by observations.
    Future works will be to expand the GRB host sample to confirm the trend observed at $z < 3$ and extend the analysis at higher redshifts. The future \textit{SVOM} \citep[Space-based multi-band astronomical Variable Objects Monitor,][]{wei2016a} mission and its dedicated follow-up network will allow us to rapidly identify high-$z$ bursts candidate and will contribute to a better controlled and homogeneous GRB host sample. Its synergy with the upcoming \textit{JWST} \citep[James Webb Space Telescope,][]{gardner2006a} offer a promising opportunity to detect and characterize GRB host galaxies at very high redshift.

\begin{acknowledgements}
    We thank the anonymous referee for insightful comments and a careful reading of the manuscript. We additionally acknowledge A. van der Wel for providing access to the PSF model of 3D-HST/CANDELS mosaics and C. Peng and B. Haüssler for useful discussions about \galfit.
    The observations are based on observations made with the NASA/ESA Hubble Space Telescope, and obtained from the Hubble Legacy Archive, which is a collaboration between the Space Telescope Science Institute (STScI/NASA), the Space Telescope European Coordinating Facility (ST-ECF/ESA) and the Canadian Astronomy Data Centre (CADC/NRC/CSA).
    This work is based on observations taken by the CANDELS Multi-Cycle Treasury Program and the 3D-HST Treasury Program (GO 12177 and 12328) with the NASA/ESA HST, which is operated by the Association of Universities for Research in Astronomy, Inc., under NASA contract NAS5-26555. 
    This research made use of Astropy\footnote{http://www.astropy.org}, a community-developed core Python package for Astronomy \citep{astropycollaboration2013a, astropycollaboration2018a}, of matplotlib, a Python library for publication quality graphics \citep{hunter2007a}, of Astroquery \citep{ginsburg2019a} and of NumPy \citep{haghi2020a}.
\end{acknowledgements}

\bibliographystyle{aa}
\bibliography{main}

\begin{appendix}
\section{Stellar mass and star formation rate of GRB hosts and field galaxies}
\label{app:main_sequence_of_galaxies}
\begin{figure}[!ht]
        \includegraphics[width=\hsize]{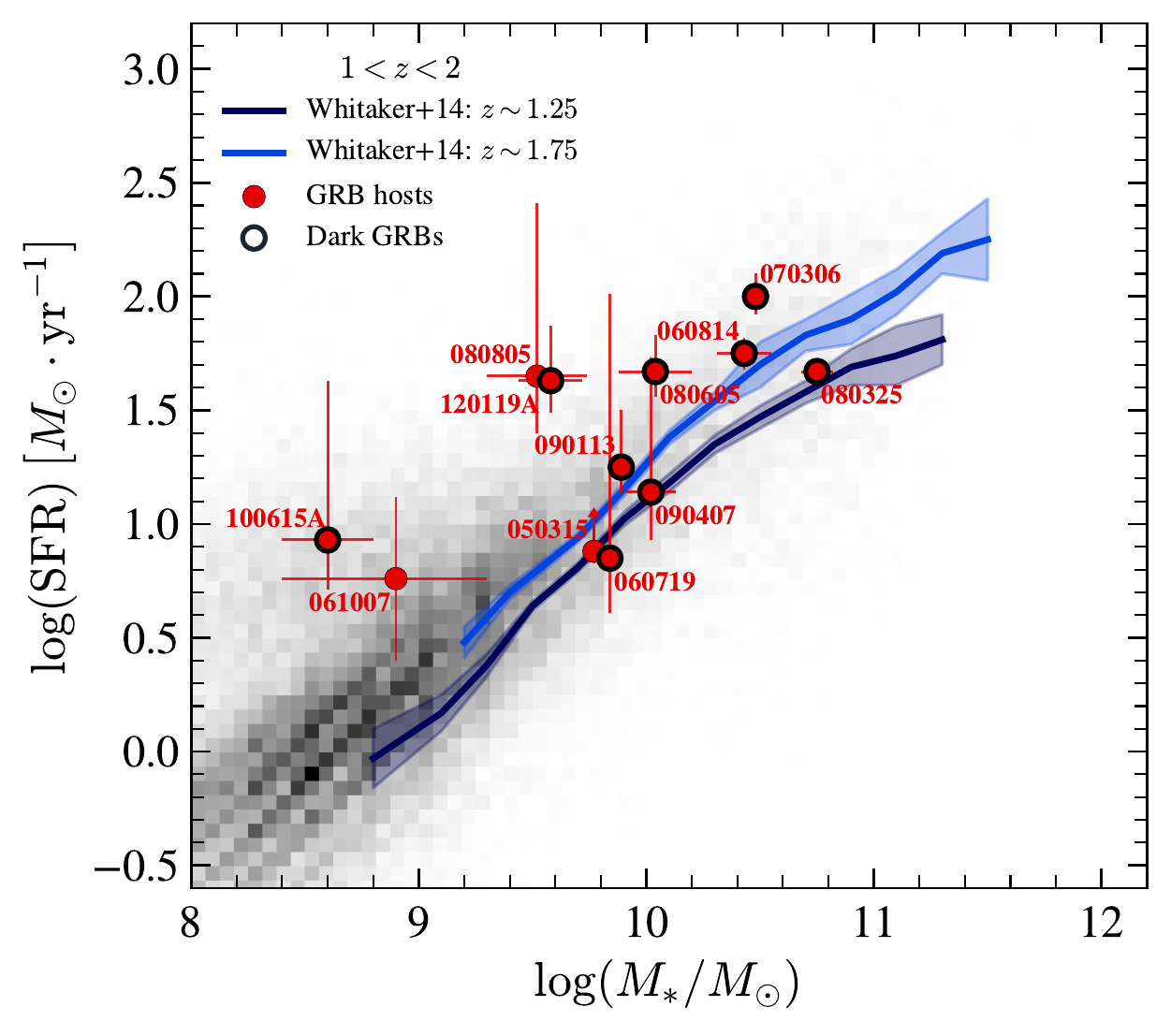}
        \includegraphics[width=\hsize]{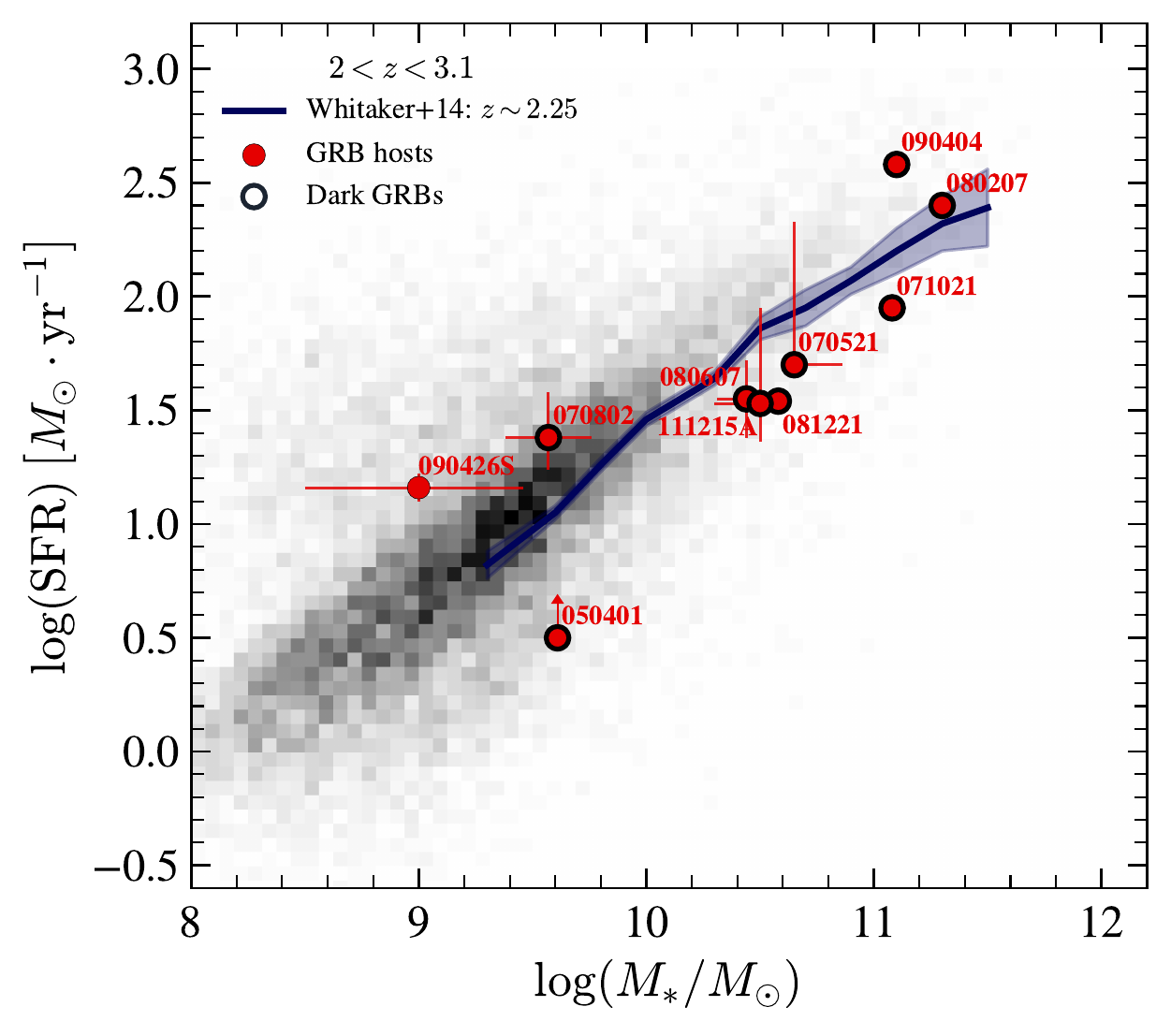}
    \caption{Star formation rate against stellar mass for GRB hosts and star-forming galaxies of the 3D-HST survey. Two redshift bins are considered, at \zmed\ (\textit{top panel}) and \zhigh\ (\textit{bottom panel}). The star-forming galaxies selected in the analysis combining \sfrtot\ and \sfruvcorr\ values are plotted in background as a 2D gray histogram. The blue and dark blue curves are the main sequence relations derived by \cite{whitaker2014a} at $1 < z < 2.5$.}
    \label{fig:main_sequence}
    \end{figure}

    For star-forming galaxies of the 3D-HST survey, we use the star formation rates (SFR) determined by \cite{whitaker2014a} and stellar masses derived from the SED fitting code \texttt{FAST} \citep{kriek2009a}.
    The \sfrtot\ of \cite{whitaker2014a} is determined by adding the rest-frame UV light (unobscured light produced by young stars) and the IR light (obscured and remitted light by dust). The total IR luminosity is estimated from the \Spitzer\ $24\ \mu$m flux density and the total UV luminosity is based on the 2800~\angstrom\ luminosity obtained from best stellar population models \citep[see][for additional details]{whitaker2014a}. \\
    For objects with no \Spitzer\ detection ($\mathrm{SNR} < 1$), we derive a UV-SFR corrected from dust extinction (\sfruvcorr). We extract for each object the observed UV luminosity at 1600~\angstrom\ available in the 3D-HST catalog. We then correct it from extinction by applying an attenuation factor $(A_{1600})$ derived from the rest-frame UV continuum slope $\beta$ and the \cite{meurer1999a} relation. Finally, we use the relation from \cite{kennicutt1998a} to convert the UV-corrected luminosity to a SFR. \\
    Because the GRB host properties used in this work were determined by SED fitting procedures differing from FAST, the impact of possible systematics arising from the different codes available in the literature should be properly considered. Fortunately, all these codes (including FAST) assume standard star formation histories (e.g., exponential declining, delayed star formation) and similar dust extinction laws, which should strongly limits the risk of large systematics. For instance, typical offsets of only $0.2-0.3$~dex for the stellar mass estimates were found from one code to another \citep{pforr2012a,mobasher2015a}. For SFR determinations, it is generally acknowledged that larger scatter can be seen when relying on SED fitting at optical and NIR wavelengths \citep{pacifici2015a,carnall2019a}. However, we favored as much as possible the use of more accurate SFR estimates, relying on determinations either based on mid-infrared photometry for the 3D-HST catalog, or using H$\alpha$ and submillimeter fluxes for the majority of GRB host galaxies. We thus believe that our comparisons between GRB hosts and field galaxies should not be hardly affected by these effects.
    In addition, the slight slope of the galaxy mass-size relation should strongly limit the impact of a systematic offset between stellar mass values on our main conclusions. \\
    In Fig.~\ref{fig:main_sequence}, we show the sample of star-forming galaxies used in the analysis as a gray-scale density plot and the GRB host galaxies as red circles at \zmed\ and \zhigh. We also overlay the main sequence (MS) of star-forming galaxies at $z \sim 1.25$, $z \sim 1.75$ and $z \sim 2.25$ from \cite{whitaker2014a}. We find good agreement between the 2D background histogram and the MS relations. This confirms that the sample of star-forming galaxies considered, combining \sfrtot\ and \sfruvcorr\ follows the trend of the MS. Finally, we note that the majority of the GRB hosts are in the typical $\sim 0.3$~dex scatter of the MS and follow its trend at both redshifts.

\section{Comparison between methods to estimate the structural parameters}
\label{app:3D-HST_comparison}

    \begin{figure}[!t]
        \includegraphics[width=\hsize]{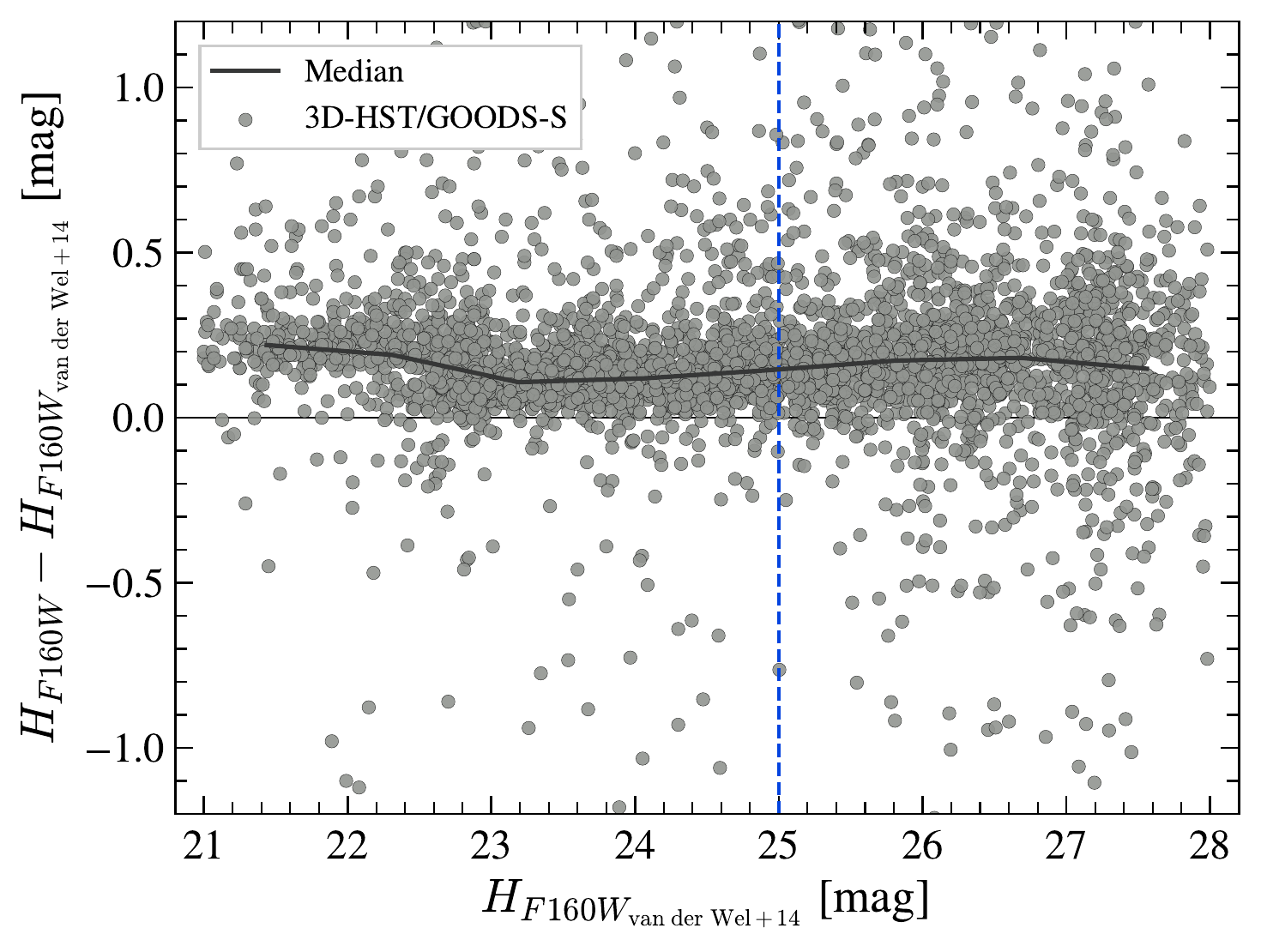}
        \includegraphics[width=\hsize]{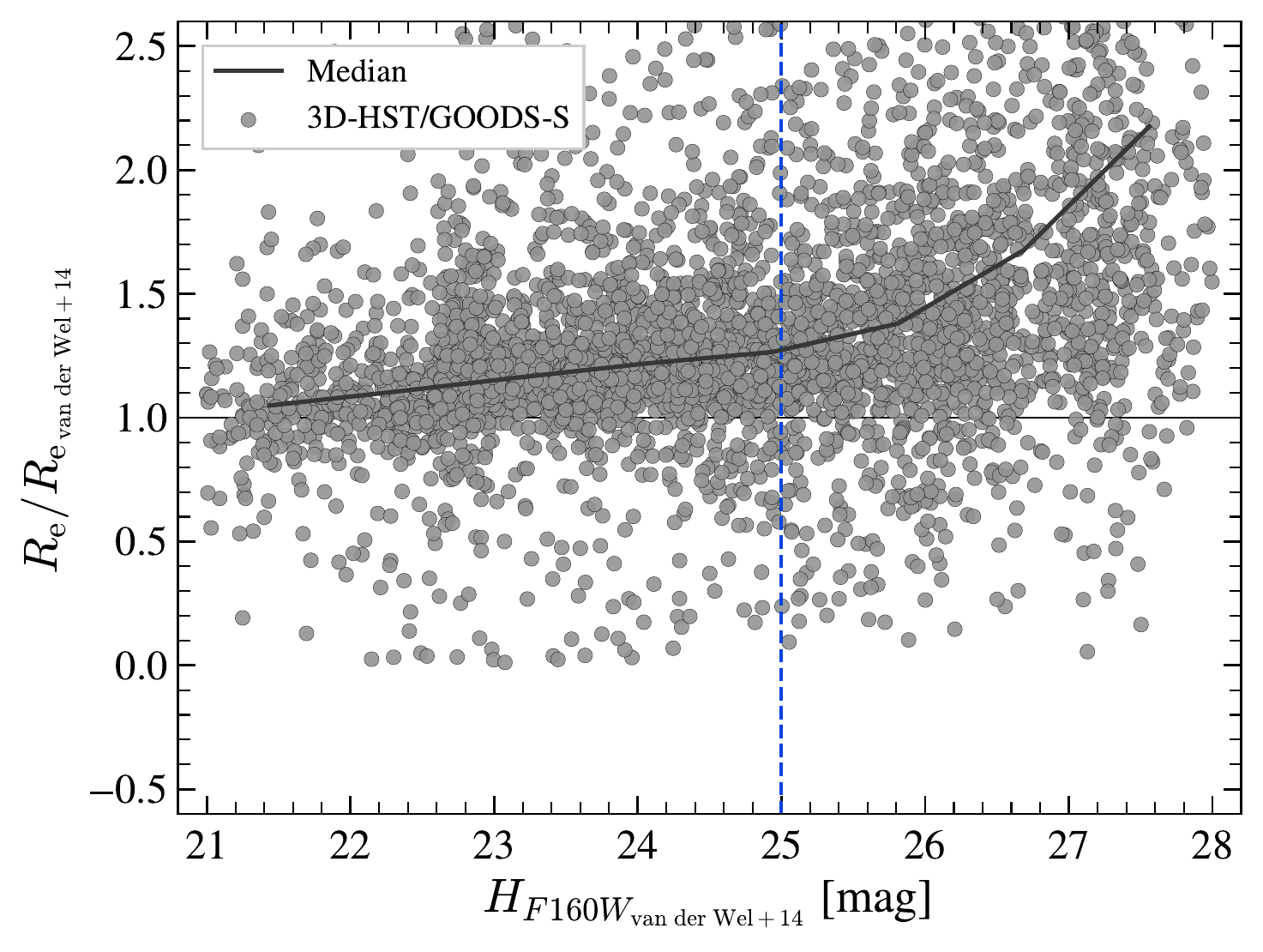}
    \caption{Comparison between $F160W$ magnitudes (\textit{top panel}) and half-light radii (\textit{bottom panel}) derived by \cite{vanderwel2014a} and our \galfit\ modeling as a function of \cite{vanderwel2014a} magnitudes. Each gray point represents an object of the 3D-HST/GOODS-S field. The gray curve is the median of the points and symbolizes the systematic offset for each parameter between the two methods. The vertical dashed blue line represents the maximum magnitude reached by GRB hosts above the 3D-HST mass-completeness limit, while our median H-band magnitude reaches $\sim 23.1$ mag.}
    \label{fig:comparison_3D-HST}
    \end{figure}

     We verify that our method of measuring GRB hosts structural parameters is consistent with the one of \cite{vanderwel2014a} used for the reference sample. This ensures that no systematic bias is present when comparing GRB hosts and 3D-HST field galaxies.
     From the catalog of \cite{vanderwel2014a}, we consider all objects in the 3D-HST/GOODS-S field with a good fit ($flag = 0$). Among them, we randomly select $\sim 4 \ 000$ objects with $21 < H_{F160W} < 28$. We then run our pipeline in a similar manner as described in Sect.~\ref{subsub:profile_fitting}.
     The pipeline failed for about 100 objects. The majority (75\%) of them are not detected by our \sx\ configuration. The source extraction method used by \cite{vanderwel2014a} is based on the ‘hot’ and ‘cold’ modes developed in \galapagos. It is optimized to extract faint sources and properly deblended bright sources in mosaics. The undetected object are probably faint galaxies captured with the optimized source extraction algorithm of \galapagos. \\
     In Fig.~\ref{fig:comparison_3D-HST}, we show the results for the half-light radius and the magnitude.
     We find a systematic offset of only $\sim 0.1$~mag between the two magnitude estimates, and note that we also tend to systematically overestimate the half-light radius with our own procedure. At H$_{F160W}=21.5$~mag, we recover \re\ values within 10\%, and the median offset then increases progressively with the magnitude until reaching 50\% at H$_{F160W}=26$~mag. This behavior is not surprising since the accuracy of the fitting process depends to first order on the S/N and that the uncertainty in the background estimate becomes dominant at H$_{F160W}>25.5$~mag \citep{vanderwel2012a}. However, the large majority (90\%) of our GRB host sample is brighter than H$_{F160W}=24$~mag, i.e., where our fitting method reveals consistent results with the one used to estimate the \re\ in the 3D-HST catalog. We have only 4 objects with a magnitude higher than 25 that are excluded once the 3D-HST mass-completeness limit is applied. We thus conclude that these small offsets should have a negligible impact on our results.
     Finally in Sect.~\ref{sec:Results}, we emphasize that the sizes of GRB hosts measured with our code are globally smaller than the size of field sources from the 3D-HST catalog. Removing the systematic effect observed in Fig.~\ref{fig:comparison_3D-HST} would thus make this difference between the two populations even more significant, since our size determination tends to overestimate the sizes constrained by \cite{vanderwel2014a}.

\section{\galfit\ models of GRB hosts}
\label{app:galfit_models}
    In Figs.~\ref{fig:galfit_models_1}, \ref{fig:galfit_models_2} and \ref{fig:galfit_models_3}, we show the best-fitting Sérsic profile derived by \galfit\ for the sample of GRB hosts. In several cases (e.g., GRB~060814), multiple objects are fit simultaneously to reduce their contamination and improve the fitting process. Masked objects from the \sx\ segmentation map are visible as black areas. For the majority of cases, the residual maps show that we manage to remove the flux of the target object. To get a more quantitative indication of the fit goodness, a common method is to use the reduced $\chi^2$ returned by \galfit. However, it can easily be misleading if the fitting process of nearby objects or the masking process is not properly performed. In order to have a better estimate of the remaining signal for each target object, we determine the fraction of pixels within $2 \times \re$ that have a residual greater than three times the sigma map in the same area. The majority of the objects (33/35) have a residual fraction $\lesssim 5 \%$ confirming the quality of the models considered. Only two objects (GRB~070306 and GRB~080605) have a residual fraction of $\sim 30 \%$. Regarding GRB~080605, a plausible explanation is that the two nearby and bright stars probably contaminate the target object and thus interfere with the fitting process. For GRB~070306, the HST observations were performed several years after the GRB detection, thus excluding a possible contamination of the GRB afterglow that could affect the fitting process. We note that adding a PSF model in addition to the Sérsic model improves the fitting process and reduces the residuals to $\sim 10 \%$. The resulting size determined by \galfit\ evolves by a factor of 2 (from 0.14" to 0.28"). This may indicate an obscured active galactic nucleus or a recent burst of star formation in the host galaxy. Because these two objects represent only a small fraction of the total sample and the models appear realistic despite the large residual fraction, `we include them in our analysis.'
    Finally, several objects (e.g., GRB~080207, GRB~111215A) would require more components to improve the residual maps. However, due to the constraint of using a single Sérsic component to model the objects, we do not add additional components to improve the residual maps. 

    \begin{figure*}
        \begin{subfigure}[ht!]{0.5\hsize}
            \centering
            \includegraphics[width=0.98\hsize]{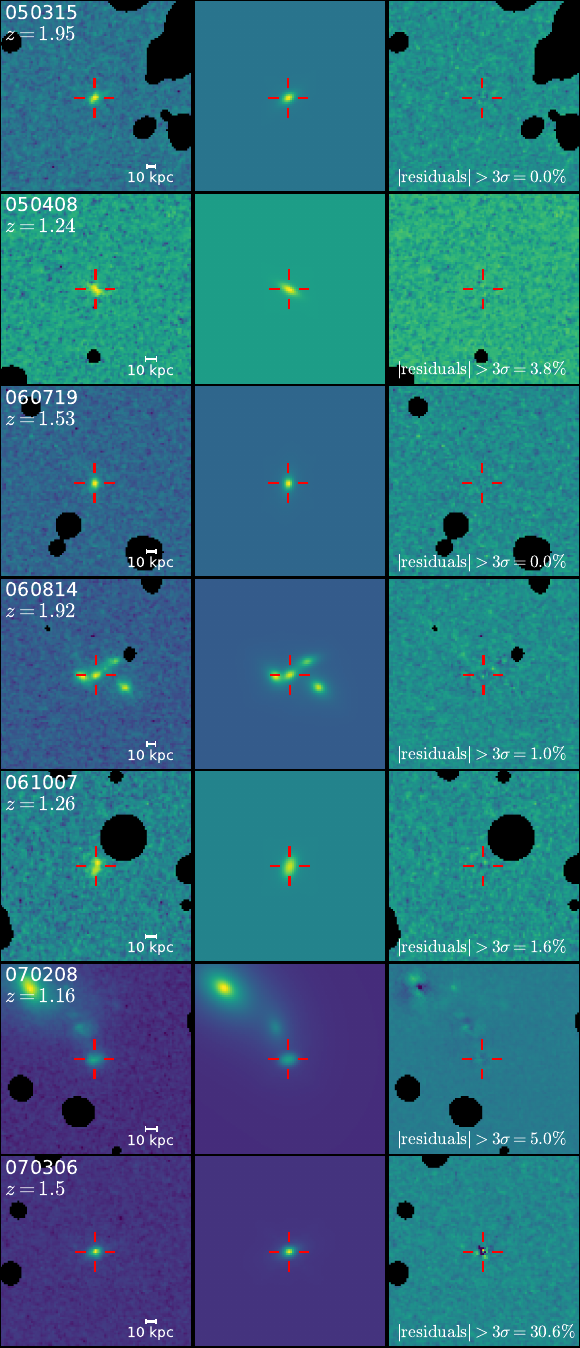}
        \end{subfigure}
        \hfill
        \begin{subfigure}[ht!]{0.5\hsize}
            \centering
            \includegraphics[width=0.98\hsize]{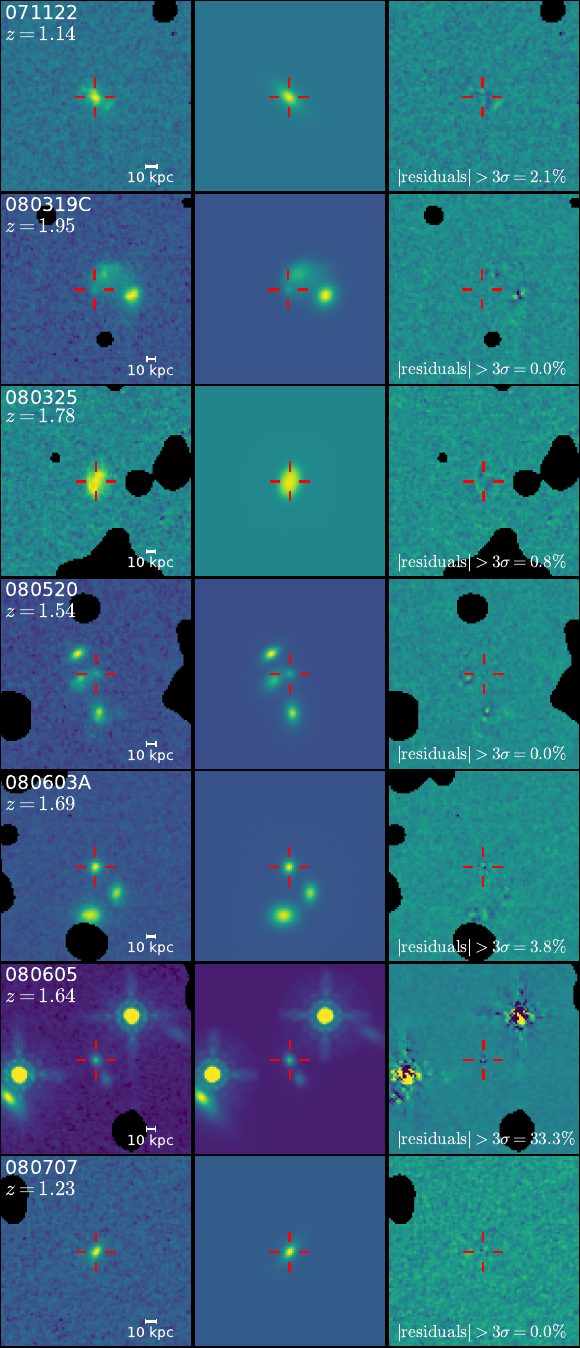}
        \end{subfigure}
        \caption{GRB hosts images from WFC3/$F160W$ observations (\textit{left}), their best \galfit\ models (\textit{middle}) and the residual maps (\textit{right}). Images are centered on the best positions of the host galaxies determined by \galfit\ and corresponds to a square region of 9 arcsec, where North is up and East is to the left. The red marks emphasize the objects considered as the GRB host galaxies. The black regions are the objects masked during the fitting processes based on the \sx\ segmentation map. The fraction of pixels with a residual greater than three times their noise within $2 \times \re$ of the target object is visible in the lower part of the residual map.}
    \label{fig:galfit_models_1}
    \end{figure*}

    \begin{figure*}[t]
        \begin{subfigure}[ht!]{0.5\hsize}
        \centering
            \includegraphics[width=0.98\hsize]{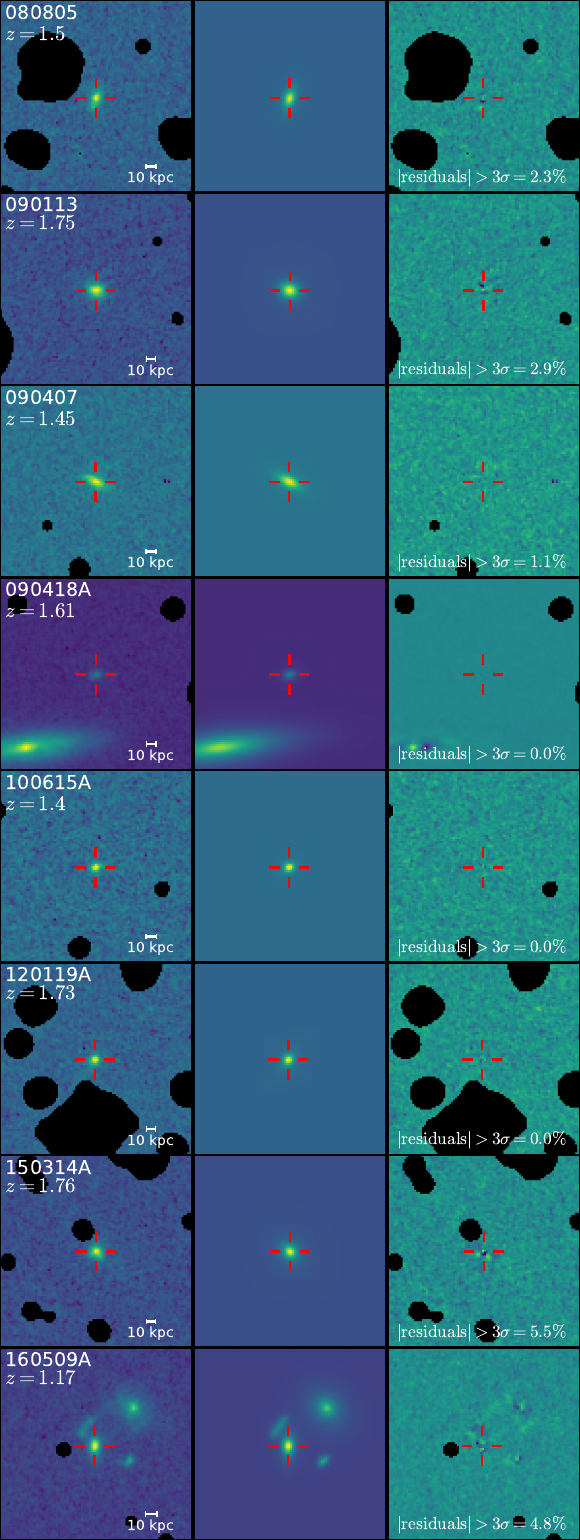}
        \end{subfigure}
    \hfill
        \begin{subfigure}[ht!]{0.5\hsize}
        \centering
            \includegraphics[width=0.98\hsize]{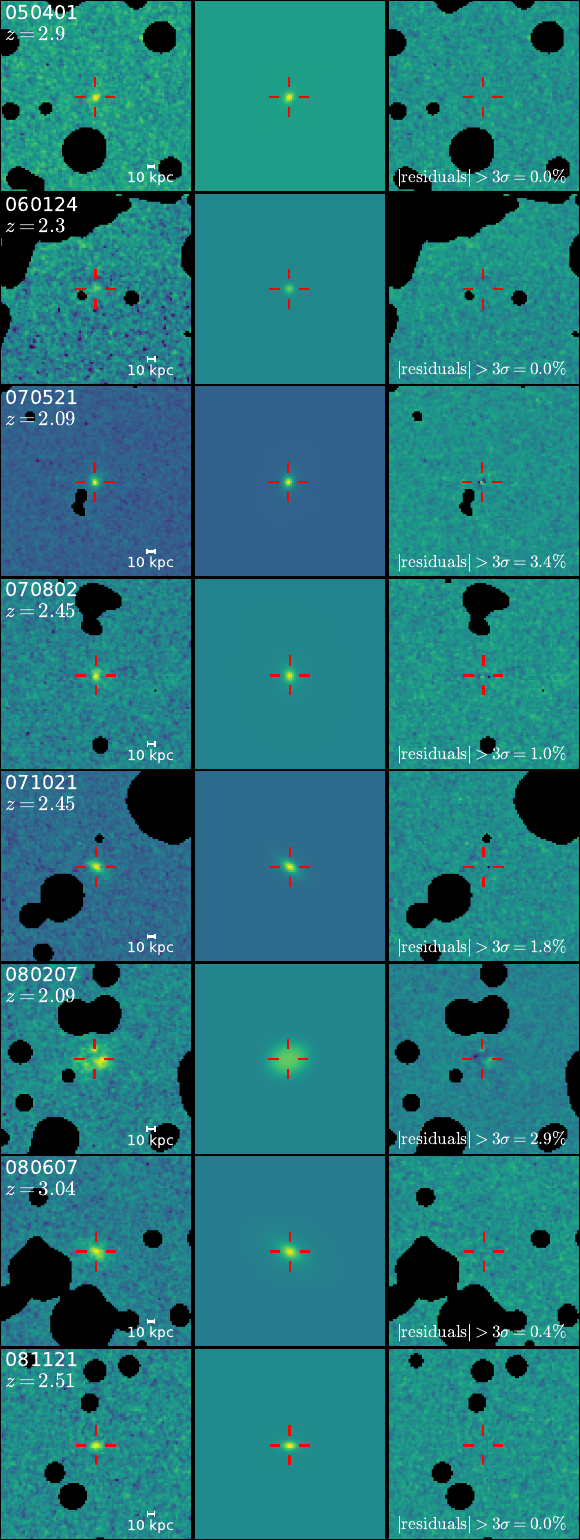}
        \end{subfigure}
    \caption{(Continued)}
    \label{fig:galfit_models_2}
    \end{figure*}

    \begin{figure}[t]
        \centering
        \includegraphics[width=0.98\hsize]{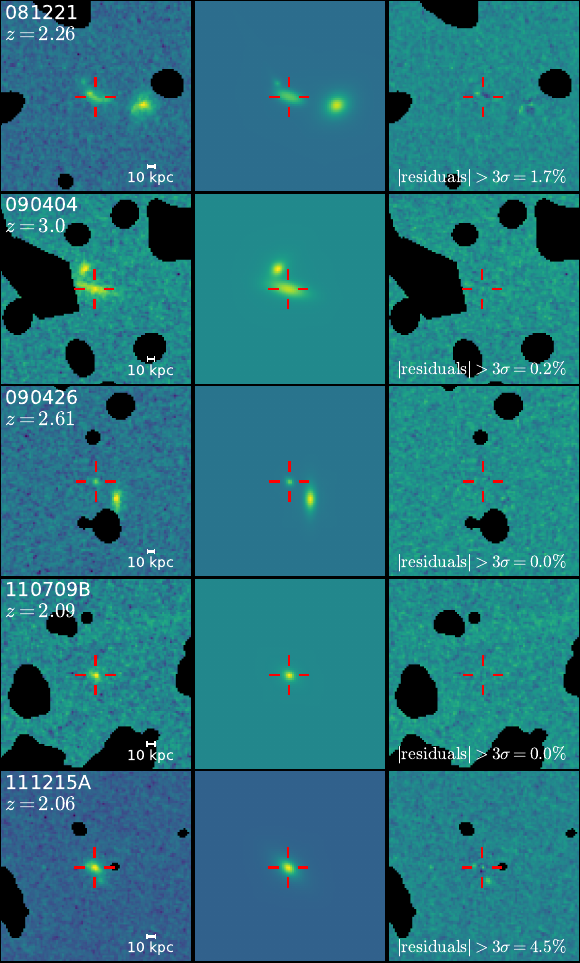}
        \caption{(Continued)}
    \label{fig:galfit_models_3}
    \end{figure}

\end{appendix}
\end{document}